\documentclass[12pt,a4paper]{article}
\usepackage{amssymb,enumitem,booktabs,subfig,xcolor,todonotes,microtype,setspace}
\usepackage[top=1.4in, bottom=1.4in, left=1.35in, right=1.35in]{geometry}
\usepackage{natbib}
\usepackage{url}
\usepackage[pdftex,colorlinks=true,hypertexnames=false]{hyperref}
\usepackage{amsmath}
\usepackage{amsfonts}
\usepackage{epsfig}
\usepackage{graphics}
\usepackage{palatino,eulervm}
\usepackage{graphicx}
\usepackage{lscape}
\usepackage{rotating}

\definecolor{darkblue}{rgb}{0,0,.6}
\hypersetup{citecolor=darkblue,linkcolor=darkblue,urlcolor=darkblue}
\allowdisplaybreaks[4]
\DeclareMathOperator*{\argmin}{arg\,min}

\providecommand{\U}[1]{\protect\rule{.1in}{.1in}}

\graphicspath{{plots/}}

\usepackage{mathrsfs}
\usepackage{amsthm,thmtools}
\declaretheorem{theorem}
\declaretheorem{lemma}
\makeatletter
\def\th@newremark{\th@remark\thm@headfont{\bfseries}}
\makeatletter
\theoremstyle{newremark}
\newtheorem{remark}{Remark}
\newtheorem{prop}{Proposition}

\newtheorem{assumption}{Assumption}
\declaretheoremstyle[
  spaceabove=6pt, spacebelow=6pt,
  headfont=\bfseries,
  notefont=\mdseries, notebraces={(}{)},
bodyfont=\normalfont,
  postheadspace=0.5em,
]{mystyle}

\begin{document}

\title{{\LARGE Nonparametric Estimation of Large Spot Volatility Matrices
for High-Frequency Financial Data}}
\author{{\normalsize Ruijun Bu\thanks{%
Management School, University of Liverpool, UK.},\ \ \ Degui Li\thanks{%
Department of Mathematics, University of York, UK. },\ \ \ Oliver Linton%
\thanks{%
Faculty of Economics, University of Cambridge, Cambridge, UK. The corresponding author, \url{obl20@cam.ac.uk}.},\ \ Hanchao
Wang\thanks{%
Zhongtai Securities Institute for Financial Studies, Shandong University,
China.}}\\
{\normalsize\em University of Liverpool,\ \ University of York,\ \ University of Cambridge,\ \ Shandong University}
}
\date{{\normalsize This version: \today}}
\maketitle

\centerline{\bf Abstract}

In this paper, we consider estimating spot/instantaneous volatility matrices
of high-frequency data collected for a large number of assets. We first
combine classic nonparametric kernel-based smoothing with a generalised
shrinkage technique in the matrix estimation for noise-free data under a
uniform sparsity assumption, a natural extension of the approximate sparsity
commonly used in the literature. The uniform consistency property is derived
for the proposed spot volatility matrix estimator with convergence rates
comparable to the optimal minimax one. For the high-frequency data
contaminated by microstructure noise, we introduce a localised
pre-averaging estimation method that reduces the effective magnitude of the noise. 
We then use the estimation tool developed in the noise-free scenario, and derive the uniform
convergence rates for the developed spot volatility matrix estimator. We further combine the kernel smoothing with the shrinkage technique to estimate the time-varying volatility matrix of the high-dimensional noise
vector. In addition, we consider large spot volatility matrix estimation in time-varying factor models with observable risk factors and derive the uniform convergence property. We provide numerical studies including simulation and empirical application to examine the performance of the proposed estimation methods in finite samples.

\smallskip


\newpage


\section{Introduction}

\label{sec1} \renewcommand{\theequation}{1.\arabic{equation}} %
\setcounter{equation}{0}

Modelling high-frequency financial data is one of the most important topics
in financial economics and has received increasing attention in recent
decades. Continuous-time econometric models such as the It\^{o}
semimartingale are often employed in the high-frequency data analysis. One
of the main components in these models is the volatility function or matrix. In the low-dimensional setting
(with a single or a small number of assets), the realised volatility is often used to estimate the integrated volatility over
a fixed time period \citep[e.g.,][]{AB98, BS02, BS04, ABDL03}. In practice,
it is not uncommon that the high-frequency financial data are contaminated
by the market microstructure noise, which leads to biased realised
volatility if the noise is ignored. Hence,
various modification techniques such as the two-scale, pre-averaging and
realised kernel have been introduced to account for the microstructure noise
and produce consistent volatility estimation \citep[e.g.,][]{ZMA05, BHLS08, KL08, JLMPV09, PV09,
CKP10, PHL16}. \cite{S05}, \cite{ABD10} and \cite{AJ14} provide
comprehensive reviews for estimating volatility with high-frequency
financial data under various settings.

\smallskip

In practical applications, financial economists often have to deal with the situation
that there are a large amount of high-frequency financial data collected for
a large number of assets. A key issue is to estimate the large volatility
structure for these assets, which has applications in various areas such as
the optimal portfolio choice and risk management. Partly motivated by 
developments in large covariance matrix estimation for low-frequency data in
the statistical literature, \cite{WZ10}, \cite{TWZ13} and \cite{KWZ16}
estimate the large volatility matrix under an approximate sparsity
assumption \citep[]{BL08}; \cite{ZL11} and \cite{XZ18} study large
volatility matrix estimation using the large-dimensional random matrix
theory \citep[]{BS10}; and \cite{LF18} propose a nonparametric
eigenvalue-regularised integrated covariance matrix for high-dimensional
asset returns. Given that there often exists co-movement between a large
number of assets and the co-movement is driven by some risk factors which
can be either observable or latent, \cite{FFX16}, \cite{AX17}, \cite{DLX19}
extend the methodologies developed by \cite{FLM11, FLM13} to estimate the large volatility matrix by imposing a
continuous-time factor model structure on the high-dimensional and
high-frequency financial data, and \cite{AX19} study the principal component
analysis of high-frequency data and derive the asymptotic distribution for
the realised eigenvalues, eigenvectors and principal
components.

\smallskip

The estimation methodologies in the aforementioned literature often rely on the
realised volatility (or covariance) matrices, measuring the integrated
volatility structure over a fixed time interval. In practice, it is often
interesting to further explore the actual spot/instantaneous volatility
structure and its dynamic change over a certain time interval, which is a
particularly important measurement for the financial assets when the market
is in a volatile period (say, the global financial crisis or COVID-19
outbreak). For a single financial asset, \cite{FW08} and \cite{K10}
introduce a kernel-based nonparametric method to estimate the spot
volatility function and establish its asymptotic properties including the
point-wise and global asymptotic distribution theory and uniform
consistency. For the noise-contaminated high-frequency data, \cite{ZB14}
combine the two-scale realised volatility with the kernel-weighted technique
to estimate the spot volatility, whereas \cite{KK16} propose a
kernel-weighted pre-averaging spot volatility estimation method. Other
nonparametric spot volatility estimation methods can be found in \cite{FFL07}
and \cite{FL20}. It seems straightforward to extend this local
nonparametric method to estimate the spot volatility matrix for a small
number of assets. However, a further extension to the setting with vast
financial assets is non-trivial. There is virtually no work on estimating
the large spot volatility matrix except \cite{K18}, which considers estimating
large spot volatility matrices and their integrated versions under the
continuous-time factor model structure for noise-free high-frequency data.

\smallskip

The main methodological and theoretical contributions of this paper are summarised as follows.

\begin{itemize}

\item {\em Large spot volatility matrix estimation with noise-free high-frequency data}. We use the nonparametric kernel-based smoothing method to estimate the volatility and co-volatility
functions as in \cite{FW08} and \cite{K10}, and then apply a generalised
shrinkage to off-diagonal estimated entries. With small off-diagonal entries
forced to be zeros, the resulting large spot volatility matrix estimate
would be non-degenerate with stable performance in finite samples. We derive the consistency property for the proposed spot volatility matrix estimator uniformly over the entire time interval under a uniform sparsity
assumption, which is also adopted by \cite{CXW13}, \cite{CL16} and \cite%
{CLL19} in the low-frequency data setting. In particular, the derived
uniform convergence rates are comparable to the optimal minimax rate in
large covariance matrix estimation \citep[e.g.,][]{CZ12}. The number of
assets is allowed to be ultra large in the sense that it can grow at an
exponential rate of $1/\Delta$ with $\Delta$ being the sampling frequency.

\item {\em Large spot volatility matrix estimation with noise-contaminated
high-frequency data and time-varying noise volatility matrix estimation}. When the high-frequency data are contaminated by the microstructure noise, we extend \cite{KK16}'s localised pre-averaging estimation method to the high-dimensional data setting. Specifically, we first pre-average the log price data via a kernel
filter and then apply the same estimation method to the kernel fitted
high-frequency data (at pseudo-sampling time points) as in the noise-free
scenario. The microstructure noise vector is assumed to be heteroskedastic with the time-varying covariance matrix satisfying the uniform sparsity assumption. We show that the existence of microstructure noises slows down the uniform convergence rates, see Theorem \ref{thm:2}. Furthermore, we combine the kernel
smoothing with generalised shrinkage to estimate the time-varying noise
volatility matrix and derive its uniform convergence property. To the best of our knowledge, there is virtually no  work on large time-varying noise volatility matrix estimation for high-frequency data.

\item {\em Large spot volatility matrix estimation with risk factors}. Since the uniform sparsity assumption is often too restrictive, we relax this restriction in Section \ref{sec4} and consider large spot volatility matrix estimation in the time-varying factor model at high frequency, i.e., a large number of asset prices are driven by a small number of observable common factors. By imposing the sparsity restriction on the spot idiosyncratic volatility matrix, we obtain the so-called ``low-rank plus sparse" spot volatility structure. A similar structure (with constant betas) is adopted by \cite{FFX16} and \cite{DLX19} in estimation of large integrated volatility matrices. We use the kernel smoothing method to estimate the spot volatility and covariance of the observed asset prices and factors as well as the time-varying betas, and apply the shrinkage technique to the estimated spot idiosyncratic volatility matrix. We derive the uniform convergence property of the developed matrix estimates, partly extending the point-wise convergence property in \cite{K18}. The developed methodology and theory can be further modified to tackle the noise-contaminated high-frequency data.

\end{itemize}

We argue that all three of the scenarios we consider above may be practically relevant. Microstructure noise is considered important in the very highest frequency of data,
whereas researchers working with five minute data, say, often ignore the noise. For the lower frequency of data there is a lot comovement in returns and the factor model is
designed to capture that comovement, whereas at the ultra high frequency, comovement is less of an issue; indeed, under the so-called Epps effect this comovement shrinks to zero
with sampling frequency.

\smallskip

The rest of the paper is organised as follows. In Section \ref{sec2}, we
estimate the large spot volatility matrix in the noise-free high-frequency
data setting and give the uniform consistency property. In Section \ref{sec3}%
, we extend the methodology and theory to the noise-contaminated data
setting and further estimate the time-varying noise volatility matrix. Section \ref%
{sec4} considers the large spot volatility matrix with systematic factors. Section \ref{sec5} reports the
simulation studies and Section \ref{sec6} provides an empirical application. Section \ref{sec7} concludes the paper. Proofs of the main theoretical results are available in Appendix A. The supplementary document contains proofs of some technical lemmas and propositions and discussions on the spot precision matrix estimation and the asynchronicity issue. Throughout the
paper, we let $\Vert\cdot\Vert_2$ be the Euclidean norm of a vector; and for
a $d\times d$ matrix ${\mathbf{A}}=(A_{ij})_{d\times d}$, we let $\Vert {%
\mathbf{A}}\Vert$ and $\Vert {\mathbf A}\Vert_F$ be the matrix spectral norm and Frobenius norm, $\vert{\mathbf{A}}%
\vert_1=\sum_{i=1}^d\sum_{i=1}^d |A_{ij}|$, $\Vert{\mathbf{A}}%
\Vert_1=\max_{1\leq j\leq d}\sum_{i=1}^d |A_{ij}|$, $\Vert{\mathbf{A}}%
\Vert_{\infty,q}=\max_{1\leq i\leq d}\sum_{j=1}^d |A_{ij}|^q$ and $\Vert {%
\mathbf{A}}\Vert_{\max}=\max_{1\leq i\leq d}\max_{1\leq j\leq d} |A_{ij}|$.


\section{Estimation with noise-free data}

\label{sec2} \renewcommand{\theequation}{2.\arabic{equation}} %
\setcounter{equation}{0}

Suppose that ${\mathbf{X}}_t=\left(X_{1,t},\cdots,X_{p,t}\right)^{^\intercal}
$ is a $p$-variate Brownian semi-martingale solving the following stochastic
differential equation: 
\begin{equation}  \label{eq2.1}
d{\mathbf{X}}_t={\boldsymbol{\mu}}_tdt+{\boldsymbol{\sigma}}_td{\mathbf{W}}%
_t,
\end{equation}
where ${\mathbf{W}}_t=\left(W_{1,t},\cdots,W_{p,t}\right)^{^\intercal}$ is a 
$p$-dimensional standard Brownian motion, ${\boldsymbol{\mu}}%
_t=(\mu_{1,t},\cdots,\mu_{p,t})^{^\intercal}$ is a $p$-dimensional drift
vector, and ${\boldsymbol{\sigma}}_t=\left(\sigma_{ij,t}\right)_{p\times p}$
is a $p\times p$ matrix. The spot volatility matrix of ${\mathbf{X}}_t$ is
defined as 
\begin{equation}  \label{eq2.2}
{\boldsymbol{\Sigma}}_t=\left(\Sigma_{ij,t}\right)_{p\times p}={\boldsymbol{%
\sigma}}_t{\boldsymbol{\sigma}}_t^{^\intercal}.
\end{equation}
Our main interest lies in estimating ${\boldsymbol{\Sigma}}_t$ when the size 
$p$ is large. As in \cite{CXW13} and \cite{CL16}, we assume that the true
spot volatility matrix satisfies the following uniform sparsity condition: $%
\left\{{\boldsymbol{\Sigma}}_t:\ 0\leq t\leq T\right\}\in \mathcal{S}%
(q,\varpi(p), T)$, where 
\begin{equation}  \label{eq2.3}
\mathcal{S}(q,\varpi(p), T)=\left\{{\boldsymbol{\Sigma}}_t=\left[\Sigma_{ij,t}\right]_{p\times p},\ t\in[%
0, T]\ \big|\ \sup_{0\leq t \leq T}\Vert{\boldsymbol{\Sigma}}%
_t\Vert_{\infty,q}\le \Lambda\varpi(p)\right\},
\end{equation}
where $0\le q<1$, $\varpi(p)$ is larger than a positive constant, $T$ is a fixed positive number and $\Lambda$ is a positive
random variable satisfying $\mathsf{E}[\Lambda]\leq C_\Lambda<\infty$. This
is a natural extension of the approximate sparsity assumption %
\citep[e.g.,][]{BL08}. Section \ref{sec4} below will relax this assumption and consider estimating large spot volatility matrices with systematic factors. The asset prices are assumed to be collected
over a fixed time interval $[0,T]$ at $0,\Delta,2\Delta,\cdots,n\Delta$,
where $\Delta$ is the sampling frequency and $n=\lfloor T/\Delta\rfloor$
with $\lfloor\cdot\rfloor$ denoting the floor function. In the main text, we
focus on the case of equidistant time points in the high-frequency data
collection. The asynchronicity issue will be discussed in Appendix C.2 of the supplement.

\smallskip

For each $1\leq i,j\leq p$, we estimate the spot co-volatility $\Sigma_{ij,t}
$ by 
\begin{equation}  \label{eq2.4}
\widehat\Sigma_{ij,t}=\sum_{k=1}^n K_h^\ast(t_k-t)\Delta X_{i,k}\Delta X_{j,k}
\end{equation}
with
\[
K_{h}^\ast(t_k-t)=K_h\left(t_k-t\right)/\left[\Delta\sum_{l=1}^nK_h\left(t_l-t\right)\right],
\]
where $t_k=k\Delta$, $K_h(u)=h^{-1}K(u/h)$, $K(\cdot)$ is a kernel function, 
$h$ is a bandwidth shrinking to zero and $\Delta
X_{i,k}=X_{i,t_k}-X_{i,t_{k-1}}$. The use of $K_h^\ast(t_k-t)$ rather than $K_h(t_k-t)$ in the estimation (\ref{eq2.4}) is to correct a constant bias when $t$ is close to the boundary points $0$ and $T$. A naive method of estimating the spot
volatility matrix ${\boldsymbol{\Sigma}}_{t}$ is to directly use $\widehat\Sigma_{ij,t}$ to form an estimated matrix. However, this
estimate often performs poorly in practice when the number of assets is very
large (say, $p>n$). To address this issue, a commonly-used technique is to
apply a shrinkage function to $\widehat\Sigma_{ij,t}$ when $i\neq j$,
forcing very small estimated off-diagonal entries to be zeros. Let $%
s_\rho(\cdot)$ denote a shrinkage function satisfying the following three
conditions: (i) $\vert s_\rho(u)\vert \leq \vert u\vert$ for $u\in\mathscr{R}
$; (ii) $s_\rho(u)=0$ if $\vert u\vert \leq \rho$; and (iii) $\vert
s_\rho(u)-u\vert\leq \rho$, where $\rho$ is a user-specified tuning
parameter. With the shrinkage function, we construct the following
nonparametric estimator of ${\boldsymbol{\Sigma}}_{t}$: 
\begin{equation}  \label{eq2.5}
\widehat{\boldsymbol{\Sigma}}_{t}=\left(\widehat\Sigma_{ij,t}^s\right)_{p%
\times p}\ \ \mathrm{with}\ \
\widehat\Sigma_{ij,t}^s=s_{\rho_1(t)}(\widehat\Sigma_{ij,t})I(i\neq
j)+\widehat\Sigma_{ii,t} I(i=j),
\end{equation}
where $\rho_1(t)$ is a tuning parameter which is allowed to change over $t$
and $I(\cdot)$ denotes the indicator function. Section \ref{sec5} discusses the choice of $\rho_1(t)$, ensuring that $\widehat{\boldsymbol\Sigma}_t$ is positive definite in finite samples. Our estimation method of the spot volatility matrix can be seen as a natural extension of the kernel-based large sparse covariance matrix estimation \citep[e.g.,][]{CXW13, CL16,CLL19} from the low-frequency data setting to the high-frequency one. We next give some technical assumptions which are
needed to derive the uniform convergence property of $\widehat{\boldsymbol{%
\Sigma}}_{t}$.

\begin{assumption}\label{ass:1}

\emph{(i)\ $\{\mu_{i,t}\}$ and $%
\{\sigma_{ij,t}\}$ are adapted locally bounded processes with continuous
sample path.}

\emph{(ii)\ With probability one, 
\begin{equation*}
\min_{1\leq i\leq p}\inf_{0\leq s\leq T}\Sigma_{ii, s}>0,\ \ \min_{1\leq
i\neq j\leq p}\inf_{0\leq s\leq T} \Sigma_{ij,s}^\ast>0, 
\end{equation*}
where $\Sigma_{ij, s}^\ast=\Sigma_{ii, s}+\Sigma_{jj, s}+2\Sigma_{ij, s}$.
For the spot covariance process $\{\Sigma_{ij,t}\}$, there exist $\gamma\in(0,1)$ and $B(t,\epsilon)$, a positive random function slowly
varying at $\epsilon=0$ and continuous with respect to $t$, such that}
\begin{equation}  \label{eq2.6}
\max_{1\leq i,j\leq p}\left\vert
\Sigma_{ij,t+\epsilon}-\Sigma_{ij,t}\right\vert\leq
B(t,\epsilon)|\epsilon|^\gamma+o(|\epsilon|^\gamma),\ \ \epsilon\rightarrow0.
\end{equation}

\end{assumption}

\begin{assumption}\label{ass:2}

 \emph{(i)\ The kernel $K(\cdot)$ is a bounded
and Lipschitz continuous function with a compact support $[-1,1]$. In
addition, $\int_{-1}^1 K(u)du=1$.} 

\emph{(ii)\ The bandwidth $h$ satisfies that $h\rightarrow0$ and $\frac{h}{%
\Delta\log (p\vee \Delta^{-1})}\rightarrow\infty$.} 

\emph{(iii) Let the time-varying tuning parameter $\rho_1(t)$ in the
generalised shrinkage be chosen as 
\begin{equation*}
\rho_1(t)=M(t)\zeta_{\Delta,p},\ \ \zeta_{\Delta,p}=h^{\gamma}+\left[\frac{%
\Delta\log (p\vee \Delta^{-1})}{h}\right]^{1/2},
\end{equation*}
where $\gamma$ is defined in (\ref{eq2.6}) and $M(t)$ is a positive function satisfying that} 
\begin{equation*}
0<\underline{C}_M\le \inf_{0\leq t\leq T}M(t)\leq\sup_{ 0\leq t\leq
T}M(t)\leq\overline{C}_M<\infty.
\end{equation*}

\end{assumption}

\begin{remark}\label{re:1}

Assumption \ref{ass:1} imposes some mild restrictions on
the drift and volatility processes. By a typical localisation procedure as
in Section 4.4.1 of \cite{JP12}, the local boundedness condition in
Assumption \ref{ass:1}(i) can be strengthened to the bounded condition over the entire
time interval, i.e., with probability one, 
\begin{equation*}
\max_{1\leq i\leq p}\sup_{0\leq s\leq T} |\mu_{i,s}|\leq C_\mu<\infty,\ \
\max_{1\leq i\leq p}\sup_{0\leq s\leq T} \Sigma_{ii,s} \leq C_\Sigma<\infty,
\end{equation*}
which are the same as Assumption A2 in \cite{TWZ13} and Assumptions (A.ii)
and (A.iii) in \cite{CHLZ20}. It may be possible to relax the uniform boundedness restriction (when $T$ is allowed to diverge) at the cost of more lengthy proofs \citep[e.g.,][]{KK16}. Assumption \ref{ass:1}(ii) gives the smoothness condition on the spot covariance process, crucial to derive the uniform
asymptotic order for the kernel estimation bias. When the spot covariance
is driven by continuous semimartingales, (\ref{eq2.6}) holds with $%
\gamma<1/2$ \citep[e.g., Ch. V, Exercise 1.20 in][]{RY99}. Assumption \ref{ass:2}(i)
contains some commonly-used conditions for the kernel function. Assumption \ref{ass:2}(ii)(iii) imposes some mild conditions on the bandwidth and time-varying shrinkage parameter. In particular, when $p$
diverges at a polynomial rate of $1/\Delta$, Assumption \ref{ass:2}(ii) reduces to the
conventional bandwidth restriction. Assumption \ref{ass:2}(iii) is comparable to that
assumed by \cite{CL16} and \cite{CLL19}. It is worthwhile to point out that the developed methodology and theory still hold when the time-varying tuning parameter in Assumption \ref{ass:2}(iii) is allowed to vary over entries in the spot volatility matrix estimation, which is expected to perform well in finite samples. For example, we set $\rho_{ij}(t)=\rho(t)(\widehat{\Sigma}_{ii,t}\widehat{\Sigma}_{jj,t})^{1/2}$ in the numerical studies and shrink the $(i,j)$-entry to zero if $\widehat{\Sigma}_{ij,t}\leq \rho(t)(\widehat{\Sigma}_{ii,t}\widehat{\Sigma}_{jj,t})^{1/2}$.  

\end{remark}

The following theorem gives the uniform convergence property (in the matrix spectral norm) for the
spot volatility matrix estimator $\widehat{\boldsymbol{\Sigma}}_{t}$ under the uniform sparsity assumption.

\begin{theorem}\label{thm:1}

Suppose that Assumptions \ref{ass:1} and \ref{ass:2} are
satisfied, and $\left\{{\boldsymbol{\Sigma}}_t:\ 0\leq t\leq T\right\}\in 
\mathcal{S}(q,\varpi(p), T)$. Then we have 
\begin{equation}  \label{eq2.7}
\sup_{0\leq t\leq T}\left\Vert\widehat{\boldsymbol{\Sigma}}_{t}-{%
\boldsymbol{\Sigma}}_t\right\Vert=O_P\left(\varpi(p)
\zeta_{\Delta,p}^{1-q}\right),
\end{equation}
where $\varpi(p)$ is defined in (\ref{eq2.3}) and $\zeta_{\Delta,p}$ is
defined in Assumption \ref{ass:2}(iii).

\end{theorem}

\begin{remark}\label{re:2}

(i) The first term of $\zeta_{\Delta,p}$ is $h^{\gamma}$, which is the bias rate due to application of the local
smoothing technique. It is slower than the conventional $h^2$-rate since we do not assume existence of smooth derivatives of $\Sigma_{ij,t}$ (with respect to $t$). The second
term of $\zeta_{\Delta,p}$ is square root of $\Delta h^{-1}\log (p\vee
\Delta^{-1})$, a typical uniform asymptotic rate for the kernel estimation
variance component. The uniform convergence rate in (\ref{eq2.7}) is
also similar to those obtained by \cite{CL16} and \cite{CLL19} in the
low-frequency data setting (disregarding the bias order). Note that the dimension $p$ affects the uniform convergence rate via $\varpi(p)$ and $\log (p\vee \Delta^{-1})$ and the
estimation consistency may be achieved in the ultra-high dimensional setting
when $p$ diverges at an exponential rate of $n=\lfloor T/\Delta\rfloor$. Treating $(nh)$ as
the ``effective" sample size in the local estimation procedure and disregarding the bias rate $h^{\gamma}$, the rate in (\ref{eq2.7}) is comparable to the optimal minimax rate in large covariance
matrix estimation \citep[e.g.,][]{CZ12}.

(ii) If we further assume that $\Sigma_{ij,t}$ is deterministic with continuous second-order derivative with respect to $t$, and $K(\cdot)$ is symmetric, we may improve the kernel estimation bias order. In fact, following the proof of Theorem \ref{thm:1}, we may show that  
\begin{equation}  \label{eq2.8}
\sup_{h\leq t\leq T-h}\left\Vert\widehat{\boldsymbol{\Sigma}}_{t}-{%
\boldsymbol{\Sigma}}_t\right\Vert=O_P\left(\varpi(p)
\zeta_{\Delta,p,\star}^{1-q}\right),
\end{equation}
where $\zeta_{\Delta,p,\star}=h^{2}+\left[\frac{\Delta\log (p\vee
\Delta^{-1})}{h}\right]^{1/2}$. The above uniform consistency property only holds over the trimmed time interval $[h, T-h]$ due to the kernel boundary effect. In practice, however, it is often important to investigate the spot volatility structure near the boundary points. For example, when we consider one trading day as a
time interval, it is particularly interesting to estimate the spot
volatility matrix near the opening and closing times which are peak times in
stock market trading. To address this issue, we may replace $K_h^\ast(t_k-t)$ in (\ref{eq2.4}) by a boundary kernel weight defined by 
\[
K_{h,t}^\star(t_k-t)=K_t\left(\frac{t_k-t}{h}\right)/\left[\Delta\sum_{l=1}^nK_t\left(\frac{t_l-t}{h}\right)\right],
\]
where $K_t(\cdot)$ is a boundary kernel satisfying $\int_{-t/h}^{(T-t)/h}uK_t(u)du=0$ (a key condition to improve the bias order near the boundary points). Examples of boundary kernels can be found in \cite{FG96} and \cite{LR07}. With this adjustment in the kernel estimation, we can extend the uniform consistency result (\ref{eq2.8}) to the entire interval $[0,T]$.

\end{remark}


\section{Estimation with contaminated high-frequency data}

\label{sec3} \renewcommand{\theequation}{3.\arabic{equation}} %
\setcounter{equation}{0}

In practice, it is not uncommon that high-frequency financial data are
contaminated by the market microstructure noise. The kernel estimation
method proposed in Section \ref{sec2} would be biased if the noise is
ignored in the estimation procedure. Consider the
following additive noise structure: 
\begin{equation}  \label{eq3.1}
{\mathbf{Z}}_{t_k}={\mathbf{X}}_{t_k}+{\boldsymbol{\xi}}_k={\mathbf{X}}%
_{t_k}+{\boldsymbol{\omega}}(t_k){\boldsymbol{\xi}}_k^\ast,
\end{equation}
where $t_k=k\Delta$, $k=1,\cdots,n$, ${\mathbf{Z}}_{t}=\left(Z_{1,t},%
\cdots,Z_{p,t}\right)^{^\intercal}$ is a vector of observed asset prices at
time $t$, and ${\boldsymbol{\xi}}_k=(\xi_{1,k},\cdots,\xi_{p,k})^{^\intercal}
$ is a $p$-dimensional vector of noises with nonlinear heteroskedasticity, ${%
\boldsymbol{\omega}}(\cdot)=\left[\omega_{ij}(\cdot)\right]_{p\times p}$ is
a $p\times p$ matrix of deterministic functions, and ${\boldsymbol{\xi}}%
_k^\ast=\left(\xi_{1,k}^\ast,\cdots,\xi_{p,k}^\ast\right)^{^\intercal}$
independently follows a $p$-variate identical distribution. The noise
structure defined in (\ref{eq3.1}) is similar to the setting considered in 
\cite{KL08} which also contains a nonlinear mean function and allows the
existence of endogeneity for a single asset. Throughout this section, we
assume that $\{{\boldsymbol{\xi}}_k^\ast\}$ is independent of the Brownian
semimartingale $\{{\mathbf{X}}_t\}$.

\subsection{Estimation of the spot volatility matrix}

\label{sec3.1}

To account for the microstructure noise and produce consistent volatility
matrix estimation, we apply the pre-averaging
technique as the realised kernel estimate \citep{BHLS08} can be seen as a
member of the pre-averaging estimation class whereas the two-scale estimate %
\citep{ZMA05} can be re-written as the realised kernel estimate with the
Bartlett-type kernel (up to the first-order approximation). The
pre-averaging method has been studied by \cite{JLMPV09}, \cite{PV09} and 
\cite{CKP10} in estimating the integrated volatility for a single asset and
is further extended by \cite{KWZ16} and \cite{DLX19} to the large
high-frequency data setting. \cite{KK16} use a
localised pre-averaging technique to estimate the spot volatility function
for a single asset and derive the uniform convergence rate for the developed
estimate. A similar technique is also used by \cite{XL02} to improve
convergence of the nonparametric spectral density estimator for time series
with general autocorrelation for low-frequency data. 

\smallskip

We first pre-average the observed high-frequency data via a kernel filter,
i.e., 
\begin{equation}  \label{eq3.2}
\widetilde{\mathbf{X}}_\tau=\frac{T}{n}\sum_{k=1}^n L_b^\dagger(t_k-\tau){\mathbf{Z}}%
_{t_k}
\end{equation}
with $L_{b}^\dagger(t_k-\tau)=L_b\left(t_k-\tau\right)/\int_0^TL_b(s-\tau)ds$, where $L_b(u)=b^{-1}L(u/b)$, $L(\cdot)$ is a kernel function and $b$ is a bandwidth. Let $\Delta 
\widetilde{X}_{i,l}=\widetilde{X}_{i,\tau_l}-\widetilde{X}_{i,\tau_{l-1}}$,
where $\widetilde{X}_{i,\tau_l}$ is the $i$-th component of $\widetilde{%
\mathbf{X}}_{\tau_l}$ and $\tau_0,\tau_1,\cdots,\tau_N$ are the
pseudo-sampling time points in the fixed interval $[0,T]$ with equal
distance $\Delta_\ast=T/N$. Replacing $\Delta X_{i,k}$ by $\Delta\widetilde{X%
}_{i,l}$ in (\ref{eq2.4}), we estimate the spot
co-volatility $\Sigma_{ij,t}$ by 
\begin{equation}  \label{eq3.3}
\widetilde{\Sigma}_{ij,t}=\sum_{l=1}^N K_h^\dagger(\tau_l-t)\Delta\widetilde{X}%
_{i,l}\Delta\widetilde{X}_{j,l},
\end{equation}
where
\[
K_{h}^\dagger(\tau_l-t)=K_h\left(\tau_l-t\right)/\left[\Delta_\ast\sum_{k=1}^NK_h\left(\tau_k-t\right)\right].
\]
Furthermore, to obtain a stable spot volatility matrix estimate in
finite samples when the dimension $p$ is large, as in (\ref{eq2.5}), we
apply shrinkage to $\widetilde{\Sigma}_{ij,t}$, $1\leq i\neq j\leq p$, and
subsequently construct 
\begin{equation}  \label{eq3.4}
\widetilde{\boldsymbol{\Sigma}}_t=\left(\widetilde{\Sigma}%
_{ij,t}^s\right)_{p\times p},\ \ \widetilde{\Sigma}_{ij,t}^s=s_{\rho_2(t)}%
\left(\widetilde{\Sigma}_{ij,t}\right)I(i\neq j)+\widetilde{\Sigma}%
_{ii,t}I(i=j),
\end{equation}
where $\rho_2(t)$ is another time-varying shrinkage parameter. We next give
some conditions needed to derive the uniform consistency property of $%
\widetilde{\boldsymbol{\Sigma}}_t$.

\begin{assumption}\label{ass:3}

\emph{(i)\ Let $\{{\boldsymbol{\xi}}_k^\ast\}$
be an independent and identically distributed (i.i.d.) sequence of
p-dimensional random vectors. Assume that $\mathsf{E}(\xi_{i,k}^\ast)=0$ and 
\begin{equation*}
\mathsf{E}\left[\exp\left(s|{\mathbf{u}}^{^\intercal}{\boldsymbol{\xi}}%
_{k}^\ast|\right)\right]\leq C_\xi<\infty,\ \ 0<s\leq s_0, 
\end{equation*}
for any $p$-dimensional vector ${\mathbf{u}}$ satisfying $\Vert{\mathbf{u}}%
\Vert_2=1$.} 

\emph{(ii)\ The deterministic functions $\omega_{ij}(\cdot)$ are bounded
uniformly over $i,j\in\{1,\cdots,p\}$, and satisfy that } 
\begin{equation*}
\max_{1\leq i\leq p}\sup_{0\leq t\leq T}\sum_{j=1}^p\omega_{ij}^2(t)\leq
C_\omega<\infty.
\end{equation*}

\end{assumption}

\begin{assumption}\label{ass:4}

\emph{(i)\ The kernel function $L(\cdot)$ is
Lipschitz continuous and has a compact support $[-1,1]$. In addition, $%
\int_{-1}^1 L(u)du=1$.} 

\emph{(ii)\ The bandwidth $b$ and the dimension $p$ satisfy that 
\begin{equation*}
b\rightarrow0,\ \ \frac{\Delta^{2\iota-1}b}{\log (p\vee \Delta^{-1})}%
\rightarrow\infty,\ \ p\Delta\exp\{-s\Delta^{-\iota}\}\rightarrow0,
\end{equation*}
where $0<\iota<1/2$ and $0<s\leq s_0$.} 

\emph{(iii) Let $\nu_{\Delta,p,N}=\sqrt{N\log(p\vee \Delta^{-1})}\left[b^{1/2}+(\Delta^{-1}b)^{-1/2}\right]\rightarrow0$ and the time-varying
tuning parameter $\rho_2(t)$ be chosen as $\rho_2(t)=M(t)\left(\zeta_{N,p}^%
\ast+\nu_{\Delta,p,N}\right)$, where $M(t)$ is defined as in Assumption \ref{ass:2}(iii) and $\zeta_{N,p}^\ast$ is defined as $\zeta_{\Delta,p}$ with $N$
replacing $\Delta^{-1}$.}
\end{assumption}

\begin{remark}\label{re:3}

We allow nonlinear heteroskedasticity on the
microstructure noise. The i.i.d. restriction on ${\boldsymbol{\xi}}_i^\ast$
may be weakened to some weak dependence conditions \citep[e.g.,][]{KWZ16,
DLX19} at the cost of more lengthy proofs. The moment condition in
Assumption \ref{ass:3}(i) is weaker than the sub-Gaussian condition %
\citep[e.g.,][]{BL08, TWZ13} which is commonly used in large covariance
matrix estimation when the dimension $p$ is ultra large. The boundedness
condition on $\omega_{ij}(\cdot)$ in Assumption \ref{ass:3}(ii) is similar to the
local boundedness restriction in Assumption \ref{ass:1}(i). Assumption \ref{ass:4}(ii) imposes
some mild restrictions on $b$ and $p$, which imply that there is a
trade-off between them. When $\iota$ is larger, $p$ diverges at a faster
exponential rate of $1/\Delta$ but the bandwidth condition becomes more
restrictive. If $p$ is divergent at a polynomial rate of $1/\Delta$, we may
let $\iota$ be sufficiently close to zero, and then the bandwidth condition
reduces to the conventional one as in Assumption \ref{ass:2}(ii). The condition $%
\nu_{\Delta,p,N}\rightarrow0$ in Assumption \ref{ass:4}(iii) is crucial to show that
the error of the kernel filter $\widetilde{\boldsymbol{X}}_\tau$ tends to
zero asymptotically, whereas the form of the time-varying shrinkage
parameter $\rho_2(t)$ is relevant to the uniform convergence rate of $%
\widetilde{\Sigma}_{ij,t}$ (see Proposition \ref{prop:A.2}).

\end{remark}

\begin{theorem}\label{thm:2}

Suppose that Assumptions \ref{ass:1}(i)(ii), \ref{ass:2}(i),
\ref{ass:3} and \ref{ass:4} are satisfied, and Assumption \ref{ass:2}(ii) holds with 
$\Delta^{-1}$ replaced by $N$. When $\left\{{\boldsymbol{\Sigma}}_t:\ 0\leq
t\leq T\right\}\in \mathcal{S}(q,\varpi(p), T)$, we have 
\begin{equation}  \label{eq3.5}
\sup_{0\leq t\leq T}\left\Vert\widetilde{\boldsymbol{\Sigma}}_{t}-{%
\boldsymbol{\Sigma}}_t\right\Vert=O_P\left(\varpi(p) \left[%
\zeta_{N,p}^\ast+\nu_{\Delta,p,N}\right]^{1-q}\right),
\end{equation}
where $\zeta_{N,p}^\ast$ and $\nu_{\Delta,p,N}$ are defined in Assumption \ref{ass:4}(iii).

\end{theorem}

\begin{remark}\label{re:4}

The uniform convergence rate in (\ref{eq3.5})
relies on $\varpi(p)$, $\zeta_{N,p}^\ast$ and $\nu_{\Delta,p,N}$. With the
high-frequency data collected at pseudo time points with sampling frequency $%
\Delta_\ast=T/N$, the rate $\zeta_{N,p}^\ast$ is comparable to $%
\zeta_{\Delta,p}$ for the noise-free kernel estimator in Section \ref{sec2}.
The rate $\nu_{\Delta,p,N}$ is due to the error of the kernel filter $%
\widetilde{\boldsymbol{X}}_\tau$ in the first step of the local
pre-averaging estimation procedure. In particular, when $q=0$, $\varpi(p)$
is bounded, $b=\Delta^{1/4}$ and $h=N^{-\frac{1}{2\gamma+1}}$ with $%
N=\Delta^{-\frac{2\gamma+1}{2\left(4\gamma+1\right)}}$, the uniform
convergence rate in (\ref{eq3.5}) becomes $\Delta^{\frac{\gamma}{%
2(4\gamma+1)}}\sqrt{\log(p\vee \Delta^{-1})}$. Furthermore, if $\gamma=1/2$, the rate is simplified to $\Delta^{1/12}\sqrt{\log(p\vee
\Delta^{-1})}$, comparable to those derived by \cite{ZB14} and \cite{KK16}
in the univariate high-frequency data setting. 

\end{remark}


\subsection{Estimation of the time-varying noise volatility matrix}

\label{sec3.2}

It is often interesting to further explore the volatility
structure of microstructure noise. \cite{CHLT21} estimate
the constant covariance matrix for high-dimensional noise and derive the
optimal convergence rates for the developed estimate. In the present paper,
we consider the time-varying noise covariance matrix defined by 
\begin{equation}  \label{eq3.6}
{\boldsymbol{\Omega}}(t)={\boldsymbol{\omega}}(t){\boldsymbol{\omega}}%
^{^\intercal}(t)=\left[\Omega_{ij}(t)\right]_{p\times p},\ \ 0\leq t\leq T.
\end{equation}
It is sensible to assume that $\left\{{\boldsymbol{\Omega}}(t):\ 0\leq t\leq
T\right\}$ satisfies the uniform sparsity condition as in (\ref{eq2.3}). For
each $1\leq i,j\leq p$, we estimate $\Omega_{ij}(t)$ by the kernel smoothing
method: 
\begin{equation}  \label{eq3.7}
\widehat{\Omega }_{ij}(t)=\frac{\Delta}{2}\sum_{k=1}^{n}K_{h_1}^\ast(t_{k}-t)%
\Delta Z_{i,t_{k}}\Delta Z_{j,t_{k}},
\end{equation}
where $h_1$ is a bandwidth, $\Delta Z_{i,t_{k}}=Z_{i,t_k}-Z_{i,t_{k-1}}$ and $K_{h_1}^\ast(t_k-t)$ is defined similarly to $K_{h}^\ast(t_k-t)$ in (\ref{eq2.4}) but with $h_1$ replacing $h$.
As in (\ref{eq2.5}) and (\ref{eq3.4}), we again apply shrinkage to $%
\widehat{\Omega}_{ij}(t)$, $1\leq i\neq j\leq p$, and construct 
\begin{equation}  \label{eq3.8}
\widehat{\boldsymbol{\Omega}}(t)=\left[\widehat{\Omega}_{ij}^s(t)\right]%
_{p\times p},\ \ \widehat{\Omega}_{ij}^s(t)=s_{\rho_3(t)}\left(\widehat{%
\Omega}_{ij}(t)\right)I(i\neq j)+\widehat{\Omega}_{ii}(t)I(i=j),
\end{equation}
where $\rho_3(t)$ is a time-varying shrinkage parameter. To derive the
uniform consistency property of $\widehat{\boldsymbol{\Omega}}(t)$, we need
to impose stronger moment condition on ${\boldsymbol{\xi}}_k^\ast$ and
smoothness restriction on $\Omega_{ij}(\cdot)$.

\begin{assumption}\label{ass:5}

\emph{(i)\ For any $p$-dimensional vector ${%
\mathbf{u}}$ satisfying $\Vert{\mathbf{u}}\Vert_2=1$,
$\mathsf{E}\left[\exp\left(s({\mathbf{u}}^{^\intercal}{\boldsymbol{\xi}}%
_{k}^\ast)^2\right)\right]\leq C_\xi^\star<\infty$, $0<s\leq s_0$.} 

\emph{(ii)\ The time-varying function $\Omega_{ij}(t)$ satisfies that 
\begin{equation*}
\max_{1\leq i,j\leq p}\left\vert
\Omega_{ij}(t)-\Omega_{ij}(s)\right\vert\leq
C_\Omega|t-s|^{\gamma_1}, 
\end{equation*}
where $C_\Omega$ is a positive constant and $0<\gamma_1<1$.} 

\emph{(iii)\ The bandwidth $h_1$ and the dimension $p$ satisfy that 
\begin{equation*}
h_1\rightarrow0,\ \ \frac{\Delta^{2\iota_\star-1}h_1}{\log (p\vee
\Delta^{-1})}\rightarrow\infty,\ \
p\Delta^{-1}\exp\{-s\Delta^{-\iota_\star}/C_\omega\}\rightarrow0,
\end{equation*}
where $0<\iota_\star<1/2$, $0<s\leq s_0$ and $C_\omega$ is defined in
Assumption \ref{ass:3}(ii).}
\end{assumption}

\begin{remark}\label{re:5}

Assumption \ref{ass:5}(i) strengthens the moment condition in Assumption \ref{ass:3}(i) and is equivalent to the sub-Gaussian condition, see Assumption A1 in \cite{TWZ13}. The smoothness condition in
Assumption \ref{ass:5}(ii) is similar to (\ref{eq2.6}), crucial to derive the
asymptotic order of the kernel estimation bias. The restrictions on $h_1$
and $p$ in Assumption \ref{ass:5}(iii) are similar to those in Assumption \ref{ass:4}(ii),
allowing $p$ to be divergent to infinity at an exponential
rate of $1/\Delta$.

\end{remark}

In the following theorem, we state the uniform consistency result for $%
\widehat{\boldsymbol{\Omega}}(t)$ with convergence rate comparable to that
in Theorem \ref{thm:1}.

\begin{theorem}\label{thm:3}

Suppose that Assumptions \ref{ass:1}, \ref{ass:2}(i), \ref{ass:3}
and \ref{ass:5} are satisfied, and Assumption \ref{ass:2}(ii)(iii) holds when $%
\rho_1(t)$, $\zeta_{\Delta,p}$ and $h$ are replaced by $\rho_3(t)$, $%
\delta_{\Delta,p}$ and $h_1$, respectively, where $\delta_{\Delta,p}=h_1^{%
\gamma_1}+\left[\frac{\Delta\log (p\vee \Delta^{-1})}{h_1}\right]^{1/2}$. If 
$\left\{{\boldsymbol{\Omega}}(t):\ 0\leq t\leq T\right\}\in \mathcal{S}%
(q,\varpi(p), T)$, we have
\begin{equation}  \label{eq3.9}
\sup_{0\leq t\leq T}\left\Vert\widehat{\boldsymbol{\Omega}}(t)-{%
\boldsymbol{\Omega}}(t)\right\Vert=O_P\left(\varpi(p)\delta_{\Delta,p}^{1-q}%
\right).
\end{equation}

\end{theorem}

\begin{remark}\label{re:6}

If the bandwidth parameter $h_1$ in (\ref{eq3.7}) is the same as $h$ in (\ref{eq2.4}), we may find that the uniform convergence rate $O_P\left(\varpi(p)\delta_{\Delta,p}^{1-q}\right)$ would be
the same as that in Theorem 1. Treating $(nh_1)$ as the ``effective" sample
size and disregarding the bias order, we may show that the uniform
convergence rate in (\ref{eq3.9}) is comparable to the optimal minimax rate
derived by \cite{CHLT21} for the constant noise covariance matrix
estimation. Meanwhile, the kernel estimation bias order $h_1^{\gamma_1}$ may be improved by strengthening the smoothness condition on $\Omega_{ij}(\cdot)$ and adopting the boundary kernel weight as suggested in Remark \ref{re:2}(ii).

\end{remark}


\section{Estimation with observed factors}\label{sec4} 
\renewcommand{\theequation}{4.\arabic{equation}} \setcounter{equation}{0}

The large spot volatility matrix estimation with the shrinkage technique developed in Sections \ref{sec2} and \ref{sec3} heavily relies on the uniform sparsity assumption (\ref{eq2.3}). However, the latter may be too restrictive in practice since the price processes of a large number of assets are often driven by some common factors such as the market factors, resulting in strong correlation among assets and failure of the sparsity condition. To address this problem, we next consider the nonparametric time-varying regression at high frequency:
\begin{equation}\label{eq4.1}
d{\mathbf Y}_t={\boldsymbol\beta}(t) d {\mathbf F}_t+d{\mathbf X}_t,
\end{equation}
where ${\boldsymbol\beta}(t)=\left[\beta_{1}(t),\cdots,\beta_p(t)\right]^{^\intercal}$ is a $p\times k$ matrix of time-varying betas (or factor loadings), ${\mathbf F}_t$ and ${\mathbf X}_t$ are $k$-variate and $p$-variate continuous semi-martingales defined by
\begin{equation}  \label{eq4.2}
d{\mathbf{F}}_t={\boldsymbol{\mu}}_t^Fdt+{\boldsymbol{\sigma}}_t^Fd{\mathbf{W}}_t^F\ \ {\rm and}\ \  d{\mathbf{X}}_t={\boldsymbol{\mu}}_t^Xdt+{\boldsymbol{\sigma}}_t^Xd{\mathbf{W}}_t^X,
\end{equation}
respectively, ${\boldsymbol{\mu}}_t^F$ and ${\boldsymbol{\mu}}_t^X$ are drift
vectors, ${\boldsymbol{\sigma}}_t^F=\left(\sigma_{ij,t}^F\right)_{k\times k}$, ${\boldsymbol{\sigma}}_t^X=\left(\sigma_{ij,t}^X\right)_{p\times p}$, ${\mathbf{W}}_t^F$ and ${\mathbf{W}}_t^X$ are 
$k$-dimensional and $p$-dimensional standard Brownian motions. For the time being, we assume that ${\mathbf Y}_t$ and ${\mathbf F}_t$ are observable and noise free but ${\mathbf X}_t$ is latent. Extension of the methodology and theory to the noise-contaminated high-frequency data will be considered later in this section. 

\smallskip

Estimation of the constant betas via the ratio of realised covariance to realised variance is proposed by \cite{BS04}, and extension to time-varying beta estimation has been studied by \cite{MZ06}, \cite{RTT15} and \cite{AKX20}, some of which allow jumps in the semi-martingale processes. The main interest of this section lies in estimating the large spot volatility structure ${\boldsymbol\Sigma}_t^Y$ of ${\mathbf Y}_t$. Letting ${\boldsymbol\Sigma}_t^F={\boldsymbol\sigma}_t^F\left({\boldsymbol\sigma}_t^F\right)^{^\intercal}$ and ${\boldsymbol\Sigma}_t^X={\boldsymbol\sigma}_t^X\left({\boldsymbol\sigma}_t^X\right)^{^\intercal}$, and assuming orthogonality between ${\mathbf X}_t$ and ${\mathbf F}_t$, see Assumption \ref{ass:6}(iii) below, it follows from (\ref{eq4.1}) that    
\begin{equation}\label{eq4.3}
{\boldsymbol\Sigma}_t^Y={\boldsymbol\beta}(t){\boldsymbol\Sigma}_t^F{\boldsymbol\beta}(t)^{^\intercal}+{\boldsymbol\Sigma}_t^X.
\end{equation}
As in \cite{FLM11, FLM13}, we impose the uniform sparsity restriction on ${\boldsymbol\Sigma}_t^X$ instead of ${\boldsymbol\Sigma}_t^Y$, i.e., $
\left\{{\boldsymbol{\Sigma}}_t^X:\ 0\leq t\leq T\right\}\in \mathcal{S}(q,\varpi(p), T)$. This is a reasonable assumption in practical applications as the asset prices, after removing the influence of systematic factors, are expected to be weakly correlated. \cite{FFX16} and \cite{DLX19} use a similar framework with constant betas to estimate large integrated volatility matrices.

\smallskip

Suppose that we observe ${\mathbf Y}_t$ and ${\mathbf F}_t$ at regular points: $t_k=k\Delta$, $k=1,\cdots,n$, as in Sections \ref{sec2} and \ref{sec3}. Let ${\boldsymbol\Sigma}_t^{YF}$ be the spot covariance between ${\mathbf Y}_t$ and ${\mathbf F}_t$. We may use the kernel smoothing method as in (\ref{eq2.4}) to estimate ${\boldsymbol\Sigma}_t^{Y}$, ${\boldsymbol\Sigma}_t^{F}$ and ${\boldsymbol\Sigma}_t^{YF}$, i.e.,
\begin{eqnarray}
&&\widehat{\boldsymbol\Sigma}_t^Y=\sum_{k=1}^nK_h^\ast(t_k-t)\Delta {\mathbf Y}_k\Delta {\mathbf Y}_k^{^\intercal},\label{eq4.4}\\
&&\widehat{\boldsymbol\Sigma}_t^F=\sum_{k=1}^nK_h^\ast(t_k-t)\Delta {\mathbf F}_k\Delta {\mathbf F}_k^{^\intercal},\label{eq4.5}\\
&&\widehat{\boldsymbol\Sigma}_t^{YF}=\sum_{k=1}^nK_h^\ast(t_k-t)\Delta {\mathbf Y}_k\Delta {\mathbf F}_k^{^\intercal},\label{eq4.6}
\end{eqnarray}
where $\Delta{\mathbf Y}_k={\mathbf Y}_{t_k}-{\mathbf Y}_{t_{k-1}}$, $\Delta{\mathbf F}_k={\mathbf F}_{t_k}-{\mathbf F}_{t_{k-1}}$, and $K_h^\ast(t_k-t)$ is defined as in (\ref{eq2.4}). Consequently, the time-varying betas ${\boldsymbol\beta}(t)$ and the spot idiosyncratic volatility matrix ${\boldsymbol\Sigma}_t^X$ are estimated by 
\begin{equation}\label{eq4.7}
\widehat{\boldsymbol\beta}(t)=\left[\widehat\beta_{1}(t),\cdots,\widehat\beta_p(t)\right]^{^\intercal}=\widehat{\boldsymbol\Sigma}_t^{YF}\left(\widehat{\boldsymbol\Sigma}_t^{F}\right)^{-1},
\end{equation}
and
\begin{equation}\label{eq4.8}
\widehat{\boldsymbol\Sigma}_t^X=\left(\widehat\Sigma_{ij,t}^X\right)_{p\times p}=\widehat{\boldsymbol\Sigma}_t^Y-\widehat{\boldsymbol\Sigma}_t^{YF}\left(\widehat{\boldsymbol\Sigma}_t^{F}\right)^{-1}\left(\widehat{\boldsymbol\Sigma}_t^{YF}\right)^{^\intercal}.
\end{equation}
With the uniform sparsity condition, it is sensible to further apply shrinkage to $\widehat\Sigma_{ij,t}^X$, i.e.,
\begin{equation}  \label{eq4.9}
\widehat{\boldsymbol{\Sigma}}_{t}^{X,s}=\left(\widehat\Sigma_{ij,t}^{X,s}\right)_{p%
\times p}\ \ \mathrm{with}\ \
\widehat\Sigma_{ij,t}^{X,s}=s_{\rho_4(t)}(\widehat\Sigma_{ij,t}^X)I(i\neq
j)+\widehat\Sigma_{ii,t}^X I(i=j),
\end{equation}
where $\rho_4(t)$ is a time-varying shrinkage parameter. We finally estimate ${\boldsymbol\Sigma}_t^Y$ as 
\begin{equation}\label{eq4.10}
\widehat{\boldsymbol\Sigma}_t^{Y, s}=\widehat{\boldsymbol\beta}(t)\widehat{\boldsymbol\Sigma}_t^F\widehat{\boldsymbol\beta}(t)^{^\intercal}+\widehat{\boldsymbol\Sigma}_t^{X,s}=\widehat{\boldsymbol\Sigma}_t^{YF}\left(\widehat{\boldsymbol\Sigma}_t^{F}\right)^{-1}\left(\widehat{\boldsymbol\Sigma}_t^{YF}\right)^{^\intercal}+\widehat{\boldsymbol\Sigma}_t^{X,s}.
\end{equation}

\smallskip

We need the following assumption to derive the uniform convergence property for $\widehat{\boldsymbol\Sigma}_t^{X,s}$ and $\widehat{\boldsymbol\Sigma}_t^{Y,s}$.

\begin{assumption}\label{ass:6}

{\em (i) Assumption \ref{ass:1} is satisfied for $\{X_t\}$ defined in (\ref{eq4.2}) (with minor notational changes).} 

{\em (ii) Let $\{{\boldsymbol{\mu}}_t^F\}$, $\{{\boldsymbol{\sigma}}_t^F\}$ and $\{{\boldsymbol{\Sigma}}_t^F\}$ satisfy the boundedness and smoothing conditions as in Assumption \ref{ass:1}.} 

{\em (iii) For any $1\leq i\leq p$ and $1\leq j\leq k$, $\left[X_{it}, F_{jt}\right]=0$ for any $t\in[0,T]$, where $X_{i,t}$ is the $i$-th element of ${\mathbf X}_t$, $F_{j,t}$ is the $j$-th element of ${\mathbf F}_t$, and $[\cdot,\cdot]$ denotes the quadratic covariation. } 

{\em (iv) The time-varying beta function $\beta_i(\cdot)$ satisfies that 
\[
\max_{1\leq i\leq p}\sup_{0\leq t\leq T}\left\Vert \beta_i(t)\right\Vert_2\leq C_\beta<\infty,\ \ \max_{1\leq i\leq p}\left\Vert \beta_i(t)-\beta_i(s)\right\Vert_2\leq C_\beta|t-s|^{\gamma},
\]
where $\gamma$ is the same as that in Assumption \ref{ass:1}(ii). In addition, there exists a positive definite matrix ${\boldsymbol\Sigma}_\beta(t)$ (with uniformly bounded eigenvalues) such that}
\begin{equation}\label{eq4.11}
\sup_{0\leq t\leq T}\left\Vert \frac{1}{p}{\boldsymbol\beta}(t)^{^\intercal}{\boldsymbol\beta}(t)-{\boldsymbol\Sigma}_\beta(t)\right\Vert=o(1).
\end{equation}

\end{assumption}

\begin{remark}\label{re:7}

The uniform boundedness and smoothness conditions imposed on the drift and spot volatility functions of ${\mathbf X}_t$ and ${\mathbf F}_t$ in Assumption \ref{ass:6}(i)(ii) are the same as those in Assumption \ref{ass:1}. This is crucial to ensure that the uniform convergence rates of $\widehat{\boldsymbol\Sigma}_t^Y$, $\widehat{\boldsymbol\Sigma}_t^F$ and $\widehat{\boldsymbol\Sigma}_t^{YF}$ (in the max norm) derived in Proposition \ref{prop:A.4} are the same as that in Proposition \ref{prop:A.1}. The orthogonality condition in Assumption \ref{ass:6}(iii) is commonly used to consistently estimate the time-varying factor model \citep[e.g.,][]{FFX16, DLX19}. Assumption \ref{ass:6}(iv) is a rather mild restriction on  time-varying betas and may be strengthened to improve the estimation bias order, see the discussion in Remark \ref{re:2}(ii). The condition (\ref{eq4.11}) indicates that all the factors are pervasive.

\end{remark}

We next present the convergence property of $\widehat{\boldsymbol\Sigma}_t^{X,s}$ and $\widehat{\boldsymbol\Sigma}_t^{Y,s}$ defined in (\ref{eq4.9}) and (\ref{eq4.10}), respectively. Due to the nonparametric factor regression model structure (\ref{eq4.1}), the largest $k$ eigenvalues of ${\boldsymbol\Sigma}_t^Y$ are spiked, diverging at a rate of $p$. Hence, ${\boldsymbol\Sigma}_t^Y$ cannot be consistently estimated in the absolute term. To address this problem, as in \cite{FLM11, FLM13}, we measure the spiked volatility matrix estimate in the following relative error:
\[\left\Vert \widehat{\boldsymbol\Sigma}_t^{Y,s}-{\boldsymbol\Sigma}_t^Y\right\Vert_{{\boldsymbol\Sigma}_t^Y}=\frac{1}{\sqrt{p}}\left\Vert \left({\boldsymbol\Sigma}_t^Y\right)^{-1/2}\left(\widehat{\boldsymbol\Sigma}_t^{Y,s}-{\boldsymbol\Sigma}_t^Y\right) \left({\boldsymbol\Sigma}_t^Y\right)^{-1/2} \right\Vert_F,\] 
where the normalisation factor $p^{-1/2}$ is used to guarantee that $\left\Vert{\boldsymbol\Sigma}_t^Y\right\Vert_{{\boldsymbol\Sigma}_t^Y}=1$.

\begin{theorem}\label{thm:4}

Suppose that Assumptions \ref{ass:2}(i)(ii) and \ref{ass:6} are satisfied, and Assumption \ref{ass:2}(iii) holds with $\rho_1(t)$ replaced by $\rho_4(t)$. When $\left\{{\boldsymbol{\Sigma}}_t^X:\ 0\leq t\leq T\right\}\in \mathcal{S}(q,\varpi(p), T)$, we have 
\begin{equation}  \label{eq4.12}
\sup_{0\leq t\leq T}\left\Vert\widehat{\boldsymbol{\Sigma}}_{t}^{X,s}-{\boldsymbol{\Sigma}}_t^X\right\Vert=O_P\left(\varpi(p)\zeta_{\Delta,p}^{1-q}\right),
\end{equation}
where $\varpi(p)$ is defined in (\ref{eq2.3}) and $\zeta_{\Delta,p}$ is defined in Assumption \ref{ass:2}(iii); and
\begin{equation}  \label{eq4.13}
\sup_{0\leq t\leq T}\left\Vert \widehat{\boldsymbol\Sigma}_t^{Y,s}-{\boldsymbol\Sigma}_t^Y\right\Vert_{{\boldsymbol\Sigma}_t^Y}=O_P\left(p^{1/2}\zeta_{\Delta,p}^2+\varpi(p)\zeta_{\Delta,p}^{1-q}\right).
\end{equation}

\end{theorem}

\begin{remark}\label{re:8}

Although ${\mathbf X}_t$ is latent in model (\ref{eq4.1}), the uniform convergence rate for $\widehat{\boldsymbol{\Sigma}}_{t}^{X,s}$ in (\ref{eq4.12}) is the same as that in Theorem \ref{thm:1} when ${\mathbf X}_t$ is observable. Treating $(nh)$ as the effective sample size in kernel estimation and disregarding the bias order in $\zeta_{\Delta,p}$, the uniform convergence rate for $\widehat{\boldsymbol{\Sigma}}_{t}^{Y,s}$ in (\ref{eq4.13}) is comparable to the convergence rates derived by \cite{FLM11} in low frequency and \cite{FFX16} in high frequency. To guarantee uniform consistency in the relative matrix estimation error, we have to further assume that $p\zeta_{\Delta,p}^4=o(1)$, limiting the divergence rate of the asset number, i.e., $p$ can only diverge at a polynomial rate of $n=\lfloor T/\Delta\rfloor$. 

\end{remark}

We next modify the above methodology and theory to accommodate microstructure noise in the asset prices and factors. Assume that 
\begin{equation}  \label{eq4.14}
{\mathbf{Z}}_{Y,t_k}={\mathbf{Y}}_{t_k}+{\boldsymbol{\omega}}_Y(t_k){\boldsymbol{\xi}}_{Y,k}^\ast,\quad
{\mathbf{Z}}_{F,t_k}={\mathbf{F}}_{t_k}+{\boldsymbol{\omega}}_F(t_k){\boldsymbol{\xi}}_{F,k}^\ast,
\end{equation}
where ${\boldsymbol{\omega}}_Y(\cdot)$ and ${\boldsymbol{\omega}}_F(\cdot)$ are matrices of deterministic functions similar to ${\boldsymbol{\omega}}(\cdot)$, and $\{{\boldsymbol{\xi}}_{Y,k}^\ast\}$ and $\{{\boldsymbol{\xi}}_{F,k}^\ast\}$ are i.i.d. sequences of random vectors similar to $\{{\boldsymbol{\xi}}_{k}^\ast\}$. Since both ${\mathbf Y}_t$ and ${\mathbf F}_t$ are latent, we need to first adopt the kernel pre-averaging technique proposed in Section \ref{sec3.1} to obtain the approximation of ${\mathbf Y}_t$ and ${\mathbf F}_t$, and then apply the kernel smoothing and generalised shrinkage as in (\ref{eq4.4})--(\ref{eq4.10}). This results in a three-stage estimation procedure which we describe as follows.

\begin{enumerate}

\item As in (\ref{eq3.2}), we pre-average the noise-contaminated ${\mathbf{Z}}_{Y,t_k}$ and ${\mathbf{Z}}_{F,t_k}$ via the kernel filter:
\begin{equation}  \label{eq4.15}
\widetilde{\mathbf{Y}}_\tau=\frac{T}{n}\sum_{k=1}^n L_b^\dagger(t_k-\tau){\mathbf{Z}}_{Y,t_k},\quad
\widetilde{\mathbf{F}}_\tau=\frac{T}{n}\sum_{k=1}^n L_b^\dagger(t_k-\tau){\mathbf{Z}}_{F,t_k},
\end{equation}
where $L_b^\dagger(t_k-\tau)$ is defined as in (\ref{eq3.2}) and we consider $\tau$ as the pseudo-sampling time points: $\tau_l=l\Delta_\ast$, $l=0,1,\cdots, N=\lfloor T/\Delta_\ast\rfloor$.

\item With $\widetilde{\mathbf Y}_{\tau_l}$ and $\widetilde{\mathbf F}_{\tau_l}$, $l=1,\cdots,N$, we estimate ${\boldsymbol\Sigma}_t^Y, {\boldsymbol\Sigma}_t^F$ and ${\boldsymbol\Sigma}_t^{YF}$ by the kernel smoothing as in (\ref{eq4.4})--(\ref{eq4.6}):
\begin{eqnarray}
&&\widetilde{\boldsymbol\Sigma}_t^Y=\sum_{l=1}^N K_h^\dagger(\tau_l-t)\Delta \widetilde{\mathbf Y}_l\Delta \widetilde{\mathbf Y}_l^{^\intercal},\notag\\
&&\widetilde{\boldsymbol\Sigma}_t^F=\sum_{l=1}^N K_h^\dagger(\tau_l-t)\Delta \widetilde{\mathbf F}_l\Delta \widetilde{\mathbf F}_l^{^\intercal},\notag\\
&&\widetilde{\boldsymbol\Sigma}_t^{YF}=\sum_{l=1}^N K_h^\dagger(\tau_l-t)\Delta \widetilde{\mathbf Y}_l\Delta \widetilde{\mathbf F}_l^{^\intercal},\notag
\end{eqnarray}
where $K_h^\dagger(\tau_l-t)$ is defined as in (\ref{eq3.3}), $\Delta \widetilde{\mathbf Y}_l=\widetilde{\mathbf Y}_{\tau_l}-\widetilde{\mathbf Y}_{\tau_{l-1}}$ and $\Delta \widetilde{\mathbf F}_l=\widetilde{\mathbf F}_{\tau_l}-\widetilde{\mathbf F}_{\tau_{l-1}}$. Furthermore, estimate ${\boldsymbol\beta}(t)$ and ${\boldsymbol\Sigma}_t^X$ by
\[
\widetilde{\boldsymbol\beta}(t)=\widetilde{\boldsymbol\Sigma}_t^{YF}\left(\widetilde{\boldsymbol\Sigma}_t^{F}\right)^{-1},\quad \widetilde{\boldsymbol\Sigma}_t^X=\left(\widetilde\Sigma_{ij,t}^X\right)_{p\times p}=\widetilde{\boldsymbol\Sigma}_t^Y-\widetilde{\boldsymbol\Sigma}_t^{YF}\left(\widetilde{\boldsymbol\Sigma}_t^{F}\right)^{-1}\left(\widetilde{\boldsymbol\Sigma}_t^{YF}\right)^{^\intercal}.
\]

\item Apply the generalised shrinkage to $\widetilde\Sigma_{ij,t}^X$, i.e.,
\[
\widetilde{\boldsymbol{\Sigma}}_{t}^{X,s}=\left(\widetilde\Sigma_{ij,t}^{X,s}\right)_{p%
\times p}\ \ \mathrm{with}\ \
\widetilde\Sigma_{ij,t}^{X,s}=s_{\rho_5(t)}(\widetilde\Sigma_{ij,t}^X)I(i\neq
j)+\widetilde\Sigma_{ii,t}^X I(i=j),
\]
where $\rho_5(t)$ is the shrinkage parameter, and then estimate ${\boldsymbol\Sigma}_t^Y$ by
\[
\widetilde{\boldsymbol\Sigma}_t^{Y, s}=\widetilde{\boldsymbol\beta}(t)\widetilde{\boldsymbol\Sigma}_t^F\widetilde{\boldsymbol\beta}(t)^{^\intercal}+\widetilde{\boldsymbol\Sigma}_t^{X,s}=\widetilde{\boldsymbol\Sigma}_t^{YF}\left(\widetilde{\boldsymbol\Sigma}_t^{F}\right)^{-1}\left(\widetilde{\boldsymbol\Sigma}_t^{YF}\right)^{^\intercal}+\widetilde{\boldsymbol\Sigma}_t^{X,s}.
\]

\end{enumerate}

As shown in Theorem \ref{thm:2}, the existence of microstructure noises slows down the uniform convergence rates. Following the proof of Lemma B.1 in Appendix B, we may show that 
\[
\max_{0\leq l\leq N}\left\vert \widetilde{\mathbf{Y}}_{\tau_l}-{\mathbf{Y}}_{\tau_l}\right\vert_{\max}+\max_{0\leq l\leq N}\left\vert \widetilde{\mathbf{F}}_{\tau_l}-{\mathbf{F}}_{\tau_l}\right\vert_{\max}=O_P\left(\nu_{\Delta,p,N}\right),
\]
where $\vert\cdot\vert_{\max}$ denotes the $L_\infty$-norm of a vector, and $\nu_{\Delta,p,N}$ is defined in Assumption \ref{ass:4}(iii). Modifying Proposition \ref{prop:A.4} and the proof of Theorem \ref{thm:4} in Appendix A, we can prove that (\ref{eq4.12}) and (\ref{eq4.13}) hold but with $\zeta_{\Delta,p}$ replaced by $\zeta_{N,p}^\ast+\nu_{\Delta,p,N}$ defined in Assumption \ref{ass:4}(iii), i.e.,
\begin{eqnarray}
&&\sup_{0\leq t\leq T}\left\Vert\widetilde{\boldsymbol{\Sigma}}_{t}^{X,s}-{\boldsymbol{\Sigma}}_t^X\right\Vert=O_P\left(\varpi(p)(\zeta_{N,p}^\ast+\nu_{\Delta,p,N})^{1-q}\right),\notag\\
&&\sup_{0\leq t\leq T}\left\Vert \widetilde{\boldsymbol\Sigma}_t^{Y,s}-{\boldsymbol\Sigma}_t^Y\right\Vert_{{\boldsymbol\Sigma}_t^Y}=O_P\left(p^{1/2}(\zeta_{N,p}^\ast+\nu_{\Delta,p,N})^2+\varpi(p)(\zeta_{N,p}^\ast+\nu_{\Delta,p,N})^{1-q}\right).\notag
\end{eqnarray}


\section{Monte-Carlo Study}\label{sec5}
\renewcommand{\theequation}{5.\arabic{equation}} \setcounter{equation}{0}

In this section, we report the Monte-Carlo simulation studies to assess the numerical performance of the proposed large spot volatility matrix and time-varying noise volatility matrix estimation methods under the sparsity condition and the factor-based spot volatility matrix estimation. Here we only consider the synchronous high-frequency data. Additional simulation results for asynchronous high-frequency data are provided in the supplement. 

\subsection{Simulation for sparse volatility matrix estimation}\label{sec5.1}


\bigskip

\noindent{\em 5.1.1.\ \ Simulation setup}

\smallskip

\noindent We generate the noise-contaminated high-frequency data according to model (\ref{eq3.1}), where ${\boldsymbol{\omega}}(t)$ is taken as the Cholesky decomposition of the noise covariance matrix ${\boldsymbol\Omega}(t)=\left[\Omega _{ij}(t)\right] _{p\times p}$, ${\boldsymbol{\xi }}_{k}^{\ast }=\left(\xi _{1,k}^{\ast },\cdots ,\xi _{p,k}^{\ast }\right)^{^{\intercal }}$ is an independent $p$-dimensional random vector of cross-sectionally independent standard normal random variables, the latent return process $\mathbf{X}_{t}$ of $p$ assets is generated from the following drift-free model: 
\begin{equation}\label{eq5.1}
d\mathbf{X}_{t}=\boldsymbol{\sigma }_{t}d\mathbf{W}_{t}^{X},\ \ t\in
\lbrack 0,T],  
\end{equation}
$\mathbf{W}_{t}^{X}=\left( W_{1,t}^{X},\cdots ,W_{p,t}^{X}\right)^{^{\intercal }}$ is a standard $p$-dimensional Brownian motion, and ${\boldsymbol{\sigma }}_{t}$ is chosen as the Cholesky decomposition of the spot covariance matrix ${\boldsymbol{\Sigma }}_{t}=\left(\Sigma_{ij,t}\right) _{p\times p}$. In the simulation, we consider the volatility matrix estimation over the time interval of a full trading day, and set the sampling interval to be $15$ seconds, i.e., $\Delta =1/(252\times 6.5\times 60\times 4)$, to generate synchronous data. We consider three structures in ${\boldsymbol{\Sigma}}_{t}$ and ${\boldsymbol\Omega}(t)$: ``banding", ``block-diagonal", and ``exponentially decaying". Following \cite{WZ10}, we generate the diagonal elements of ${\boldsymbol\Sigma}_{t}$ from the following geometric Ornstein-Uhlenbeck model \citep[e.g.,][]{BS02}:
\begin{equation*}
d\log \Sigma _{ii,t}=-0.6\left(0.157+\log \Sigma _{ii,t}\right)dt+0.25dW_{i,t}^\Sigma,\ \ W_{i,t}^\Sigma=\iota_{i}W_{i,t}^{X}+\sqrt{1-\iota_{i}^{2}}W_{i,t}^\ast,
\end{equation*}
where $\mathbf{W}_{t}^\ast=\left(W_{1,t}^\ast,\cdots,W_{p,t}^\ast\right)^{^\intercal}$ is a standard $p$-dimensional Brownian motion independent of $\mathbf{W}_{t}^{X}$, and $\iota_{i}$ is a random number generated uniformly between $-0.62$ and $-0.30$, reflecting the leverage effects. The diagonal elements of ${\boldsymbol\Omega}(t)$ are defined as daily cyclical deterministic functions of time: 
\[
\Omega _{ii}\left( t\right) =c_{i}\left\{ \frac{1}{2}\left[ \cos \left(2\pi t/T\right) +1\right] \times \left( \overline\omega -\underline\omega\right) +\underline\omega\right\}, 
\]
where $\overline\omega=1$ and $\underline\omega=0.1$ reflect the observation by \cite{KL08} that the noise level is high at both the opening and the closing times of a trading day and is low in the middle of the day, and the scalar $c_{i}$ controls the noise ratio for each asset which is chosen to match the highest noise ratio considered by \cite{WZ10}. As in \cite{BS02, BS04}, we define a continuous-time stochastic process $\kappa^{\Sigma}_t$ by 
\begin{eqnarray}
\kappa^{\Sigma}_t&=&\frac{e^{2\kappa_{t}}-1}{e^{2\kappa_{t}}+1},\ \ 
d\kappa_{t}=0.03\left( 0.64-\kappa_{t}\right) dt+0.118\kappa_{t}dW_{t}^\kappa,\notag\\
W_{t}^\kappa &=&\sqrt{0.96}W_{t}^{\diamond}-0.2\sum_{i=1}^{p}W_{i,t}^{X}/\sqrt{p}  \notag
\end{eqnarray}%
where $W_{t}^\diamond$ is a standard univariate Brownian motion independent of $\mathbf{W}_{t}^{X}$ and $\mathbf{W}_{t}^\ast$. Let
\[
\kappa^\Omega_t =\frac{\overline\kappa-\underline\kappa}{2}\left[ \cos \left( 2\pi t/T\right) +1\right] +\underline\kappa,
\]
where $\overline\kappa=0.5$ and $\underline\kappa=-0.5$. We will use $\kappa^{\Sigma}_t$ and $\kappa^{\Omega}_t$ to define the off-diagonal elements in ${\boldsymbol{\Sigma}}_{t}$ and ${\boldsymbol\Omega}(t)$, respectively, which are specified as follows.

\begin{itemize}

\item Banding structure for ${\boldsymbol{\Sigma}}_{t}$ and ${\boldsymbol\Omega}(t)$: The off-diagonal elements are defined by
\[
\Sigma _{ij,t} =\left(\kappa^\Sigma_{t}\right)^{\vert i-j\vert }\sqrt{\Sigma_{ii,t}\Sigma_{jj,t}}\cdot I\left( \left\vert i-j\right\vert \leq 2\right),\]
and
\[\Omega_{ij}(t) =\left(\kappa^\Omega_t\right) ^{\vert i-j\vert }\sqrt{\Omega _{ii}( t) \Omega _{jj}(t) }\cdot I\left( \left\vert i-j\right\vert \leq 2\right),
\]
for $1\leq i\neq j\leq p$.

\item Block-diagonal structure for ${\boldsymbol{\Sigma}}_{t}$ and ${\boldsymbol\Omega}(t)$: The off-diagonal elements are defined by
\[
\Sigma _{ij,t} =\left(\kappa^\Sigma_{t}\right)^{\vert i-j\vert }\sqrt{\Sigma_{ii,t}\Sigma_{jj,t}}\cdot I\left( (i,j)\in {\cal B}\right),
\]
\[
\Omega _{ij}(t)=\left(\kappa^\Omega_t\right) ^{\vert i-j\vert }\sqrt{\Omega _{ii}( t) \Omega _{jj}(t) }\cdot I\left((i,j)\in {\cal B}\right) ,
\]
for $1\leq i\neq j\leq p$, where ${\cal B}$ is a collection of row and column indices $(i,j)$ located within our randomly generated diagonal blocks \footnote{As in \cite{DLX19}, to generate blocks with random sizes, we fix the largest block size at $20$ when $p=200$ and randomly generate the sizes of the remaining blocks from a random integer uniformly picked between $5$ and $20$. When $p=500$, the largest size is $40$, and the random integer is uniformly picked between $10$ and $40$. Block sizes are randomly generated but fixed across all Monte Carlo repetitions.}.

\item Exponentially decaying structure for ${\boldsymbol{\Sigma}}_{t}$ and ${\boldsymbol\Omega}(t)$: The off-diagonal elements are defined by
\begin{equation}\label{eq5.2}
\Sigma _{ij,t} =\left(\kappa^\Sigma_{t}\right)^{\vert i-j\vert }\sqrt{\Sigma_{ii,t}\Sigma_{jj,t}},\ \ \Omega_{ij}(t) =\left(\kappa^\Omega_t\right) ^{\vert i-j\vert }\sqrt{\Omega _{ii}( t) \Omega _{jj}(t) },\ \ 1\leq i\neq j\leq p.
\end{equation}

\end{itemize}

It is clear that the sparsity condition is not satisfied when the off-diagonal elements of ${\boldsymbol{\Sigma}}_{t}$ and ${\boldsymbol\Omega}(t)$ are exponentially decaying as in (\ref{eq5.2}). The number of assets $p$ is set as $p=200$ and $500$ and the replication number is $R=200$

\bigskip


\noindent{\em 5.1.2.\ \ Volatility matrix estimation}

\smallskip

\noindent In the simulation studies, we consider the following volatility matrix estimates.

\begin{itemize}

\item Noise-free spot volatility matrix estimate $\widehat{\boldsymbol\Sigma }_{t}$. This infeasible estimate serves as a benchmark in comparing the numerical performance of various estimation methods. As in Section \ref{sec2}, we apply the kernel smoothing method to estimate $\Sigma_{ij,t}$ by directly using the latent return process ${\mathbf X}_t$, where the bandwidth is determined by the leave-one-out cross validation. We apply four shrinkage methods to $\widehat{\Sigma}_{ij,t}$ for $i\neq j$: hard thresholding (Hard), soft thresholding (Soft), adaptive LASSO (AL) and smoothly clipped absolute deviation (SCAD). For comparison, we also compute the naive estimate without applying any regularisation technique.

\item Noise-contaminated spot volatility matrix estimate $\widetilde{\boldsymbol\Sigma}_{t}$. We combine the kernel smoothing with pre-averaging in Section \ref{sec3.1} to estimate $\Sigma_{ij,t}$ by using the noise-contaminated process ${\mathbf Z}_t$. As in the noise-free estimation, we apply four shrinkage methods to $\widetilde{\Sigma}_{ij,t}$ for $i\neq j$ and also compute the naive estimate without applying the shrinkage.

\item Time-varying noise volatility matrix estimate $\widehat{\boldsymbol\Omega}(t)$. We combine the kernel smoothing with four shrinkage techniques in the estimation as in Section \ref{sec3.2} and also the naive estimate without shrinkage.

\end{itemize}

The choice of tuning parameter in shrinkage is similar to that in \cite{DLX19}. For example, in the noise-free spot volatility estimate, we set the tuning parameter as $\rho_{ij}(t)=\rho(t)(\widehat{\Sigma}_{ii,t}\widehat{\Sigma}_{jj,t})^{1/2}$ where $\rho(t)$ is chosen as the minimum value among the grid of values on $[0,1]$ such that the shrinkage estimate of the spot volatility matrix is positive definite. To evaluate the estimation performance of $\widehat{\boldsymbol\Sigma }_{t}$, we consider $21$ equidistant time points on $[0,T]$ and compute the following Mean Frobenius Loss (MFL) and Mean Spectral Loss (MSL) over $200$ repetitions: 
\begin{eqnarray*}
\text{MFL} &=&\frac{1}{200}\sum_{m=1}^{200}\left( \frac{1}{21}\sum_{j=1}^{21}\left\Vert \boldsymbol{\widehat{\Sigma}}_{t_{j}}^{(m)}-\boldsymbol{\Sigma }_{t_{j}}^{(m)}\right\Vert _{F}\right), \\
\text{MSL} &=&\frac{1}{200}\sum_{m=1}^{200}\left( \frac{1}{21}\sum_{j=1}^{21}\left\Vert \widehat{\boldsymbol\Sigma}_{t_{j}}^{(m)}-\boldsymbol{\Sigma}_{t_{j}}^{(m)}\right\Vert\right),
\end{eqnarray*}%
where $t_{j}$, $j=1,2,\cdots,21$ are the equidistant time points on the interval $[0,T]$, and $\boldsymbol{\widehat{\Sigma}}_{t_{j}}^{(m)}$ and $\boldsymbol{\Sigma }_{t_{j}}^{(m)}$ are respectively the estimated and true spot volatility matrices at $t_{j}$ for the $m$-th repetition. The ``\text{MFL}" and ``\text{MSL}" can be similarly defined for $\widetilde{\boldsymbol\Sigma}_{t}$ and $\widehat{\boldsymbol\Omega}(t)$.

\bigskip

\noindent{\em 5.1.3.\ \ Simulation results}

\smallskip

\noindent Table 1 reports the simulation results when the dimension is $p=200$. The three panels in the table (from top to bottom) report the results where the true volatility matrix structures are banding, block-diagonal, and exponentially decaying, respectively. In each panel, the MFL results are reported on the left, whereas the MSL results are on the right. The first two rows of each panel contain the MFL and MSL results for the spot volatility matrix estimation whereas the third row contains the results for the time-varying noise volatility matrix estimation. 

\smallskip

For the noise-free estimate $\widehat{\boldsymbol\Sigma }_{t}$, when the volatility matrix structure is banding, the performance of the four shrinkage estimators are substantially better than that of the naive estimate (without any shrinkage). In particular, the results of the soft thresholding, adaptive LASSO and SCAD are very similar and their MFL and MSL values are approximately one third of those of the naive estimator. Meanwhile, the performance of the hard thresholding is less accurate (despite the much stronger level of shrinking used), but is still much better than the naive estimate. These results show that the shrinkage technique is an effective tool in estimating the sparse volatility matrix. Similar results are obtained for the noise-contaminated estimate $\widetilde{\boldsymbol\Sigma}_{t}$. Unsurprisingly, due to the microstructure noise, the MFL and MSL values of the local pre-averaging estimates are noticeably higher than the corresponding values of the noise-free estimates. We next turn the attention to the time-varying noise volatility matrix estimate $\widehat{\boldsymbol\Omega}(t)$. As in the spot volatility matrix estimation, the naive method again produces the highest MFL and MSL values. The performance of the four shrinkage estimators are similar with the adaptive LASSO and SCAD being slightly better than the hard and soft thresholding. The simulation results for the block-diagonal and exponentially decaying covariance matrix settings, reported in the middle and bottom panels of Table 1, are fairly close to those for the banding setting. Overall, the results in Table 1 show that the shrinkage methods perform well not only in the sparse covariance matrix settings but also in the non-sparse one (i.e., the exponentially decaying setting). 

\newpage

\begin{center}
{\footnotesize Table 1: Estimation results for the spot volatility and time-varying noise covariance matrices when $p=200$ }
{\footnotesize 
\begin{tabular}{lllllllclllll}
\hline\hline
&  & \multicolumn{11}{c}{``Banding"} \\ \cline{3-7}\cline{9-13}
&  & {\ Naive} & {\ Hard} & {\ Soft} & {\ AL} & {\ SCAD} & \multicolumn{1}{l}{
} & {\ Naive} & {\ Hard} & {\ Soft} & {\ AL} & {\ SCAD} \\ 
\cline{3-7}\cline{9-13}
$\boldsymbol{\widehat{\Sigma}}_{t}$ & {\ MFL} & \multicolumn{1}{r}{14.396} & 
\multicolumn{1}{r}{11.407} & \multicolumn{1}{r}{5.490} & \multicolumn{1}{r}{
4.038} & \multicolumn{1}{r}{4.830} & \multicolumn{1}{l}{MSL} & 
\multicolumn{1}{r}{3.963} & \multicolumn{1}{r}{1.799} & \multicolumn{1}{r}{
1.073} & \multicolumn{1}{r}{0.867} & \multicolumn{1}{r}{0.987} \\ 
$\widetilde{\boldsymbol\Sigma}_{t}$ & {\ MFL} & \multicolumn{1}{r}{18.497} & 
\multicolumn{1}{r}{12.899} & \multicolumn{1}{r}{12.196} & \multicolumn{1}{r}{
12.064} & \multicolumn{1}{r}{12.177} & \multicolumn{1}{l}{MSL} & 
\multicolumn{1}{r}{4.796} & \multicolumn{1}{r}{2.347} & \multicolumn{1}{r}{
2.260} & \multicolumn{1}{r}{2.255} & \multicolumn{1}{r}{2.262} \\ 
$\widehat{\boldsymbol\Omega}(t)$ & {\ MFL} & \multicolumn{1}{r}{11.714} & 
\multicolumn{1}{r}{4.226} & \multicolumn{1}{r}{4.740} & \multicolumn{1}{r}{
3.237} & \multicolumn{1}{r}{3.960} & \multicolumn{1}{l}{MSL} & 
\multicolumn{1}{r}{3.281} & \multicolumn{1}{r}{0.682} & \multicolumn{1}{r}{
1.039} & \multicolumn{1}{r}{0.571} & \multicolumn{1}{r}{0.753} \\ 
\hline
&  & \multicolumn{11}{c}{``Block-diagonal"} \\ \cline{3-7}\cline{9-13}
&  & {\ Naive} & {\ Hard} & {\ Soft} & {\ AL} & {\ SCAD} & \multicolumn{1}{l}{
} & {\ Naive} & {\ Hard} & {\ Soft} & {\ AL} & {\ SCAD} \\ 
\cline{3-7}\cline{9-13}
$\widehat{\boldsymbol\Sigma}_{t}$ & {\ MFL} & \multicolumn{1}{r}{14.398} & 
\multicolumn{1}{r}{11.277} & \multicolumn{1}{r}{5.818} & \multicolumn{1}{r}{
4.786} & \multicolumn{1}{r}{5.424} & \multicolumn{1}{l}{MSL} & 
\multicolumn{1}{r}{4.000} & \multicolumn{1}{r}{2.293} & \multicolumn{1}{r}{
1.310} & \multicolumn{1}{r}{1.233} & \multicolumn{1}{r}{1.386} \\ 
$\widetilde{\boldsymbol\Sigma}_{t}$ & {\ MFL} & \multicolumn{1}{r}{18.475} & 
\multicolumn{1}{r}{12.811} & \multicolumn{1}{r}{12.192} & \multicolumn{1}{r}{
12.059} & \multicolumn{1}{r}{12.158} & \multicolumn{1}{l}{MSL} & 
\multicolumn{1}{r}{4.915} & \multicolumn{1}{r}{2.777} & \multicolumn{1}{r}{
2.663} & \multicolumn{1}{r}{2.669} & \multicolumn{1}{r}{2.662} \\ 
$\widehat{\boldsymbol\Omega}(t)$ & {\ MFL} & \multicolumn{1}{r}{11.713} & 
\multicolumn{1}{r}{4.076} & \multicolumn{1}{r}{4.875} & \multicolumn{1}{r}{
3.240} & \multicolumn{1}{r}{3.964} & \multicolumn{1}{l}{MSL} & 
\multicolumn{1}{r}{3.274} & \multicolumn{1}{r}{0.741} & \multicolumn{1}{r}{
1.098} & \multicolumn{1}{r}{0.606} & \multicolumn{1}{r}{0.816} \\ \hline
&  & \multicolumn{11}{c}{``Exponentially decaying"} \\ \cline{3-7}\cline{9-13}
&  & {\ Naive} & {\ Hard} & {\ Soft} & {\ AL} & {\ SCAD} & \multicolumn{1}{l}{
} & {\ Naive} & {\ Hard} & {\ Soft} & {\ AL} & {\ SCAD} \\ 
\cline{3-7}\cline{9-13}
$\widehat{\boldsymbol\Sigma}_{t}$ & {\ MFL} & \multicolumn{1}{r}{14.402} & 
\multicolumn{1}{r}{12.033} & \multicolumn{1}{r}{6.091} & \multicolumn{1}{r}{
5.287} & \multicolumn{1}{r}{5.976} & \multicolumn{1}{l}{MSL} & 
\multicolumn{1}{r}{4.078} & \multicolumn{1}{r}{2.456} & \multicolumn{1}{r}{
1.410} & \multicolumn{1}{r}{1.348} & \multicolumn{1}{r}{1.510} \\ 
$\widetilde{\boldsymbol\Sigma}_{t}$ & {\ MFL} & \multicolumn{1}{r}{18.738} & 
\multicolumn{1}{r}{13.464} & \multicolumn{1}{r}{12.748} & \multicolumn{1}{r}{
12.655} & \multicolumn{1}{r}{12.739} & \multicolumn{1}{l}{MSL} & 
\multicolumn{1}{r}{4.977} & \multicolumn{1}{r}{2.934} & \multicolumn{1}{r}{
2.810} & \multicolumn{1}{r}{2.819} & \multicolumn{1}{r}{2.815} \\ 
$\widehat{\boldsymbol\Omega}(t)$ & {\ MFL} & \multicolumn{1}{r}{11.715} & 
\multicolumn{1}{r}{4.330} & \multicolumn{1}{r}{4.860} & \multicolumn{1}{r}{
3.355} & \multicolumn{1}{r}{4.077} & \multicolumn{1}{l}{MSL} & 
\multicolumn{1}{r}{3.297} & \multicolumn{1}{r}{0.774} & \multicolumn{1}{r}{
1.085} & \multicolumn{1}{r}{0.626} & \multicolumn{1}{r}{0.833} \\  \hline\hline
\end{tabular}
}
\end{center}

\noindent{\scriptsize The selected bandwidths are $h^\ast=90$ for $\widehat{\boldsymbol\Sigma}_{t}$, $h^\ast=90$ and $b^\ast=4$ for $\widetilde{\boldsymbol\Sigma}_{t}$, and $h_1^{\ast}=90$ for $\widehat{\boldsymbol\Omega}(t)$, where $h^{\ast}=h/\Delta$, $b^{\ast}=b/\Delta$, and $h_{1}^{\ast}=h_1/\Delta$.}

\medskip

The simulation results when the dimension is $p=500$ are reported in Table 2. Overall the results are very similar to those in Table 1, so we omit the detailed discussion and comparison to save the space.

\medskip

\begin{center}
{\footnotesize Table 2: Estimation results for the spot volatility and time-varying noise covariance matrices when $p=500$ }
{\footnotesize 
\begin{tabular}{lllllllllllll}
\hline\hline
&  & \multicolumn{11}{c}{``Banding"} \\ \cline{3-7}\cline{9-13}
&  & {\ Naive} & {\ Hard} & {\ Soft} & {\ AL} & {\ SCAD} &  & {\ Naive} & {\
Hard} & {\ Soft} & {\ AL} & {\ SCAD} \\ \cline{3-7}\cline{9-13}
$\widehat{\boldsymbol\Sigma}_{t}$ & {\ MFL} & \multicolumn{1}{r}{21.971} & 
\multicolumn{1}{r}{4.067} & \multicolumn{1}{r}{5.167} & \multicolumn{1}{r}{
4.916} & \multicolumn{1}{r}{3.954} & {\ MSL} & \multicolumn{1}{r}{3.907} & 
\multicolumn{1}{r}{0.621} & \multicolumn{1}{r}{0.715} & \multicolumn{1}{r}{
0.698} & \multicolumn{1}{r}{0.568} \\ 
$\widetilde{\boldsymbol\Sigma}_{t}$ & {\ MFL} & \multicolumn{1}{r}{28.479} & 
\multicolumn{1}{r}{19.193} & \multicolumn{1}{r}{18.617} & \multicolumn{1}{r}{
17.930} & \multicolumn{1}{r}{18.466} & {\ MSL} & \multicolumn{1}{r}{4.767} & 
\multicolumn{1}{r}{2.339} & \multicolumn{1}{r}{2.281} & \multicolumn{1}{r}{
2.228} & \multicolumn{1}{r}{2.281} \\ 
$\widehat{\boldsymbol\Omega}(t)$ & {\ MFL} & \multicolumn{1}{r}{18.269} & 
\multicolumn{1}{r}{4.045} & \multicolumn{1}{r}{4.826} & \multicolumn{1}{r}{
5.532} & \multicolumn{1}{r}{4.547} & {\ MSL} & \multicolumn{1}{r}{3.307} & 
\multicolumn{1}{r}{0.461} & \multicolumn{1}{r}{0.540} & \multicolumn{1}{r}{
0.675} & \multicolumn{1}{r}{0.519} \\ 
\hline
&  & \multicolumn{11}{c}{``Block-diagonal"} \\  \cline{3-7}\cline{9-13}
&  & {\ Naive} & {\ Hard} & {\ Soft} & {\ AL} & {\ SCAD} &  & {\ Naive} & {\
Hard} & {\ Soft} & {\ AL} & {\ SCAD} \\ \cline{3-7}\cline{9-13}
$\widehat{\boldsymbol\Sigma}_{t}$ & {\ MFL} & \multicolumn{1}{r}{21.973} & 
\multicolumn{1}{r}{5.703} & \multicolumn{1}{r}{6.429} & \multicolumn{1}{r}{
5.928} & \multicolumn{1}{r}{5.480} & {\ MSL} & \multicolumn{1}{r}{3.999} & 
\multicolumn{1}{r}{0.855} & \multicolumn{1}{r}{1.134} & \multicolumn{1}{r}{
0.895} & \multicolumn{1}{r}{0.886} \\ 
$\widetilde{\boldsymbol\Sigma}_{t}$ & {\ MFL} & \multicolumn{1}{r}{28.682} & 
\multicolumn{1}{r}{19.685} & \multicolumn{1}{r}{19.155} & \multicolumn{1}{r}{
18.539} & \multicolumn{1}{r}{19.029} & {\ MSL} & \multicolumn{1}{r}{4.917} & 
\multicolumn{1}{r}{2.854} & \multicolumn{1}{r}{2.782} & \multicolumn{1}{r}{
2.736} & \multicolumn{1}{r}{2.798} \\ 
$\widehat{\boldsymbol\Omega}(t)$ & {\ MFL} & \multicolumn{1}{r}{18.271} & 
\multicolumn{1}{r}{4.208} & \multicolumn{1}{r}{4.935} & \multicolumn{1}{r}{
5.686} & \multicolumn{1}{r}{4.684} & {\ MSL} & \multicolumn{1}{r}{3.312} & 
\multicolumn{1}{r}{0.522} & \multicolumn{1}{r}{0.603} & \multicolumn{1}{r}{
0.751} & \multicolumn{1}{r}{0.572} \\ \hline
&  & \multicolumn{11}{c}{``Exponentially decaying"} \\ \cline{3-7}\cline{9-13}
&  & {\ Naive} & {\ Hard} & {\ Soft} & {\ AL} & {\ SCAD} &  & {\ Naive} & {\
Hard} & {\ Soft} & {\ AL} & {\ SCAD} \\ \cline{3-7}\cline{9-13}
$\widehat{\boldsymbol\Sigma}_{t}$ & {\ MFL} & \multicolumn{1}{r}{21.973} & 
\multicolumn{1}{r}{6.069} & \multicolumn{1}{r}{6.697} & \multicolumn{1}{r}{
6.120} & \multicolumn{1}{r}{5.739} & {\ MSL} & \multicolumn{1}{r}{4.035} & 
\multicolumn{1}{r}{0.894} & \multicolumn{1}{r}{1.173} & \multicolumn{1}{r}{
0.927} & \multicolumn{1}{r}{0.921} \\ 
$\widetilde{\boldsymbol\Sigma}_{t}$ & {\ MFL} & \multicolumn{1}{r}{28.867} & 
\multicolumn{1}{r}{20.195} & \multicolumn{1}{r}{19.561} & \multicolumn{1}{r}{
18.950} & \multicolumn{1}{r}{19.454} & {\ MSL} & \multicolumn{1}{r}{4.938} & 
\multicolumn{1}{r}{2.914} & \multicolumn{1}{r}{2.836} & \multicolumn{1}{r}{
2.788} & \multicolumn{1}{r}{2.850} \\ 
$\widehat{\boldsymbol\Omega}(t)$ & {\ MFL} & \multicolumn{1}{r}{18.275} & 
\multicolumn{1}{r}{4.335} & \multicolumn{1}{r}{5.001} & \multicolumn{1}{r}{
5.763} & \multicolumn{1}{r}{4.745} & {\ MSL} & \multicolumn{1}{r}{3.322} & 
\multicolumn{1}{r}{0.533} & \multicolumn{1}{r}{0.610} & \multicolumn{1}{r}{
0.757} & \multicolumn{1}{r}{0.578} \\ 
\hline\hline
\end{tabular}
}
\end{center}

\noindent{\scriptsize The selected bandwidths are $h^{\ast }=240$ for $\widehat{\boldsymbol\Sigma}_{t}$, $h^{\ast }=240$, $b^{\ast }=4$ for $\widetilde{\boldsymbol\Sigma}_{t}$ and $h_1^{\ast }=240$ for $\widehat{\boldsymbol\Omega}(t)$, where $h^{\ast}=h/\Delta$, $b^{\ast}=b/\Delta$, and $h_{1}^{\ast}=h_1/\Delta$.}

\subsection{Simulation for factor-based spot volatility matrix estimation}\label{sec5.2}


\bigskip

\noindent{\em 5.2.1.\ \ Simulation setup}

\smallskip

\noindent We generate ${\mathbf Y}_t$ via (\ref{eq4.1}), where the $p$-dimensional idiosyncratic returns follow the dynamics of $d\mathbf{X}_{t}$ defined in (\ref{eq5.1}). In this simulation, we only consider $p=500$. As in \cite{AKX20}, we adopt a three-factor model, where the factors $\mathbf{F}_{t}=\left(F_{1,t},F_{2,t},F_{3,t}\right) ^{^\intercal}$ are generated by
\begin{equation*}
\left( 
\begin{array}{c}
dF_{1,t} \\ 
dF_{2,t} \\ 
dF_{3,t}%
\end{array}%
\right) =\left( 
\begin{array}{c}
\mu _{1}^{F} \\ 
\mu _{2}^{F} \\ 
\mu _{3}^{F}%
\end{array}%
\right) dt+\left( 
\begin{array}{ccc}
\sigma _{1,t} & 0 & 0 \\ 
0 & \sigma _{2,t} & 0 \\ 
0 & 0 & \sigma _{3,t}%
\end{array}%
\right) \left( 
\begin{array}{ccc}
1 & \rho_{12} & \rho_{13} \\ 
\rho_{12} & 1 & \rho_{23} \\ 
\rho_{13} & \rho_{23} & 1%
\end{array}%
\right) \left( 
\begin{array}{c}
dW_{1,t}^{F} \\ 
dW_{2,t}^{F} \\ 
dW_{3,t}^{F}%
\end{array}%
\right). 
\end{equation*}%
The factor volatilities are driven by 
\begin{equation*}
d\sigma _{k,t}^{2}=\tilde{\kappa}_{k}\left( \tilde{\alpha}_{k}-\sigma
_{k,t}^{2}\right) dt+\tilde{\nu}_{k}\sigma _{k,t}d\widetilde{W}_{k,t},\text{ \ \
\ \ \ }k=1,2,3,
\end{equation*}%
where ${\sf E}[ dW_{k,t}^{F}d\widetilde{W}_{k,t}] =\rho_{k}dt$, allowing for potential leverage effects in the factor dynamics. Both $W_{k,t}^{F}$ and $\widetilde{W}_{k,t}$ are standard univariate Brownian motions. In the simulation, we set $\left( \tilde{\kappa}_{1},\tilde{\kappa}_{2},\tilde{%
\kappa}_{3}\right) =\left( 3,4,5\right) $, $\left( \tilde{\alpha}_{1},\tilde{%
\alpha}_{2},\tilde{\alpha}_{3}\right) =\left( 0.09,0.04,0.06\right) $, $%
\left( \tilde{\nu}_{1},\tilde{\nu}_{2},\tilde{\nu}_{3}\right) =\left(
0.3,0.4,0.3\right) $, $\left( \mu _{1}^{F},\mu _{2}^{F},\mu _{3}^{F}\right)
=\left( 0.05,0.03,0.02\right) $, $\left( \rho _{1},\rho _{2},\rho
_{3}\right) =\left( -0.6,-0.4,-0.25\right) $ and $\left( \rho _{12},\rho
_{13},\rho _{23}\right) =\left( 0.05,0.10,0.15\right) $.

\smallskip

We consider the following three cases for generating the time-varying beta processes: $\beta_{i}(t)=\left[\beta_{i,1}(t), \beta_{i,2}(t), \beta_{i,3}(t)\right]^{^\intercal}$, $i=1,\cdots,p$. 

\begin{itemize}

\item Constant betas. The factor loadings are constants over time, i.e., $\beta _{i,l}(t)=\beta _{i,l}$, $i=1,\cdots,p$ and $l=1,2,3$. For each $i$, we set $\beta _{i,1}\sim {\sf U}\left(
0.25,2.25\right)$ and $\beta _{i,2},\beta _{i,3}\sim {\sf U}\left(-0.5,0.5\right)$.

\item Deterministic time-varying betas. Consider the following deterministic function: 
\begin{equation*}
\beta _{i,l}\left( t\right) =\frac{1}{2}\left[ \cos \left( \pi (
t-\omega _{i,l})/T \right) +1\right] \times \left( \overline{\beta} _{i,l}-\underline{\beta}_{i,l}\right) +\underline{\beta} _{i,l},\ \ i=1,\cdots,p,\ \ l=1,2,3,
\end{equation*}%
where $\omega _{i,1},\omega_{i,2},\omega _{i,3}\sim {\sf U}(0, 2T)$, $(\underline{\beta} _{i,1},\overline{\beta}_{i,1})$ is a pair of two random numbers from ${\sf U}\left( 0.25,2.25\right)$ whereas
$(\underline{\beta}_{i,2},\overline{\beta}_{i,2})$ and $(\underline{\beta}_{i,3},\overline{\beta}_{i,3})$ are pairs of random numbers from ${\sf U}\left( -0.5,0.5\right)$.

\item Stochastic time-varying betas. As in \cite{AKX20}, we consider the following diffusion process: 
\[
d\beta _{i,l}(t)=\kappa _{i,l}^{\beta }\left( \alpha _{i,l}^{\beta }-\beta
_{i,l}(t)\right) dt+\upsilon _{i,l}^{\beta }dW_{i,l,t}^{\beta},\ \ i=1,\cdots,p,\ \ l=1,2,3,
\]
where $W_{i,l,t}^{\beta}$ are standard Brownian motions independently over $i$ and $l$, $\kappa _{i,1}^{\beta},\kappa_{i,2}^{\beta},\kappa _{i,3}^{\beta}\sim {\sf U}(1,3)$, $\alpha _{i,1}^{\beta }\sim{\sf U}( 0.25,2.25)$, $\alpha _{i,2}^{\beta},\alpha_{i,3}^{\beta }\sim {\sf U}(-0.5, 0.5)$ and $\upsilon _{i,1}^{\beta },\upsilon _{i,2}^{\beta},\upsilon _{i,3}^{\beta } \sim {\sf U}(2,4)$.

\end{itemize}

\noindent{\em 5.2.2.\ \ Simulation Results}

\smallskip

\noindent The spot idiosyncratic volatility matrix is estimated via (\ref{eq4.9}). For ease of comparison, we use exactly the same bandwidth as in our first experiment. The results for the noise-free and noise-contaminated spot idiosyncratic volatility matrix estimates $\widehat{\boldsymbol{\Sigma }}_{t}$ and $\widetilde{\boldsymbol{\Sigma }}_{t}$ measured by MFL and MSL are reported in Table 3, which reveal some desirable observations. Firstly, we note that our estimation results in terms of MFL and MSL are almost identical across different types of dynamics of factor loadings, indicating that the developed estimation procedure is robust in finite samples to different assumption of the factor loading dynamics as long as they satisfy our smooth restriction, see Assumption \ref{ass:6}(iv). Secondly, the MFL and MSL values are similar to those reported in Table 2 which were obtained based on data generating model without common factors. This means that the proposed nonparametric time-varying high frequency regression can effectively remove common factors, resulting in accurate estimation of the spot idiosyncratic volatility matrix. 

\smallskip

The factor-based spot volatility matrix of ${\mathbf Y}_t$ is estimated via (\ref{eq4.10}). As discussed in Section \ref{sec4}, we measure the accuracy of the spiked volatility matrix estimate by the relative error defined above Theorem \ref{thm:4}, i.e., consider the following Mean Relative Loss (MRL):
\[
	\text{MRL}=\frac{1}{200}\sum_{m=1}^{200}\left( \frac{1}{21}%
	\sum_{j=1}^{21}\left\Vert \widehat{\boldsymbol{\Sigma }}_{t_{j}}^{Y,(m)}-%
	\boldsymbol{\Sigma }_{t_{j}}^{Y,(m)}\right\Vert _{\boldsymbol{\Sigma }%
		_{t_{j}}^{Y,(m)}}\right) .
\]
The relevant results are reported in Table 4, where $\widehat{\boldsymbol\Sigma }_{t}^{Y}$ and $\widetilde{\boldsymbol\Sigma }_{t}^{Y}$ denote the noise-free and noise-contaminated factor-based spot volatility matrix estimates, respectively. We can see that the performance of the shrinkage estimates is substantially better than that of the naive estimate. Unsurprisingly, due to the presence of microstructure noise, the MRL results of $\widetilde{\boldsymbol\Sigma }_{t}^{Y}$ are much higher than those of $\widehat{\boldsymbol\Sigma }_{t}^{Y}$. As in Table 3, our proposed estimation is robust to different factor loading dynamics.

\begin{landscape}
\begin{center}

Table 3: Estimation results for the spot idiosyncratic volatility matrices

{\footnotesize 
\begin{tabular}{llllllllclllll}
\hline\hline
&  &  & \multicolumn{11}{c}{\textquotedblleft Banding"} \\ \cline{4-14}
$\beta $ Dynamics &  &  & \multicolumn{5}{c}{Frobenius Norm} &  & 
\multicolumn{5}{c}{Spectral Norm} \\ \cline{4-8}\cline{10-14}
&  &  & \ Naive & \ Hard & \ Soft & \ AL & \ SCAD & \multicolumn{1}{l}{} & \
Naive & \ Hard & \ Soft & \ AL & \ SCAD \\ \cline{4-8}\cline{10-14}
Constant & $\widehat{\Sigma }_{t}$ & \ MFL & \multicolumn{1}{r}{21.9037} & 
\multicolumn{1}{r}{4.2461} & \multicolumn{1}{r}{5.2485} & \multicolumn{1}{r}{
4.9880} & \multicolumn{1}{r}{3.9910} & \multicolumn{1}{l}{MSL} & 
\multicolumn{1}{r}{3.8887} & \multicolumn{1}{r}{0.6359} & \multicolumn{1}{r}{
0.7291} & \multicolumn{1}{r}{0.7154} & \multicolumn{1}{r}{0.5720} \\ 
& $\widetilde{\boldsymbol{\Sigma }}_{t}$ & \ MFL & \multicolumn{1}{r}{30.6646
} & \multicolumn{1}{r}{19.3752} & \multicolumn{1}{r}{18.3036} & 
\multicolumn{1}{r}{17.7552} & \multicolumn{1}{r}{18.1388} & 
\multicolumn{1}{l}{MSL} & \multicolumn{1}{r}{11.1910} & \multicolumn{1}{r}{
2.3576} & \multicolumn{1}{r}{2.2653} & \multicolumn{1}{r}{2.2160} & 
\multicolumn{1}{r}{2.2554} \\ 
Deterministic & $\widehat{\Sigma }_{t}$ & \ MFL & \multicolumn{1}{r}{21.9127}
& \multicolumn{1}{r}{4.1916} & \multicolumn{1}{r}{5.2503} & 
\multicolumn{1}{r}{4.9898} & \multicolumn{1}{r}{3.9842} & \multicolumn{1}{l}{
MSL} & \multicolumn{1}{r}{3.8901} & \multicolumn{1}{r}{0.6313} & 
\multicolumn{1}{r}{0.7288} & \multicolumn{1}{r}{0.7144} & \multicolumn{1}{r}{
0.5712} \\ 
& $\widetilde{\boldsymbol{\Sigma }}_{t}$ & \ MFL & \multicolumn{1}{r}{30.5672
} & \multicolumn{1}{r}{19.3633} & \multicolumn{1}{r}{18.2947} & 
\multicolumn{1}{r}{17.7267} & \multicolumn{1}{r}{18.1284} & 
\multicolumn{1}{l}{MSL} & \multicolumn{1}{r}{10.9662} & \multicolumn{1}{r}{
2.3571} & \multicolumn{1}{r}{2.2636} & \multicolumn{1}{r}{2.2128} & 
\multicolumn{1}{r}{2.2536} \\ 
Stochastic & $\widehat{\Sigma }_{t}$ & \ MFL & \multicolumn{1}{r}{21.9099} & 
\multicolumn{1}{r}{4.2123} & \multicolumn{1}{r}{5.2498} & \multicolumn{1}{r}{
4.9893} & \multicolumn{1}{r}{3.9872} & \multicolumn{1}{l}{MSL} & 
\multicolumn{1}{r}{3.8896} & \multicolumn{1}{r}{0.6331} & \multicolumn{1}{r}{
0.7289} & \multicolumn{1}{r}{0.7149} & \multicolumn{1}{r}{0.5717} \\ 
& $\widetilde{\boldsymbol{\Sigma }}_{t}$ & \ MFL & \multicolumn{1}{r}{30.7323
} & \multicolumn{1}{r}{19.3896} & \multicolumn{1}{r}{18.3164} & 
\multicolumn{1}{r}{17.7839} & \multicolumn{1}{r}{18.1538} & 
\multicolumn{1}{l}{MSL} & \multicolumn{1}{r}{11.3262} & \multicolumn{1}{r}{
2.3603} & \multicolumn{1}{r}{2.2708} & \multicolumn{1}{r}{2.2203} & 
\multicolumn{1}{r}{2.2602} \\ \hline
&  &  & \multicolumn{11}{c}{\textquotedblleft Block-diagonal"} \\ 
\cline{4-14}
$\beta $ Dynamics &  &  & \multicolumn{5}{c}{Frobenius Norm} &  & 
\multicolumn{5}{c}{Spectral Norm} \\ \cline{4-8}\cline{10-14}
&  &  & \ Naive & \ Hard & \ Soft & \ AL & \ SCAD & \multicolumn{1}{l}{} & \
Naive & \ Hard & \ Soft & \ AL & \ SCAD \\ \cline{4-8}\cline{10-14}
Constant & $\widehat{\Sigma }_{t}$ & \ MFL & \multicolumn{1}{r}{21.9047} & 
\multicolumn{1}{r}{5.6802} & \multicolumn{1}{r}{6.4718} & \multicolumn{1}{r}{
5.9421} & \multicolumn{1}{r}{5.4710} & \multicolumn{1}{l}{MSL} & 
\multicolumn{1}{r}{3.9741} & \multicolumn{1}{r}{0.8722} & \multicolumn{1}{r}{
1.1481} & \multicolumn{1}{r}{0.9106} & \multicolumn{1}{r}{0.9014} \\ 
& $\widetilde{\boldsymbol{\Sigma }}_{t}$ & \ MFL & \multicolumn{1}{r}{30.7266
} & \multicolumn{1}{r}{19.8195} & \multicolumn{1}{r}{18.8114} & 
\multicolumn{1}{r}{18.3162} & \multicolumn{1}{r}{18.6638} & 
\multicolumn{1}{l}{MSL} & \multicolumn{1}{r}{10.9751} & \multicolumn{1}{r}{
2.8701} & \multicolumn{1}{r}{2.7656} & \multicolumn{1}{r}{2.7097} & 
\multicolumn{1}{r}{2.7551} \\ 
Deterministic & $\widehat{\boldsymbol{\Sigma }}_{t}$ & \ MFL & 
\multicolumn{1}{r}{21.9137} & \multicolumn{1}{r}{5.6821} & 
\multicolumn{1}{r}{6.4738} & \multicolumn{1}{r}{5.9436} & \multicolumn{1}{r}{
5.4729} & \multicolumn{1}{l}{MSL} & \multicolumn{1}{r}{3.9754} & 
\multicolumn{1}{r}{0.8718} & \multicolumn{1}{r}{1.1479} & \multicolumn{1}{r}{
0.9103} & \multicolumn{1}{r}{0.9012} \\ 
& $\widetilde{\boldsymbol{\Sigma }}_{t}$ & \ MFL & \multicolumn{1}{r}{30.6284
} & \multicolumn{1}{r}{19.8161} & \multicolumn{1}{r}{18.8043} & 
\multicolumn{1}{r}{18.2953} & \multicolumn{1}{r}{18.6547} & 
\multicolumn{1}{l}{MSL} & \multicolumn{1}{r}{10.7452} & \multicolumn{1}{r}{
2.8706} & \multicolumn{1}{r}{2.7663} & \multicolumn{1}{r}{2.7092} & 
\multicolumn{1}{r}{2.7559} \\ 
Stochastic & $\widehat{\boldsymbol{\Sigma }}_{t}$ & \ MFL & 
\multicolumn{1}{r}{21.9108} & \multicolumn{1}{r}{5.6811} & 
\multicolumn{1}{r}{6.4732} & \multicolumn{1}{r}{5.9433} & \multicolumn{1}{r}{
5.4722} & \multicolumn{1}{l}{MSL} & \multicolumn{1}{r}{3.9751} & 
\multicolumn{1}{r}{0.8721} & \multicolumn{1}{r}{1.1480} & \multicolumn{1}{r}{
0.9104} & \multicolumn{1}{r}{0.9013} \\ 
& $\widetilde{\boldsymbol{\Sigma }}_{t}$ & \ MFL & \multicolumn{1}{r}{30.7955
} & \multicolumn{1}{r}{19.8314} & \multicolumn{1}{r}{18.8237} & 
\multicolumn{1}{r}{18.3434} & \multicolumn{1}{r}{18.6767} & 
\multicolumn{1}{l}{MSL} & \multicolumn{1}{r}{11.1149} & \multicolumn{1}{r}{
2.8719} & \multicolumn{1}{r}{2.7691} & \multicolumn{1}{r}{2.7142} & 
\multicolumn{1}{r}{2.7584} \\ \hline
&  &  & \multicolumn{11}{c}{\textquotedblleft Exponentially decaying"} \\ 
\cline{4-14}
$\beta $ Dynamics &  &  & \multicolumn{5}{c}{Frobenius Norm} & 
\multicolumn{1}{l}{} & \multicolumn{5}{c}{Spectral Norm} \\ 
\cline{4-8}\cline{10-14}
&  &  & \ Naive & \ Hard & \ Soft & \ AL & \ SCAD & \multicolumn{1}{l}{} & \
Naive & \ Hard & \ Soft & \ AL & \ SCAD \\ \cline{4-8}\cline{10-14}
Constant & $\widehat{\Sigma }_{t}$ & \ MFL & \multicolumn{1}{r}{21.9057} & 
\multicolumn{1}{r}{6.0626} & \multicolumn{1}{r}{6.7715} & \multicolumn{1}{r}{
6.1573} & \multicolumn{1}{r}{5.7617} & \multicolumn{1}{l}{MSL} & 
\multicolumn{1}{r}{4.0142} & \multicolumn{1}{r}{0.9106} & \multicolumn{1}{r}{
1.1898} & \multicolumn{1}{r}{0.9453} & \multicolumn{1}{r}{0.9388} \\ 
& $\widetilde{\boldsymbol{\Sigma }}_{t}$ & \ MFL & \multicolumn{1}{r}{30.8728
} & \multicolumn{1}{r}{20.3802} & \multicolumn{1}{r}{19.2709} & 
\multicolumn{1}{r}{18.7715} & \multicolumn{1}{r}{19.1262} & 
\multicolumn{1}{l}{MSL} & \multicolumn{1}{r}{10.8858} & \multicolumn{1}{r}{
2.9381} & \multicolumn{1}{r}{2.8260} & \multicolumn{1}{r}{2.7707} & 
\multicolumn{1}{r}{2.8154} \\ 
Deterministic & $\widehat{\boldsymbol{\Sigma }}_{t}$ & \ MFL & 
\multicolumn{1}{r}{21.9147} & \multicolumn{1}{r}{6.0709} & 
\multicolumn{1}{r}{6.7737} & \multicolumn{1}{r}{6.1589} & \multicolumn{1}{r}{
5.7637} & \multicolumn{1}{l}{MSL} & \multicolumn{1}{r}{4.0156} & 
\multicolumn{1}{r}{0.9112} & \multicolumn{1}{r}{1.1896} & \multicolumn{1}{r}{
0.9450} & \multicolumn{1}{r}{0.9387} \\ 
& $\widetilde{\boldsymbol{\Sigma }}_{t}$ & \ MFL & \multicolumn{1}{r}{30.7746
} & \multicolumn{1}{r}{20.3564} & \multicolumn{1}{r}{19.2632} & 
\multicolumn{1}{r}{18.7460} & \multicolumn{1}{r}{19.1173} & 
\multicolumn{1}{l}{MSL} & \multicolumn{1}{r}{10.6538} & \multicolumn{1}{r}{
2.9354} & \multicolumn{1}{r}{2.8247} & \multicolumn{1}{r}{2.7673} & 
\multicolumn{1}{r}{2.8140} \\ 
Stochastic & $\widehat{\boldsymbol{\Sigma }}_{t}$ & \ MFL & 
\multicolumn{1}{r}{21.9118} & \multicolumn{1}{r}{6.0636} & 
\multicolumn{1}{r}{6.7730} & \multicolumn{1}{r}{6.1585} & \multicolumn{1}{r}{
5.7630} & \multicolumn{1}{l}{MSL} & \multicolumn{1}{r}{4.0151} & 
\multicolumn{1}{r}{0.9106} & \multicolumn{1}{r}{1.1897} & \multicolumn{1}{r}{
0.9451} & \multicolumn{1}{r}{0.9388} \\ 
& $\widetilde{\boldsymbol{\Sigma }}_{t}$ & \ MFL & \multicolumn{1}{r}{30.9430
} & \multicolumn{1}{r}{20.3820} & \multicolumn{1}{r}{19.2839} & 
\multicolumn{1}{r}{18.8017} & \multicolumn{1}{r}{19.1405} & 
\multicolumn{1}{l}{MSL} & \multicolumn{1}{r}{11.0295} & \multicolumn{1}{r}{
2.9381} & \multicolumn{1}{r}{2.8291} & \multicolumn{1}{r}{2.7745} & 
\multicolumn{1}{r}{2.8173} \\ \hline\hline
\end{tabular}%
}
\end{center}
\end{landscape}

\medskip

\begin{center}

Table 4: Mean relative loss for the factor-based spot volatility matrix estimation

{\footnotesize 
	\begin{tabular}{lllllll}
		\hline\hline
		&  &  \multicolumn{5}{c}{\textquotedblleft Banding"} \\ 
		\cline{3-7}\cline{3-7}
		$\beta $ Dynamics &  &  \ Naive & \ Hard & \ Soft & \ AL & \ SCAD \\ \cline{3-7}\cline{3-7}
		Constant & $\widehat{\boldsymbol\Sigma }_{t}^{Y}$ &  \multicolumn{1}{r}{1.1192}
		& \multicolumn{1}{r}{0.5417} & \multicolumn{1}{r}{0.7802} & 
		\multicolumn{1}{r}{0.7762} & \multicolumn{1}{r}{0.4391} \\ 
		& $\widetilde{\boldsymbol\Sigma }_{t}^{Y}$  & \multicolumn{1}{r}{2.2280} & 
		\multicolumn{1}{r}{1.7243} & \multicolumn{1}{r}{1.4939} & \multicolumn{1}{r}{
			1.4478} & \multicolumn{1}{r}{1.4654} \\ 
		Deterministic & $\widehat{\boldsymbol\Sigma }_{t}^{Y}$  & \multicolumn{1}{r}{
			1.1207} & \multicolumn{1}{r}{0.5257} & \multicolumn{1}{r}{0.7823} & 
		\multicolumn{1}{r}{0.7775} & \multicolumn{1}{r}{0.4371} \\ 
		& $\widetilde{\boldsymbol\Sigma }_{t}^{Y}$  & \multicolumn{1}{r}{2.2287} & 
		\multicolumn{1}{r}{1.7182} & \multicolumn{1}{r}{1.4882} & \multicolumn{1}{r}{
			1.4385} & \multicolumn{1}{r}{1.4586} \\ 
		Stochastic & $\widehat{\boldsymbol\Sigma }_{t}^{Y}$  & \multicolumn{1}{r}{1.1208}
		& \multicolumn{1}{r}{0.5389} & \multicolumn{1}{r}{0.7829} & 
		\multicolumn{1}{r}{0.7780} & \multicolumn{1}{r}{0.4406} \\ 
		& $\widetilde{\boldsymbol\Sigma }_{t}^{Y}$  & \multicolumn{1}{r}{2.2273} & 
		\multicolumn{1}{r}{1.7279} & \multicolumn{1}{r}{1.4986} & \multicolumn{1}{r}{
			1.4544} & \multicolumn{1}{r}{1.4719} \\ \hline
		&  &  \multicolumn{5}{c}{\textquotedblleft Block-diagonal"} \\
		\cline{3-7}\cline{3-7}
		$\beta $ Dynamics &  &  \ Naive & \ Hard & \ Soft & \ AL & \ SCAD \\ \cline{3-7}\cline{3-7}
		Constant & $\widehat{\boldsymbol\Sigma }_{t}^{Y}$ &  \multicolumn{1}{r}{1.1192}
		& \multicolumn{1}{r}{0.3842} & \multicolumn{1}{r}{0.3650} & 
		\multicolumn{1}{r}{0.3962} & \multicolumn{1}{r}{0.3249} \\ 
		& $\widetilde{\boldsymbol\Sigma }_{t}^{Y}$ & \multicolumn{1}{r}{1.7146} & 
		\multicolumn{1}{r}{0.8421} & \multicolumn{1}{r}{0.7938} & \multicolumn{1}{r}{
			0.7176} & \multicolumn{1}{r}{0.7514} \\ 
		Deterministic & $\widehat{\boldsymbol\Sigma }_{t}^{Y}$  & \multicolumn{1}{r}{
			1.1201} & \multicolumn{1}{r}{0.3840} & \multicolumn{1}{r}{0.3651} & 
		\multicolumn{1}{r}{0.3958} & \multicolumn{1}{r}{0.3241} \\ 
		& $\widetilde{\boldsymbol\Sigma }_{t}^{Y}$ & \multicolumn{1}{r}{1.7152} & 
		\multicolumn{1}{r}{0.8410} & \multicolumn{1}{r}{0.7911} & \multicolumn{1}{r}{
			0.7155} & \multicolumn{1}{r}{0.7486} \\ 
		Stochastic & $\widehat{\boldsymbol\Sigma }_{t}^{Y}$  & \multicolumn{1}{r}{1.1202}
		& \multicolumn{1}{r}{0.3868} & \multicolumn{1}{r}{0.3678} & 
		\multicolumn{1}{r}{0.3983} & \multicolumn{1}{r}{0.3272} \\ 
		& $\widetilde{\boldsymbol\Sigma }_{t}^{Y}$ & \multicolumn{1}{r}{1.7146} & 
		\multicolumn{1}{r}{0.8435} & \multicolumn{1}{r}{0.7949} & \multicolumn{1}{r}{
			0.7188} & \multicolumn{1}{r}{0.7528} \\ \hline
		&    & \multicolumn{5}{c}{\textquotedblleft Exponentially decaying"}  \\ 
		\cline{3-7}\cline{3-7}
		$\beta $ Dynamics &  &   \ Naive & \ Hard & \ Soft & \ AL & \ SCAD \\ \cline{3-7}\cline{3-7}
		Constant & $\widehat{\boldsymbol\Sigma }_{t}^{Y}$  & \multicolumn{1}{r}{1.1192}
		& \multicolumn{1}{r}{0.4086} & \multicolumn{1}{r}{0.3726} & 
		\multicolumn{1}{r}{0.4055} & \multicolumn{1}{r}{0.3347} \\ 
		& $\widetilde{\boldsymbol\Sigma }_{t}^{Y}$ & \multicolumn{1}{r}{1.7338} & 
		\multicolumn{1}{r}{0.8636} & \multicolumn{1}{r}{0.8047} & \multicolumn{1}{r}{
			0.7272} & \multicolumn{1}{r}{0.7619} \\ 
		Deterministic & $\widehat{\boldsymbol\Sigma }_{t}^{Y}$  & \multicolumn{1}{r}{
			1.1201} & \multicolumn{1}{r}{0.4079} & \multicolumn{1}{r}{0.3727} & 
		\multicolumn{1}{r}{0.4051} & \multicolumn{1}{r}{0.3339} \\ 
		& $\widetilde{\boldsymbol\Sigma }_{t}^{Y}$  & \multicolumn{1}{r}{1.7344} & 
		\multicolumn{1}{r}{0.8614} & \multicolumn{1}{r}{0.8016} & \multicolumn{1}{r}{
			0.7249} & \multicolumn{1}{r}{0.7589} \\ 
		Stochastic & $\widehat{\boldsymbol\Sigma }_{t}^{Y}$  & \multicolumn{1}{r}{1.1203}
		& \multicolumn{1}{r}{0.4111} & \multicolumn{1}{r}{0.3754} & 
		\multicolumn{1}{r}{0.4075} & \multicolumn{1}{r}{0.3370} \\ 
		& $\widetilde{\boldsymbol\Sigma }_{t}^{Y}$  & \multicolumn{1}{r}{1.7338} & 
		\multicolumn{1}{r}{0.8645} & \multicolumn{1}{r}{0.8058} & \multicolumn{1}{r}{
			0.7283} & \multicolumn{1}{r}{0.7631} \\ \hline\hline
	\end{tabular}%
}

\end{center}


\section{Empirical Study}\label{sec6}
\renewcommand{\theequation}{6.\arabic{equation}} \setcounter{equation}{0}

We apply the proposed methods to the intraday returns of the S\&P 500 component stocks to demonstrate the effectiveness of our nonparametric spot volatility matrix estimation in revealing time-varying patterns. We consider the 5-minute returns of the S\&P 500 stocks collected in September 2008. On September 15 Lehman Brothers filed for bankruptcy, causing shockwaves throughout the global financial system. Hence, it is interesting to examine how the spot volatility structure of the returns evolved during this one-month period. In addition, to demonstrate the effectiveness of our model with observed risk factors in explaining the systemic component of the dependence structure, we also collect the 5-minute returns of twelve factors. The first three factors are constructed in \cite{AKX20} as our proxy for the market (MKT), small-minus-big market capitalisation (SMB), and high-minus-low price-earning ratio (HML). The other nine factors are the widely available sector SDPR ETFs, which are intended to tract the following nine largest S\&P sectors: Energy (XLE), Materials (XLB), Industrials (XLI), Consumer Discretionary (XLY), Consumer Staples (XLP), Health Care (XLV), Financial (XLF), Information Technology (XLK), Utilities (XLU). We sort our stocks according to their GICS (Global Industry Classification Standard) codes, so that they are grouped by sectors in the above order. Consequently, the correlation (sub)matrix for stocks within each sector corresponds to a block on the diagonal of the full correlation matrix \citep[e.g.,][]{FFX16}.

We only use stocks that are included in the S\&P 500 index and whose GICS codes are unchanged in September 2008. We also exclude stocks that do not belong to any of the above nine sectors. This leaves us with a total of $p=482$ stocks. All the returns are synchronised via the previous-tick subsampling technique \citep{Zh11}, and overnight returns are removed because of potential dividends and stock splits. Consequently, we have $1638$ time series
observations for each of the $482$ stocks. For the 5-minute returns, we may assume that the potential impact of microstructure noises are negligible. The smoothing parameter in our kernel estimation is chosen as $h=2/252$ (equivalent to 2 trading days)\footnote{We experimented three bandwidth choices, namely, 1 day, 2 days and 3 days.
	We found that $h=1/252$ ($h=3/252$) produced clearly undersmoothed
	(oversmoothed) time series of estimated deciles of the cross sectional
	distribution of the variances and pairwise correlations of our returns,
	whereas $h=2/252$ seems to be most reasonable. Our qualitative
	conclusion is unaffected by the choices of $h$ within the range of 1 to 3
	days.}.

\begin{figure}
	\centering
	\includegraphics[width=1.0\textwidth]{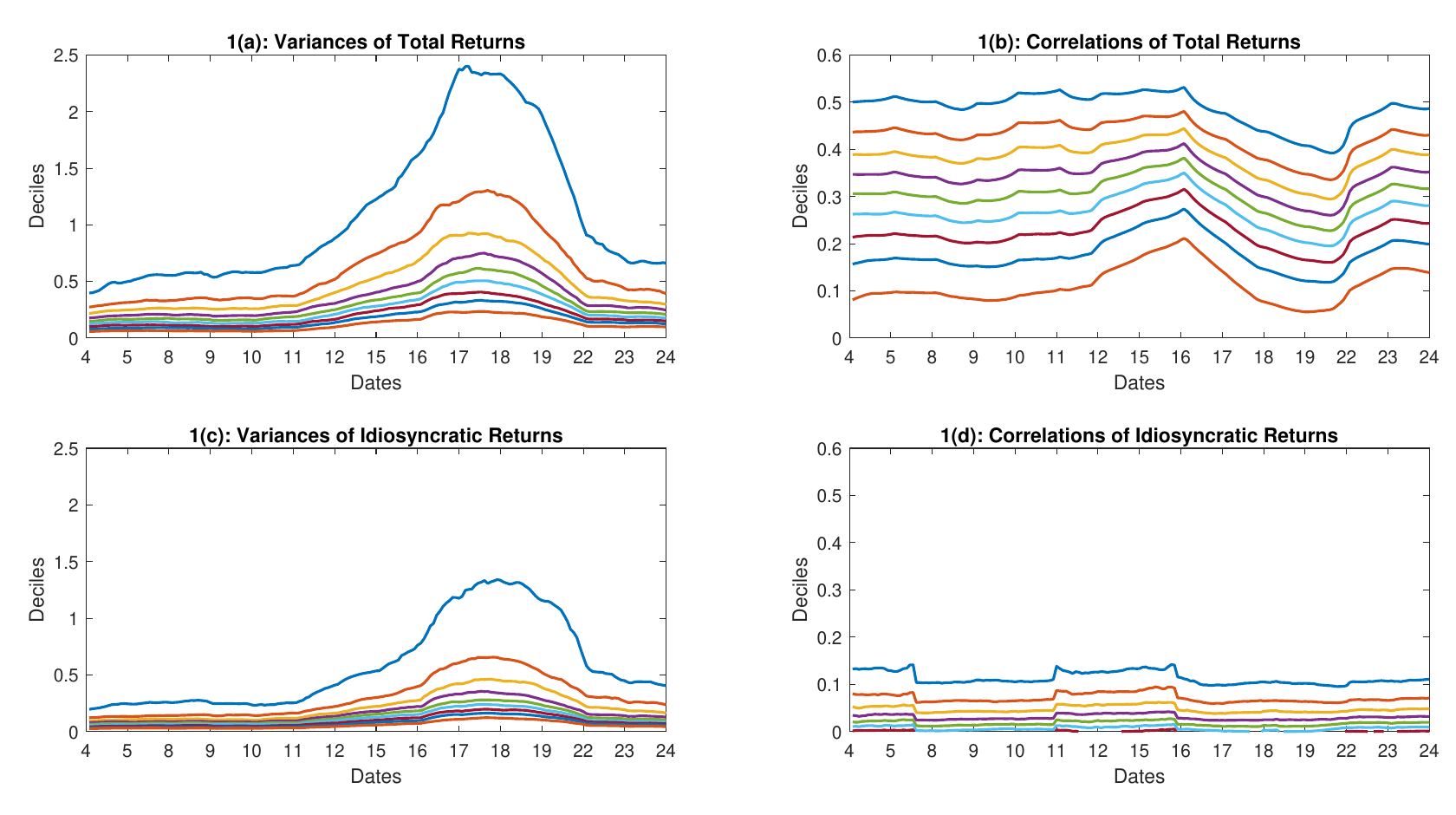}
	\caption{Deciles of spot variances and pairwise correlations in September 2008}\label{fig:figure_1}
\end{figure}

We start with estimating the spot volatility matrices of the total returns (i.e., the observed returns) without incorporating the observed factors or applying any shrinkage. To visualise the potential time variation of the estimated spot matrices, as in \cite{BHMR19}, we plot the time series of deciles of the distribution of the estimated spot variances and the pairwise correlations\footnote{The nine decile levels we use in this study are the 10th, 20th,..., and 90th percentiles.}. The patterns of the spot variances and correlations in Figure \ref{fig:figure_1}(a) and (b) reveal some clear evidence of time variation in our sampling period. We note that the distributions of the variances are relatively narrow and stay low on the first few days of the month. However, close to Lehman Brothers' announcement on the 15th, they start to rise and get wider quite rapidly and reach the peak around the 17th and the 18th. The spot variances at the peak are much higher than those on the earlier days of the month. The distributions return to the earlier level in the following week. In contrast, the distributions of pairwise spot correlations also start to shift up around the same time, but quickly reach the peak on the 16th (only one day after the bankruptcy news), and then dip to a relatively low point around the 19th before returning to the earlier level. Such time-varying features in the dynamics covariance structure are quite interesting and sensible, reflecting the impact of market news. Hence, our proposed spot volatility matrix estimation methodology provides a useful tool for revealing such dynamics.%

\begin{figure}
	\centering
	\includegraphics[width=1.0\textwidth]{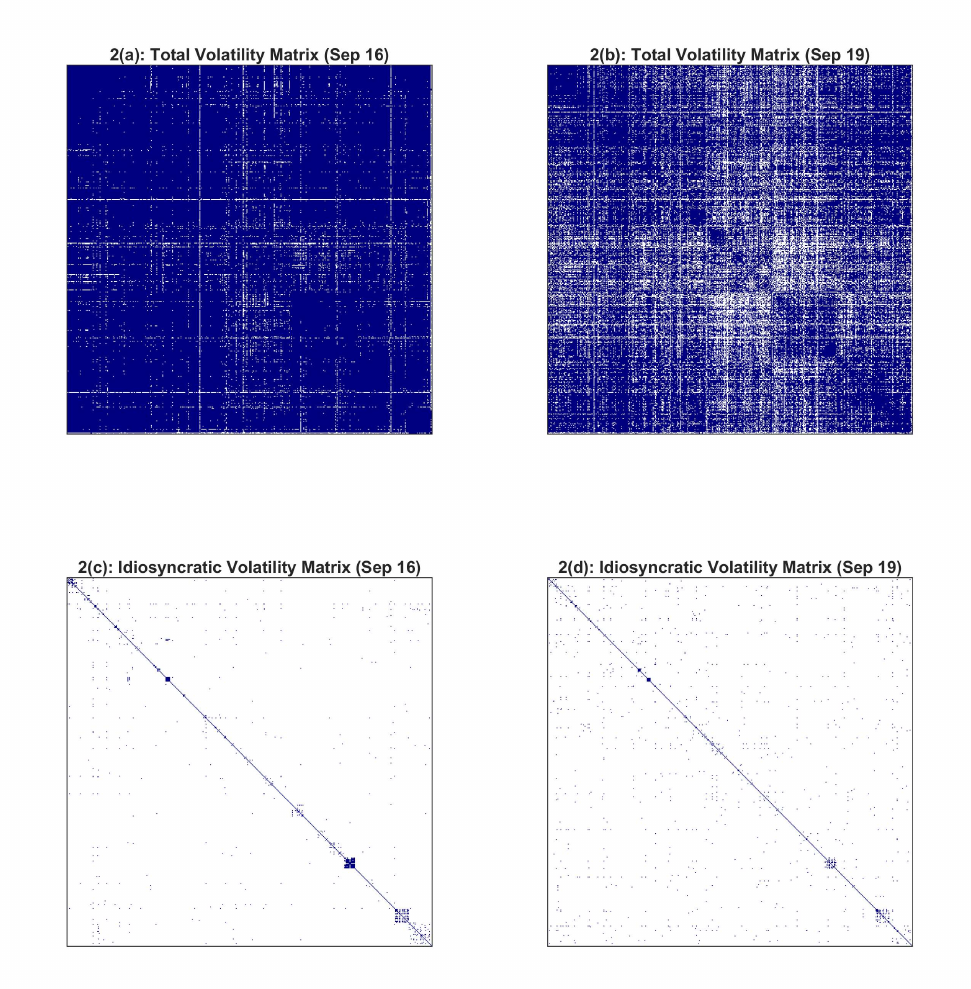}
	\caption{Sparsity patterns of the total and idiosyncratic volatility matrix estimates on September 16 and 19, 2008}\label{fig:figure_2}
\end{figure}

To examine whether it is appropriate to directly apply shrinkage techniques to the spot volatility matrices of the total returns, following \cite{FFX16} we plot in Figure \ref{fig:figure_2}(a) and (b) their sparsity patterns on the 16th and the 19th of September \footnote{Recall that as in \cite{DLX19} our tuning parameter used for each pairwise spot covariance in our shrinkage method is proportional to the product of the spot standard deviations of the returns of that pair of assets. Therefore, the sparsity pattern is effectively determined by the spot correlation matrix.}.  The deep blue dots correspond to the locations of pairwise correlations that are at least $0.15$, whereas the white dots correspond to those smaller than $0.15$. Note that the covariance structure of the total returns is very dense on these two days. Therefore, it is not appropriate to directly apply the shrinkage technique as in Sections \ref{sec2} and \ref{sec3}. Meanwhile, although both are quite dense, we can still clearly see their differences. Consistent with our observation from the decile plots of the correlations, we can see that the plot for the 16th is almost completely covered by blue dots, but the plot for the 19th in contrast has significantly more areas covered in white.

We next incorporate the twelve observed factors in the large spot volatility matrix estimation as suggested in Section \ref{sec4}. In particular, we are interested on estimating the spot idiosyncratic volatility matrix, which is expected to satisfy the sparsity restriction. To save space, we choose to only report results using the SCAD shrinkage due to its satisfactory performance in our simulations. In Figure \ref{fig:figure_1}(c) and (d), we plot the deciles of the estimated spot idiosyncratic variances and correlations over trading days. In Figure \ref{fig:figure_1}(c), we observe a significant upward shift of the distribution of the spot variances of the idiosyncratic returns around the time of Lehman Brothers' bankruptcy, indicating that the observed factors may not fully capture the time variation of the spot variances. In contrast, the deciles of the spot correlations in Figure \ref{fig:figure_1}(d) seem to be quite flat throughout the entire month, suggesting that the systematic factors may explain the time variation in the distribution of the pairwise correlations better than that of the variances.

We finally plot the sparsity patterns of the two estimated spot idiosyncratic volatility matrices on 16 and 19 September in Figure \ref{fig:figure_2}(c) and (d), respectively. Unlike Figure \ref{fig:figure_2}(a) and (b), we note that the estimated spot idiosyncratic volatility matrices are highly sparse on the two days. This is consistent with our observation from Figure \ref{fig:figure_1}(d), confirming that the observed factors can effectively account for the time variation in the spot covariance structure of the returns. Meanwhile, we also note that the two idiosyncratic volatility matrices are clearly not diagonal and still carry some visible time variation. Lastly, it is worth mentioning that the estimated spot idiosyncratic volatility matrices do not exhibit significant correlations within the blocks along the diagonal lines, except for some very limited actions in the lower right corner of the two matrices (the lower right corner corresponds to the XLU sector according to our sorting).



\section{Conclusion}\label{sec7}

We developed nonparametric estimation methods for large spot volatility matrices under the uniform sparsity assumption. We allowed for microstructure noise and observed common risk factors and employed kernel smoothing and generalised shrinkage. In each scenario we obtained the uniform convergence rates for the large estimated covariance matrices and these reflect the smoothness and sparsity assumptions we made. The simulation results show that the proposed estimation methods work well in finite samples for both the noise-free and noise-contaminated data. The empirical study demonstrated the effectiveness on S\&P 500 stocks five minute data. Several issues can be further explored. For example, it is worthwhile to further study the spot precision matrix estimation which is briefly discussed in Appendix C.1 of the supplement and explore its application to optimal portfolio choice. 


\section*{Acknowledgements}

The authors would like to thank a Co-Editor and two reviewers for the constructive comments, which helped to improve the article. The first author's research was partly supported by the BA Talent Development Award (No. TDA21$\backslash$210027). The second author’s research was partly supported by the BA/Leverhulme Small Research Grant funded by the Leverhulme Trust (No. SRG1920/ 100603).



\appendix

\section*{Appendix A:\ Proofs of the main results}
\label{app:A} \renewcommand{\theequation}{A.\arabic{equation}} %
\setcounter{equation}{0}

In this appendix, we give the proofs of the main theorems. We start with four propositions whose proofs are available in Appendix B of the supplement.

\renewcommand{\theprop}{A.\arabic{prop}}\setcounter{prop}{0}

\begin{prop}\label{prop:A.1}

\emph{Suppose that Assumptions \ref{ass:1} and \ref{ass:2}(i)(ii) are satisfied.
Then, we have 
\begin{equation}  \label{eqA.1}
\max_{1\leq i,j\leq p}\sup_{0\leq t\leq
T}\left\vert\widehat\Sigma_{ij,t}-\Sigma_{ij,t}\right\vert=O_P\left(%
\zeta_{\Delta,p}\right),
\end{equation}
where $\zeta_{\Delta,p}=h^{\gamma}+\left[\frac{\Delta\log (p\vee
\Delta^{-1})}{h}\right]^{1/2}$.}

\end{prop}

\begin{prop}\label{prop:A.2}

\emph{Suppose that Assumptions \ref{ass:1}, \ref{ass:2}(i),
\ref{ass:3} and \ref{ass:4}(i)(ii) are satisfied, and Assumption \ref{ass:2}(ii)
holds with $\Delta^{-1}$ replaced by $N$. 
\begin{equation}  \label{eqA.2}
\max_{1\leq i,j\leq p}\sup_{0\leq t\leq
T}\left\vert\widetilde\Sigma_{ij,t}-\Sigma_{ij,t}\right\vert=O_P\left(%
\zeta_{N,p}^\ast+\nu_{\Delta,p,N}\right),
\end{equation}
where $\zeta_{N,p}^\ast$ and $\nu_{\Delta,p,N}$ are defined in Assumption
\ref{ass:4}(iii).}

\end{prop}

\begin{prop}\label{prop:A.3}

\emph{Suppose that Assumptions \ref{ass:1}, \ref{ass:2}(i),
\ref{ass:3} and \ref{ass:5} are satisfied. Then, we have 
\begin{equation}  \label{eqA.3}
\max_{1\leq i,j\leq p}\sup_{0\leq t\leq
T}\left\vert\widehat\Omega_{ij}(t)-\Omega_{ij}(t)\right\vert=O_P\left(%
\delta_{\Delta,p}\right),
\end{equation}
where $\delta_{\Delta,p}=h_1^{\gamma_1}+\left[\frac{\Delta\log (p\vee
\Delta^{-1})}{h_1}\right]^{1/2}$.}

\end{prop}

\begin{prop}\label{prop:A.4}

\emph{Suppose that Assumptions \ref{ass:2}(i)(ii) and \ref{ass:6} are satisfied.
Then, we have 
\begin{eqnarray}
&&\sup_{0\leq t\leq T}\left\Vert\widehat{\boldsymbol\Sigma}_{t}^Y-{\boldsymbol\Sigma}_{t}^Y\right\Vert_{\max}=O_P\left(\zeta_{\Delta,p}\right),\notag\\
&&\sup_{0\leq t\leq T}\left\Vert\widehat{\boldsymbol\Sigma}_{t}^F-{\boldsymbol\Sigma}_{t}^F\right\Vert_{\max}=O_P\left(\zeta_{\Delta,p}\right),\notag\\
&&\sup_{0\leq t\leq T}\left\Vert\widehat{\boldsymbol\Sigma}_{t}^{YF}-{\boldsymbol\Sigma}_{t}^{YF}\right\Vert_{\max}=O_P\left(\zeta_{\Delta,p}\right).\notag
\end{eqnarray}
where $\zeta_{\Delta,p}$ is defined as in Proposition \ref{prop:A.1}.}

\end{prop}

\noindent\textbf{Proof of Theorem \ref{thm:1}}.\ By the definition of $\widehat{%
\boldsymbol{\Sigma}}_{t}^s$ and the property of $s_{\rho}(\cdot)$, we
readily have that 
{\footnotesize\begin{eqnarray}
\sup_{0\leq t\leq T}\left\Vert\widehat{\boldsymbol{\Sigma}}_{t}^s-{%
\boldsymbol{\Sigma}}_t\right\Vert  
&\le&\sup_{0\leq t\leq T} \max_{1\le i\le
p}\sum_{j=1}^{p}\left|\widehat\Sigma_{ij,t}^s-\Sigma_{ij, t}\right|  \notag
\\
&=&\sup_{0\leq t\leq T} \max_{1\le i\le
p}\left|\widehat\Sigma_{ii,t}-\Sigma_{ii, t}\right|+\sup_{0\leq t\leq T } \max_{1\le i\le
p}\sum_{j=1,\neq i}^{p}\left|s_{\rho_1(t)}\left(\widehat\Sigma_{ij,t}\right)I\left(%
\left\vert\widehat\Sigma_{ij,t}\right\vert>\rho_1(t)\right)-\Sigma_{ij,
t}\right|\notag \\
&=&\sup_{0\leq t\leq T} \max_{1\le i\le
p}\left|\widehat\Sigma_{ii,t}-\Sigma_{ii, t}\right|+\sup_{0\leq t\leq T } \max_{1\le i\le
p}\sum_{j=1,\neq i}^{p}\left|s_{\rho_1(t)}\left(\widehat\Sigma_{ij,t}\right)I\left(%
\left\vert\widehat\Sigma_{ij,t}\right\vert>\rho_1(t)\right)-\right.
\notag \\
&&\left.\Sigma_{ij,
t}I\left(\left\vert\widehat\Sigma_{ij,t}\right\vert>\rho_1(t)\right)-\Sigma_{ij,
t}I\left(\left\vert\widehat\Sigma_{ij,t}\right\vert\le\rho_1(t)\right)\right|
\notag \\
&\le & \sup_{0\leq t\leq T} \max_{1\le i\le
p}\left|\widehat\Sigma_{ii,t}-\Sigma_{ii, t}\right|+\sup_{0\leq t\leq T } \max_{1\le i\le
p}\sum_{j=1,\neq i}^{p}\left|s_{\rho_1(t)}\left(\widehat\Sigma_{ij,t}\right)-%
\widehat\Sigma_{ij,t}\right|
I\left(\left\vert\widehat\Sigma_{ij,t}\right\vert>\rho_1(t)\right)+  \notag
\\
& &\sup_{0\leq t\leq T } \max_{1\le i\le
p}\sum_{j=1,\neq i}^{p}\left|\widehat\Sigma_{ij,t}-\Sigma_{ij,
t}\right|I\left(\left\vert\widehat\Sigma_{ij,t}\right\vert>\rho_1(t)\right)+
\sup_{0\leq t\leq T}\max_{1\le i\le p}\sum_{j=1,\neq i}^{p}\left|\Sigma_{ij,
t}\right|
I\left(\left\vert\widehat\Sigma_{ij,t}\right\vert\le\rho_1(t)\right)  \notag
\\
& =:& \Pi_1+\Pi_2+\Pi_3+\Pi_4.  \label{eqA.4}
\end{eqnarray}}

Define the event 
\begin{equation*}
\mathcal{G}(M)=\left\{\max_{1\le i,j\le p}\sup_{0\leq t\leq T
}\left|\widehat\Sigma_{ij,t}-\Sigma_{ij, t}\right|\le
M\zeta_{\Delta,p}\right\}
\end{equation*}
where $M$ is a positive constant. For any small $\epsilon>0$, by (\ref{eqA.1}%
), we may find a sufficiently large constant $M_\epsilon>0$ such that 
\begin{equation}  \label{eqA.5}
\mathsf{P}\left(\mathcal{G}(M_\epsilon)\right)\geq 1-\epsilon.
\end{equation}
By property (iii) of the shrinkage function and (\ref{eqA.5}), we have 
\begin{equation*}
\Pi_2\le \sup_{0\leq t\leq T }\rho_1(t)\left[ \max_{1\le i\le
p}\sum_{j=1}^{p}I\left(\left\vert\widehat\Sigma_{ij,t}\right\vert>\rho_1(t)%
\right)\right]
\end{equation*}
and 
\begin{equation*}
\Pi_3 \le M_\epsilon\zeta_{\Delta,p}\left[\sup_{0\leq t\leq T}\max_{1\le
i\le
p}\sum_{j=1}^{p}I\left(\left\vert\widehat\Sigma_{ij,t}\right\vert>\rho_1(t)%
\right)\right]
\end{equation*}
conditional on the event $\mathcal{G}(M_\epsilon)$. By the reverse triangle
inequality and Proposition \ref{prop:A.1}, 
\begin{equation*}
\left|\widehat\Sigma_{ij,t}\right|\le \left|\Sigma_{ij,
t}\right|+M_\epsilon\zeta_{\Delta,p}
\end{equation*}
on $\mathcal{G}(M_\epsilon)$. Letting $\underline{C}_M=2M_\epsilon$ in
Assumption \ref{ass:2}(iii), as $\left\{{\boldsymbol{\Sigma}}_t:\ 0\leq t\leq
T\right\}\in \mathcal{S}(q,\varpi(p), T)$, we have 
\begin{eqnarray}
\Pi_2+\Pi_3 &\le&\zeta_{\Delta,p}(\overline{C}_M+M_\epsilon)\left[
\sup_{0\leq t\leq T }\max_{1\le i\le
p}\sum_{j=1}^{p}I\left(\left\vert\widehat\Sigma_{ij,t}\right\vert>\underline{%
C}_M\zeta_{\Delta,p}\right)\right]  \notag \\
&\le&\zeta_{\Delta,p}(\overline{C}_M+M_\epsilon)\left[ \sup_{0\leq t\leq T
}\max_{1\le i\le
p}\sum_{j=1}^{p}I\left(\left\vert\widehat\Sigma_{ij,t}\right\vert>M_\epsilon%
\zeta_{\Delta,p}\right)\right]  \notag \\
&=& O_P\left(\zeta_{\Delta,p}\right)\left[\sup_{0\leq t\leq T }\max_{1\le
i\le p}\sum_{j=1}^{p}\frac{|\Sigma_{ij, t}|^q}{\left(M_\epsilon\zeta_{%
\Delta,p}\right)^q}\right]  \notag \\
&=&O_P\left(\Lambda\varpi(p)\zeta_{\Delta,p}^{1-q}\right)=O_P\left(\varpi(p)
\zeta_{\Delta,p}^{1-q}\right).  \label{eqA.6}
\end{eqnarray}
on the event $\mathcal{G}(M_\epsilon)$, where $\overline{C}_M$ is defined in
Assumption \ref{ass:2}(iii). Note that the events $\left\{\left\vert\widehat%
\Sigma_{ij,t}\right\vert\le \rho_1(t)\right\}$ and $\mathcal{G}(M_\epsilon)$
jointly imply that $\left\{|\Sigma_{ij, t}|\le \left(\overline{C}%
_M+M_\epsilon\right)\zeta_{\Delta,p}\right\}$. Then, we may show that 
\begin{eqnarray}
\Pi_4 &\le&\sup_{0\leq t\leq T }\max_{1\le i\le
p}\sum_{j=1}^{p}|\Sigma_{ij, t}|I\left(|\Sigma_{ij, t}|\le \left(\overline{C}%
_M+M_\epsilon\right)\zeta_{\Delta,p}\right)  \notag \\
&\le& \left(\overline{C}_M+M_\epsilon\right)^{1-q}\zeta_{\Delta,p}^{1-q}%
\sup_{0\leq t\leq T }\max_{1\le i\le p}\sum_{j=1}^{p}|\Sigma_{ij, t}|^q 
\notag \\
&=&O_P\left(\Lambda \varpi(p)
\zeta_{\Delta,p}^{1-q}\right)=O_P\left(\varpi(p)
\zeta_{\Delta,p}^{1-q}\right).  \label{eqA.7}
\end{eqnarray}
By Proposition \ref{prop:A.1}, we readily have that
\begin{equation}\label{eqA.8}
\Pi_{1}=O_P\left(\zeta_{\Delta,p}\right)=O_P\left(\varpi(p)
\zeta_{\Delta,p}^{1-q}\right).
\end{equation}
By (\ref{eqA.6})--(\ref{eqA.8}), and letting $\epsilon\rightarrow0$ in (%
\ref{eqA.5}), we complete the proof of Theorem \ref{thm:1}. \hfill$\blacksquare$

\smallskip

\noindent\textbf{Proof of Theorem \ref{thm:2}}.\ \ The proof is similar to the proof
of Theorem \ref{thm:1} with Proposition \ref{prop:A.2} replacing Proposition \ref{prop:A.1}. Details are
omitted to save the space.\hfill$\blacksquare$

\smallskip

\noindent\textbf{Proof of Theorem \ref{thm:3}}.\ \ The proof is similar to the proof
of Theorem \ref{thm:1} with Proposition \ref{prop:A.3} replacing Proposition \ref{prop:A.1}. Details are
omitted to save the space.\hfill$\blacksquare$

\smallskip

\noindent\textbf{Proof of Theorem \ref{thm:4}}.\ \ By Proposition \ref{prop:A.4} and the definition of $\widehat\Sigma_{ij,t}^X$ in (\ref{eq4.8}), we may show that
\begin{equation}  \label{eqA.9}
\max_{1\leq i,j\leq p}\sup_{0\leq t\leq T}\left\vert\widehat\Sigma_{ij,t}^X-\Sigma_{ij,t}^X\right\vert=O_P\left(\zeta_{\Delta,p}\right).
\end{equation}
With (\ref{eqA.9}), following the proof of Theorem \ref{thm:1}, we complete the proof of (\ref{eq4.12}).

We next turn to the proof of (\ref{eq4.13}). Note that
\[
\sup_{0\leq t\leq T}\left\Vert \widehat{\boldsymbol\Sigma}_t^{Y,s}- {\boldsymbol\Sigma}_t^Y\right\Vert_{{\boldsymbol\Sigma}_t^Y}^2\le 2\sup_{0\leq t\leq T}\left[\left\Vert \widehat{\boldsymbol\Sigma}_t^{X,s}- {\boldsymbol\Sigma}_t^X\right\Vert_{{\boldsymbol\Sigma}_t^Y}^2+\left\Vert \widehat{\boldsymbol\beta}(t)\widehat{\boldsymbol\Sigma}_t^F\widehat{\boldsymbol\beta}(t)^{^\intercal}-{\boldsymbol\beta}(t){\boldsymbol\Sigma}_t^F{\boldsymbol\beta}(t)^{^\intercal}\right\Vert_{{\boldsymbol\Sigma}_t^Y}^2\right].
\]
For any $p\times p$ matrix ${\boldsymbol\Sigma}$, since all the eigenvalues of ${\boldsymbol\Sigma}_t^Y$ are strictly larger than a positive constant, 
\begin{equation}\label{eqA.10}
\Vert {\boldsymbol\Sigma}\Vert_{{\boldsymbol\Sigma}_t^Y}^2=\frac{1}{p}\left\Vert\left({\boldsymbol\Sigma}_t^Y\right)^{-1/2}{\boldsymbol\Sigma}\left({\boldsymbol\Sigma}_t^Y\right)^{-1/2}\right\Vert_{F}^2\leq \frac{C}{p}\Vert {\boldsymbol\Sigma}\Vert_F^2,
\end{equation}
where $C>0$ is a generic constant whose value may change from line to line. By (\ref{eq4.12}) and (\ref{eqA.10}), we prove 
\begin{equation}\label{eqA.11}
\sup_{0\leq t\leq T}\left\Vert \widehat{\boldsymbol\Sigma}_t^{X,s}- {\boldsymbol\Sigma}_t^X\right\Vert_{{\boldsymbol\Sigma}_t^Y}^2\leq\frac{C}{p}\sup_{0\leq t\leq T}\left\Vert \widehat{\boldsymbol\Sigma}_t^{X,s}- {\boldsymbol\Sigma}_t^X\right\Vert_F^2\leq C\sup_{0\leq t\leq T}\left\Vert \widehat{\boldsymbol\Sigma}_t^{X,s}- {\boldsymbol\Sigma}_t^X\right\Vert^2=O_P\left([\varpi(p)\zeta_{\Delta,p}^{1-q}]^2\right).
\end{equation}

By the definition of $\widehat{\boldsymbol\beta}(t)$ in (\ref{eq4.7}) and Proposition \ref{prop:A.4}, we readily have that
\begin{equation}\label{eqA.12}
\max_{1\leq i\leq p}\sup_{0\leq t\leq T}\left\Vert \widehat\beta_i(t)-\beta_i(t)\right\Vert_2=O_P\left(\zeta_{\Delta,p}\right).
\end{equation}
Write ${\mathbf D}_t^\beta=\widehat{\boldsymbol\beta}(t)-{\boldsymbol\beta}(t)$ and ${\mathbf D}_t^F=\widehat{\boldsymbol\Sigma}_t^F-{\boldsymbol\Sigma}_t^F$. Note that 
\begin{eqnarray*}
\widehat{\boldsymbol\beta}(t)\widehat{\boldsymbol\Sigma}_t^F\widehat{\boldsymbol\beta}(t)^{^\intercal}-{\boldsymbol\beta}(t){\boldsymbol\Sigma}_t^F{\boldsymbol\beta}(t)^{^\intercal}&=& {\mathbf D}_t^\beta{\mathbf D}_t^F{\mathbf D}_t^{\beta{^\intercal}}+ {\mathbf D}_t^\beta{\boldsymbol\Sigma}_t^F{\mathbf D}_t^{\beta{^\intercal}}+ {\mathbf D}_t^\beta{\mathbf D}_t^F{\boldsymbol\beta}(t)^{^\intercal}+\\
&&{\mathbf D}_t^\beta{\boldsymbol\Sigma}_t^F{\boldsymbol\beta}(t)^{^\intercal}+{\boldsymbol\beta}(t){\mathbf D}_t^F{\mathbf D}_t^{\beta{^\intercal}}+ {\boldsymbol\beta}(t){\boldsymbol\Sigma}_t^F{\mathbf D}_t^{\beta{^\intercal}}+{\boldsymbol\beta}(t){\mathbf D}_t^F{\boldsymbol\beta}(t)^{^\intercal}.
\end{eqnarray*}
By (\ref{eqA.10}), (\ref{eqA.12}) and Proposition \ref{prop:A.4}, we have
\begin{eqnarray}
\sup_{0\leq t\leq T}\left\Vert {\mathbf D}_t^\beta{\mathbf D}_t^F{\mathbf D}_t^{\beta{^\intercal}}\right\Vert_{{\boldsymbol\Sigma}_t^Y}^2&\leq& C\sup_{0\leq t\leq T} \frac{1}{p}\left\Vert {\mathbf D}_t^\beta{\mathbf D}_t^F{\mathbf D}_t^{\beta{^\intercal}}\right\Vert_F^2\notag\\
&\leq&\frac{C}{p}\sup_{0\leq t\leq T} \left\Vert {\mathbf D}_t^\beta\right\Vert_F^4 \sup_{0\leq t\leq T}\left\Vert {\mathbf D}_t^F\right\Vert^2\notag\\
&=&O_{P}\left( p\zeta_{\Delta,p}^6\right).\label{eqA.13}
\end{eqnarray}
Similarly, we can show that
\begin{equation}\label{eqA.14}
\sup_{0\leq t\leq T}\left\Vert{\mathbf D}_t^\beta{\boldsymbol\Sigma}_t^F{\mathbf D}_t^{\beta{^\intercal}}\right\Vert_{{\boldsymbol\Sigma}_t^Y}^2\leq C\sup_{0\leq t\leq T} \frac{1}{p}\left\Vert {\mathbf D}_t^\beta\right\Vert_F^4=O_{P}\left( p\zeta_{\Delta,p}^4\right).
\end{equation}
By (\ref{eq4.3}), Assumption \ref{ass:6}(iv) and Sherman-Morrison-Woodbury formula, we may show that  
\begin{equation}\label{eqA.15}
\sup_{0\leq t\leq T}\left\Vert {\boldsymbol\beta}(t)^{^\intercal}\left({\boldsymbol\Sigma}_t^Y\right)^{-1}{\boldsymbol\beta}(t)\right\Vert=O_P(1).
\end{equation}
Using (\ref{eqA.12}), (\ref{eqA.15}) and Proposition \ref{prop:A.4}, we have
\begin{eqnarray}
\sup_{0\leq t\leq T}\left\Vert {\mathbf D}_t^\beta{\mathbf D}_t^F{\boldsymbol\beta}(t)^{^\intercal}\right\Vert_{{\boldsymbol\Sigma}_t^Y}^2
&=&\frac{1}{p}\sup_{0\leq t\leq T}{\sf trace}\left\{{\mathbf D}_t^F{\mathbf D}_t^{\beta{^\intercal}}\left({\boldsymbol\Sigma}_t^Y\right)^{-1}{\mathbf D}_t^\beta{\mathbf D}_t^F{\boldsymbol\beta}(t)^{^\intercal}\left({\boldsymbol\Sigma}_t^Y\right)^{-1}{\boldsymbol\beta}(t)\right\}\notag\\
&\le & \frac{C}{p}\sup_{0\leq t\leq T} \left\Vert {\mathbf D}_t^\beta\right\Vert_F^2 \sup_{0\leq t\leq T} \left\Vert {\mathbf D}_t^F\right\Vert^2 \sup_{0\leq t\leq T} \left\Vert {\boldsymbol\beta}(t)^{^\intercal}\left({\boldsymbol\Sigma}_t^Y\right)^{-1}{\boldsymbol\beta}(t)\right\Vert\notag\\
&\le&  \frac{C}{p}\sup_{0\leq t\leq T} \left\Vert {\mathbf D}_t^\beta\right\Vert_F^2\sup_{0\leq t\leq T} \left\Vert {\mathbf D}_t^F\right\Vert^2=O_P\left(\zeta_{\Delta,p}^4\right),\label{eqA.16}
\end{eqnarray}
and 
\begin{equation}\label{eqA.17}
\sup_{0\leq t\leq T}\left\Vert {\boldsymbol\beta}(t){\mathbf D}_t^F{\mathbf D}_t^{\beta{^\intercal}}\right\Vert_{{\boldsymbol\Sigma}_t^Y}^2=O_P\left(\zeta_{\Delta,p}^4\right).
\end{equation}
Similar to the proof of (\ref{eqA.16}), we also have
\begin{equation}\label{eqA.18}
\sup_{0\leq t\leq T}\left\Vert {\mathbf D}_t^\beta{\boldsymbol\Sigma}_t^F{\boldsymbol\beta}(t)^{^\intercal}\right\Vert_{{\boldsymbol\Sigma}_t^Y}^2
\le  \frac{C}{p}\sup_{0\leq t\leq T} \left\Vert {\mathbf D}_t^\beta\right\Vert_F^2 \sup_{0\leq t\leq T} \left\Vert {\boldsymbol\Sigma}_t^F\right\Vert^2 =O_P\left(\zeta_{\Delta,p}^2\right),
\end{equation}
and
\begin{equation}\label{eqA.19}
\sup_{0\leq t\leq T}\left\Vert {\boldsymbol\beta}(t){\boldsymbol\Sigma}_t^F{\mathbf D}_t^{\beta{^\intercal}}\right\Vert_{{\boldsymbol\Sigma}_t^Y}^2
=O_P\left(\zeta_{\Delta,p}^2\right).
\end{equation}
By (\ref{eqA.15}) and Proposition \ref{prop:A.4}, we may show that 
\begin{eqnarray}
\sup_{0\leq t\leq T}\left\Vert{\boldsymbol\beta}(t){\mathbf D}_t^F{\boldsymbol\beta}(t)^{^\intercal}\right\Vert_{{\boldsymbol\Sigma}_t^Y}^2&=&\frac{1}{p}\sup_{0\leq t\leq T}{\sf trace}\left\{{\mathbf D}_t^F{\boldsymbol\beta}(t)^{^\intercal}\left({\boldsymbol\Sigma}_t^Y\right)^{-1}{\boldsymbol\beta}(t){\mathbf D}_t^F{\boldsymbol\beta}(t)^{^\intercal}\left({\boldsymbol\Sigma}_t^Y\right)^{-1}{\boldsymbol\beta}(t)\right\}\notag\\
&\le&\frac{C}{p}\sup_{0\leq t\leq T} \left\Vert {\mathbf D}_t^F\right\Vert^2\sup_{0\leq t\leq T} \left\Vert {\boldsymbol\beta}(t)^{^\intercal}\left({\boldsymbol\Sigma}_t^Y\right)^{-1}{\boldsymbol\beta}(t)\right\Vert^2=O_P\left(\zeta_{\Delta,p}^2/p\right). \label{eqA.20}
\end{eqnarray}
With (\ref{eqA.13}), (\ref{eqA.14}) and (\ref{eqA.16})--(\ref{eqA.20}), we have 
\begin{equation}\label{eqA.21}
\sup_{0\leq t\leq T}\left\Vert \widehat{\boldsymbol\beta}(t)\widehat{\boldsymbol\Sigma}_t^F\widehat{\boldsymbol\beta}(t)^{^\intercal}-{\boldsymbol\beta}(t){\boldsymbol\Sigma}_t^F{\boldsymbol\beta}(t)^{^\intercal}\right\Vert_{{\boldsymbol\Sigma}_t^Y}^2=O_P\left(p\zeta_{\Delta,p}^4+\zeta_{\Delta,p}^2\right).
\end{equation}
By virtue of (\ref{eqA.11}) and (\ref{eqA.21}), we complete the proof of (\ref{eq4.13}).\hfill$\blacksquare$

\newpage


\begin{center}

{\Large Supplement to ``Nonparametric Estimation of Large Spot Volatility Matrices
for High-Frequency Financial Data"}

\end{center}

\bigskip


In this supplement, we provide the detailed proofs of the propositions stated in Appendix A, discuss the spot precision matrix estimation, address the asynchronicity issue and report additional simulation results.


\section*{Appendix B:\ Proofs of technical results}

\label{app:B} \renewcommand{\theequation}{B.\arabic{equation}} %
\setcounter{equation}{0}

{\small As discussed in Remark 1, the local boundedness condition in Assumption 1(i) can be strengthened to the following uniform boundedness condition:
\begin{equation}\label{bound}
\max_{1\leq i\leq p}\sup_{0\leq s\leq T} |\mu_{i,s}|\leq C_\mu<\infty,\ \
\max_{1\leq i\leq p}\sup_{0\leq s\leq T} \Sigma_{ii,t} \leq C_\Sigma<\infty,
\end{equation}
with probability one. Throughout this appendix, we let $C$ denote a generic positive constant
whose value may change from line to line.

\smallskip

\noindent\textbf{Proof of Proposition A.1}.\ Throughout this proof, we let $%
\zeta_{\Delta,p}^\ast=\left[\frac{\Delta\log (p\vee \Delta^{-1})}{h}\right]%
^{1/2}$. By (2.1), we have 
\begin{eqnarray}
(\Delta X_{i,k})(\Delta
X_{j,k})&=&\left(\int_{t_{k-1}}^{t_k}\mu_{i,s}ds+\sum_{l=1}^{p}%
\int_{t_{k-1}}^{t_k}\sigma_{il,s}dW_{l,s}\right)\left(\int_{t_{k-1}}^{t_k}%
\mu_{j,u}du+\sum_{l=1}^{p}\int_{t_{k-1}}^{t_k}\sigma_{jl,u}dW_{l,u}\right) 
\notag \\
&=&\left(\int_{t_{k-1}}^{t_k}\mu_{i,s}ds\int_{t_{k-1}}^{t_k}\mu_{j,u}du%
\right)+\left(\int_{t_{k-1}}^{t_k}\sum_{l=1}^{p}\sigma_{il,s}dW_{l,s}%
\int_{t_{k-1}}^{t_k}\mu_{j,u}du\right)+  \notag \\
&&\left(\int_{t_{k-1}}^{t_k}\mu_{i,s}ds\int_{t_{k-1}}^{t_k}\sum_{l=1}^{p}%
\sigma_{jl,u}dW_{l,u}\right)+\left(\int_{t_{k-1}}^{t_k}\sum_{l=1}^{p}%
\sigma_{il,s}dW_{l,s}\int_{t_{k-1}}^{t_k}\sum_{l=1}^{p}\sigma_{jl,u}dW_{l,u}%
\right)  \notag \\
&=&M_{ij,k}(1)+M_{ij,k}(2)+M_{ij,k}(3)+M_{ij,k}(4).  \notag
\end{eqnarray}
This leads to the following decomposition: 
\begin{eqnarray}
\sum_{k=1}^n K_h(t_k-t)\Delta X_{i,k}\Delta X_{j,k}&=&\sum_{k=1}^n K_h(t_{k}-t)M_{ij,k}(1)+\sum_{k=1}^n
K_h(t_{k}-t)M_{ij,k}(2)+\notag\\
&&\sum_{k=1}^n K_h(t_{k}-t)M_{ij,k}(3)+\sum_{k=1}^n
K_h(t_{k}-t)M_{ij,k}(4).\notag
\end{eqnarray}

By (\ref{bound}) and Assumption 2(i)(ii), we readily have that 
\begin{eqnarray}
\max_{1\leq i,j\leq p}\sup_{0\leq t\leq T}\left\vert\sum_{k=1}^n
K_h(t_{k}-t)M_{ij,k}(1)\right\vert&\leq& \max_{1\leq i,j\leq p}\max_{1\leq
k\leq n} \vert M_{ij,k}(1)\vert \sup_{0\leq t\leq T}\sum_{k=1}^n
K_h(t_{k}-t)  \notag \\
&\leq&C\Delta \sup_{0\leq t\leq T}\Delta\sum_{k=1}^n K_h(t_{k}-t)  \notag
\\
&=&O_P\left(\Delta\right)=o_P\left(\zeta_{\Delta,p}^\ast\right),
\label{eqB.1}
\end{eqnarray}
as $\Delta\sum_{k=1}^n K_h(t_{k}-t)$ is bounded uniformly over $t$.

We next show that 
\begin{equation}  \label{eqB.2}
\max_{1\leq i,j\leq p}\sup_{0\leq t\leq T}\left\vert\sum_{k=1}^n
K_h(t_{k}-t)M_{ij,k}(4)-\sum_{k=1}^n
K_h(t_{k}-t)\int_{t_{k-1}}^{t_k}\Sigma_{ij,
s}ds\right\vert=O_P\left(\zeta_{\Delta,p}^\ast\right).
\end{equation}
Let $dX_{i,t}^\ast=\sum_{l=1}^p \sigma_{il,t} dW_{l,t}$, $\Delta
X_{i,k}^\ast=\int_{t_{k-1}}^{t_k}\sum_{l=1}^p \sigma_{il,s} dW_{l,s}=X_{i,t_k}^\ast-X_{i,t_{k-1}}^\ast$ and $%
X_{i,t}^\ast$ be adapted to the underlying filtration $(\mathcal{F}%
_t)_{t\geq0}$. Note that 
\begin{eqnarray}
M_{ij,k}(4)&=&\Delta X_{i,k}^\ast\Delta X_{j,k}^\ast=\frac{1}{2}\left[%
\left(\Delta X_{i,k}^\ast+\Delta X_{j,k}^\ast\right)\left(\Delta
X_{i,k}^\ast+\Delta X_{j,k}^\ast\right)-\left(\Delta
X_{i,k}^\ast\right)^2-\left(\Delta X_{j,k}^\ast\right)^2\right]  \notag \\
&=:&\frac{1}{2}\left[M_{ij,k}^\ast(4)-\left(\Delta
X_{i,k}^\ast\right)^2-\left(\Delta X_{j,k}^\ast\right)^2\right].  \notag
\end{eqnarray}
Hence, to show (\ref{eqB.2}), it is sufficient to prove that 
\begin{equation}  \label{eqB.3}
\max_{1\leq i\leq p}\sup_{0\leq t\leq T}\left\vert\sum_{k=1}^n
K_h(t_{k}-t)\left(\Delta X_{i,k}^\ast\right)^2-\sum_{k=1}^n
K_h(t_{k}-t)\int_{t_{k-1}}^{t_k}\Sigma_{ii,
s}ds\right\vert=O_P\left(\zeta_{\Delta,p}^\ast\right)
\end{equation}
and 
\begin{equation}  \label{eqB.4}
\max_{1\leq i,j\leq p}\sup_{0\leq t\leq T}\left\vert\sum_{k=1}^n
K_h(t_{k}-t)M_{ij,k}^\ast(4)-\sum_{k=1}^n
K_h(t_{k}-t)\int_{t_{k-1}}^{t_k}\Sigma_{ij, s}^\ast
ds\right\vert=O_P\left(\zeta_{\Delta,p}^\ast\right),
\end{equation}
where $\Sigma_{ij, s}^\ast$ is defined in Assumption 1(ii). 

We next only prove (\ref{eqB.3}) as the proof of (\ref{eqB.4}) is analogous. Consider covering the interval $[0,T]$ by some disjoint intervals $%
\mathcal{T}_v$ with centre $\tau_v^\ast$ and length $d=h^2\zeta_{\Delta,p}^%
\ast$, $v=1,2,\cdots, V$. Observe that 
\begin{eqnarray}
&&\max_{1\leq i\leq p}\sup_{0\leq t\leq T}\left\vert\sum_{k=1}^n
K_h(t_{k}-t)\left(\Delta X_{i,k}^\ast\right)^2-\sum_{k=1}^n
K_h(t_{k}-t)\int_{t_{k-1}}^{t_k}\Sigma_{ii, s}ds\right\vert  \notag \\
&\leq&\max_{1\leq i\leq p}\max_{1\leq v\leq V}\left\vert\sum_{k=1}^n
K_h(t_{k}-\tau_v^\ast)\left(\Delta X_{i,k}^\ast\right)^2-\sum_{k=1}^n
K_h(t_{k}-\tau_v^\ast)\int_{t_{k-1}}^{t_k}\Sigma_{ii, s}ds\right\vert+ 
\notag \\
&&\max_{1\leq i\leq p}\max_{1\leq v\leq V}\sup_{t\in\mathcal{T}_v}
\left\vert\sum_{k=1}^n \left[K_h(t_{k}-t)-K_h(t_{k}-\tau_v^\ast)\right]%
\left(\Delta X_{i,k}^\ast\right)^2\right\vert+  \notag \\
&&\max_{1\leq i\leq p}\max_{1\leq v\leq V}\sup_{t\in\mathcal{T}_v}
\left\vert\sum_{k=1}^n \left[K_h(t_{k}-t)-K_h(t_{k}-\tau_v^\ast)\right]%
\int_{t_{k-1}}^{t_k}\Sigma_{ii,s}ds\right\vert.  \label{eqB.5}
\end{eqnarray}
As the kernel function has the compact support $[-1,1]$, we have, for any $t%
\in[0, T]$, 
\begin{equation*}
\sum_{k=1}^n K_h(t_{k}-t)\left[\left(\Delta
X_{i,k}^\ast\right)^2-\int_{t_{k-1}}^{t_k}\Sigma_{ii, s}ds\right]%
=\sum_{k=l(t)}^{u(t)}K_h(t_{k}-t)\left[\left(\Delta
X_{i,k}^\ast\right)^2-\int_{t_{k-1}}^{t_k}\Sigma_{ii, s}ds\right],
\end{equation*}
where $l(t)=\lfloor (t-h)/\Delta\rfloor\vee 1$ and $u(t)=\lfloor
(t+h)/\Delta\rfloor\wedge n$. Letting $\mathcal{N}$ be a standard normal random variable, by Lemma 1 in 
\cite{FLY12}, we have 
\begin{equation}  \label{eqB.6}
\mathsf{E}\left(\exp\{\psi(\mathcal{N}^2-1)\}\right)\leq
\exp\left\{2\psi^2\right\}\ \ \mathrm{for}\ |\psi|\leq1/4.
\end{equation}
Following the argument in the proof of Lemma 3 in \cite{FLY12} and using (%
\ref{eqB.6}), for $k=l(\tau_v^\ast),l(\tau_v^\ast)+1,\cdots,u(\tau_v^\ast)$, 
\begin{eqnarray}
&&\mathsf{E}\left(\exp\left\{\theta
\left(\Delta^{-1}h\right)^{1/2}K_h(t_{k}-\tau_v^\ast)\left[\left(\Delta
X_{i,k}^\ast\right)^2-\int_{t_{k-1}}^{t_k}\Sigma_{ii, s}ds\right]\right\} |%
\mathcal{F}_{t_{k-1}}\right)  \notag \\
&\leq&\exp\left\{\frac{2\Delta}{h}\theta^2 C_\Sigma^2 K^2\left(\frac{%
t_{k}-\tau_v^\ast}{h}\right)\right\},  \notag
\end{eqnarray}
where $\theta$ satisfies that $\left\vert\theta C_\Sigma (\Delta
h^{-1})^{1/2}K\left(\frac{t_{k}-\tau_v^\ast}{h}\right)\right\vert\leq 1/4$
and $C_\Sigma$ is defined in (\ref{bound}). Consequently, we have 
\begin{eqnarray}
&&\mathsf{E}\left(\exp\left\{\theta
\left(\Delta^{-1}h\right)^{1/2}\sum_{k=1}^nK_h(t_{k}-\tau_v^\ast)\left[%
\left(\Delta X_{i,k}^\ast\right)^2-\int_{t_{k-1}}^{t_k}\Sigma_{ii, s}ds%
\right]\right\} \right)  \notag \\
&=&\mathsf{E}\left(\exp\left\{\theta
\left(\Delta^{-1}h\right)^{1/2}\sum_{k=l(\tau_v^\ast)}^{u(\tau_v^\ast)}K_h(t_{k}-\tau_v^\ast)\left[\left(\Delta
X_{i,k}^\ast\right)^2-\int_{t_{k-1}}^{t_k}\Sigma_{ii, s}ds\right]\right\}
\right)  \notag \\
&\leq&\exp\left\{2\theta^2 C_\Sigma^2 \nu(\tau_v^\ast)\right\},  \label{eqB.7}
\end{eqnarray}
where $\nu(\tau_v^\ast)=(\Delta/h)\sum_{k=l(\tau_v^\ast)}^{u(\tau_v^\ast)}K^2(t_{k}-\tau_v^\ast)$. By (\ref{eqB.7}), using the Markov
inequality and choosing $\theta=\frac{\sqrt{\log (p\vee \Delta^{-1})}}{%
C_\Sigma^2\nu(\tau_v^\ast)}$, we can prove that 
\begin{equation*}
\mathsf{P}\left(\left\vert\sum_{k=1}^n K_h(t_{k}-\tau_v^\ast)\left(\Delta
X_{i,k}^\ast\right)^2-\sum_{k=1}^n
K_h(t_{k}-\tau_v^\ast)\int_{t_{k-1}}^{t_k}\Sigma_{ii, s}ds\right\vert>M
\zeta_{\Delta,p}^\ast\right)\leq 2\exp\{-C(M) \log (p\vee \Delta^{-1})\}, 
\end{equation*}
where $C(M)$ is positive and becomes sufficiently large if we choose $M$ to
be large enough. Then, by the Bonferroni inequality, we have 
\begin{eqnarray}
&&\mathsf{P}\left(\max_{1\leq i\leq p}\max_{1\leq v\leq
V}\left\vert\sum_{k=1}^n K_h(t_{k}-\tau_v^\ast)\left(\Delta
X_{i,k}^\ast\right)^2-\sum_{k=1}^n
K_h(t_{k}-\tau_v^\ast)\int_{t_{k-1}}^{t_k}\Sigma_{ii, s}ds\right\vert>M
\zeta_{\Delta,p}^\ast\right)  \notag \\
&\leq&\sum_{i=1}^p\sum_{v=1}^V 2\exp\{-C(M) (\log (p\vee
\Delta^{-1}))\}\rightarrow0,  \notag
\end{eqnarray}
where the convergence is due to the fact $pV=o\left(\exp\{C_M \log (p\vee
\Delta^{-1})\}\right)$ as $V$ is divergent at a polynomial rate of $1/\Delta$
and $C(M)$ is sufficiently large, which implies that 
\begin{equation}  \label{eqB.8}
\max_{1\leq i\leq p}\max_{1\leq v\leq V}\left\vert\sum_{k=1}^n
K_h(t_{k}-\tau_v^\ast)\left(\Delta X_{i,k}^\ast\right)^2-\sum_{k=1}^n
K_h(t_{k}-\tau_v^\ast)\int_{t_{k-1}}^{t_k}\Sigma_{ii,
s}ds\right\vert=O_P(\zeta_{\Delta,p}^\ast).
\end{equation}

By the smoothness condition on the kernel function in Assumption 2(i), we
have 
\begin{eqnarray}
&&\max_{1\leq i\leq p}\max_{1\leq v\leq V}\sup_{t\in\mathcal{T}_v}
\left\vert\sum_{k=1}^n \left[K_h(t_{k}-t)-K_h(t_{k}-\tau_v^\ast)\right]%
\left(\Delta X_{i,k}^\ast\right)^2\right\vert  \notag \\
&\leq&\max_{1\leq v\leq V}\sup_{t\in\mathcal{T}_v}\left\vert
K_h(t_{k}-t)-K_h(t_{k}-\tau_v^\ast)\right\vert \max_{1\leq i\leq
p}\sum_{k=1}^n \left(\Delta X_{i,k}^\ast\right)^2  \notag \\
&=&O\left(dh^{-2}\right)\max_{1\leq i\leq p}\sum_{k=1}^n \left(\Delta
X_{i,k}^\ast\right)^2.  \notag
\end{eqnarray}
Similar to the proof of (\ref{eqB.8}), we may show that 
\begin{equation*}
\max_{1\leq i\leq p}\sum_{k=1}^n \left(\Delta X_{i,k}^\ast\right)^2\leq
\max_{1\leq i\leq p} \int_{0}^{T}\Sigma_{ii, s}ds +o_P(1)=O_P(1) 
\end{equation*}
as $T$ is fixed and $\Sigma_{ii,t}$ is uniformly bounded by $C_\Sigma$.
Hence, by the choice of $d$, we have 
\begin{equation}  \label{eqB.9}
\max_{1\leq i\leq p}\max_{1\leq v\leq V}\sup_{t\in\mathcal{T}_v}
\left\vert\sum_{k=1}^n \left[K_h(t_{k}-t)-K_h(t_{k}-\tau_v^\ast)\right]%
\left(\Delta X_{i,k}^\ast\right)^2\right\vert=O_P(\zeta_{\Delta,p}^\ast).
\end{equation}
Analogously, we also have 
\begin{equation}  \label{eqB.10}
\max_{1\leq i\leq p}\max_{1\leq v\leq V}\sup_{t\in\mathcal{T}_v}
\left\vert\sum_{k=1}^n \left[K_h(t_{k}-t)-K_h(t_{k}-\tau_v^\ast)\right]%
\int_{t_{k-1}}^{t_k}\Sigma_{ii,s}ds\right\vert=O_P(\zeta_{\Delta,p}^\ast).
\end{equation}
By (\ref{eqB.5}) and (\ref{eqB.8})--(\ref{eqB.10}), we complete the proof of
(\ref{eqB.3}).

By (\ref{eqB.1}), (\ref{eqB.2}) and the Cauchy-Schwarz inequality, we have 
\begin{eqnarray}
&&\max_{1\leq i,j\leq p}\sup_{0\leq t\leq T}\left\vert\sum_{k=1}^n
K_h(t_{k}-t)M_{ij,k}(2)\right\vert^2  \notag \\
&\leq&\max_{1\leq i\leq p}\sup_{0\leq t\leq T}\sum_{k=1}^n
K_h(t_{k}-t)\left(\Delta X_{i,k}^\ast\right)^2\max_{1\leq j\leq
p}\sup_{0\leq t\leq T} \sum_{k=1}^n
K_h(t_{k}-t)\left(\int_{t_{k-1}}^{t_k}\mu_{j,u}du\right)^2  \notag \\
&=&O_P\left(\Delta \right)\cdot O_P(1)=O_P\left(\Delta \right),  \notag
\end{eqnarray}
indicating that 
\begin{equation}  \label{eqB.11}
\max_{1\leq i,j\leq p}\sup_{0\leq t\leq T}\left\vert\sum_{k=1}^n
K_h(t_{k}-t)M_{ij,k}(2)\right\vert=O_P\left(\Delta^{1/2}\right)=o_P\left(%
\zeta_{\Delta,p}^\ast\right),
\end{equation}
and similarly, 
\begin{equation}  \label{eqB.12}
\max_{1\leq i,j\leq p}\sup_{0\leq t\leq T}\left\vert\sum_{k=1}^n
K_h(t_{k}-t)M_{ij,k}(3)\right\vert=O_P\left(\Delta^{1/2}\right)=o_P\left(%
\zeta_{\Delta,p}^\ast\right).
\end{equation}
With (\ref{eqB.1}), (\ref{eqB.2}), (\ref{eqB.11}) and (\ref{eqB.12}), we
prove that 
\begin{equation}  \label{eqB.13}
\max_{1\leq i,j\leq p}\sup_{0\leq t\leq T}\left\vert \sum_{k=1}^n K_h(t_k-t)\Delta X_{i,k}\Delta X_{j,k}-\sum_{k=1}^n
K_h(t_{k}-t)\int_{t_{k-1}}^{t_k}\Sigma_{ij,
s}ds\right\vert=O_P\left(\zeta_{\Delta,p}^\ast\right).
\end{equation}
Since $\Delta\sum_{k=1}^n K_h(t_{k}-t)$ is strictly larger than a positive constant uniformly over $t$, by (\ref{eqB.13}), we readily have that
\begin{equation}  \label{eqB.13.1}
\max_{1\leq i,j\leq p}\sup_{0\leq t\leq T}\left\vert \widehat\Sigma_{ij,t}-\sum_{k=1}^n
K_h^\ast(t_{k}-t)\int_{t_{k-1}}^{t_k}\Sigma_{ij,
s}ds\right\vert=O_P\left(\zeta_{\Delta,p}^\ast\right).
\end{equation}

On the other hand, by (2.6) in Assumption 1(ii), we may show that 
\begin{equation}  \label{eqB.14}
\max_{1\leq i,j\leq p}\sup_{0\leq t\leq T}\left\vert \sum_{k=1}^n
K_h^\ast(t_{k}-t)\int_{t_{k-1}}^{t_k}\Sigma_{ij, s}ds-\Sigma_{ij,
t}\right\vert=O_P\left(h^{\gamma}\right).
\end{equation}
Then we complete the proof of (A.1) by virtue of (\ref{eqB.13.1}) and (\ref{eqB.14}).
\hfill$\blacksquare$

\smallskip

We next turn to the proof of Proposition A.2, in which a crucial step is to
derive a uniform consistency for $\widetilde{X}_{i,\tau}$. The latter is
stated in Lemma \ref{le:B.1} below.

\renewcommand{\thelemma}{B.\arabic{lemma}}\setcounter{lemma}{0}

\begin{lemma}\label{le:B.1}

Suppose that Assumptions 1(i), 3 and 4(i)(ii) are satisfied. Then we have
\begin{equation}  \label{eqB.15}
\max_{1\leq i\leq p}\max_{0\leq l\leq N}\left\vert \widetilde{X}%
_{i,\tau_l}-X_{i,\tau_l}\right\vert=O_P\left(\sqrt{\log(p\vee \Delta^{-1})}%
\left[b^{1/2}+\left(\Delta^{-1}b\right)^{-1/2}\right]\right).
\end{equation}

\end{lemma}

\noindent\textbf{Proof of Lemma \ref{le:B.1}}.\ \ By the definition of $\widetilde{%
\mathbf{X}}_{\tau}$ in (3.2), we write 
\begin{eqnarray}
\widetilde{X}_{i,\tau_l}-X_{i,\tau_l}&=&\frac{T}{n}\sum_{k=1}^n
L_b^\dagger(t_k-\tau_l)Z_{i,t_k}-X_{i,\tau_l}\notag\\
&=&\Pi_{i,l}^\dagger(1)+\Pi_{i,l}^\dagger(2)+%
\Pi_{i,l}^\dagger(3)+\Pi_{i,l}^\dagger(4),\label{eqB.16}
\end{eqnarray}
where 
\begin{eqnarray}
\Pi_{i,l}^\dagger(1)&=&\frac{T}{n}\sum_{k=1}^n L_b^\dagger(t_k-\tau_l)\xi_{i,k},  \notag \\
\Pi_{i,l}^\dagger(2)&=&\sum_{k=1}^n
L_b^\dagger(t_k-\tau_l)\int_{(k-1)\Delta}^{k\Delta}(X_{i,t_k}-X_{i,s})ds,  \notag \\
\Pi_{i,l}^\dagger(3)&=&\sum_{k=1}^n \int_{(k-1)\Delta}^{k\Delta}\left[%
L_b^\dagger(t_k-\tau_l)-L_b^\dagger(s-\tau_l)\right]X_{i,s}ds,  \notag \\
\Pi_{i,l}^\dagger(4)&=&\int_{0}^{T}L_b^\dagger(s-\tau_l)X_{i,s}ds-X_{i,\tau_l}.  \notag
\end{eqnarray}

Let $\nu_{\Delta,p}^\ast=\left[\frac{\Delta\log (p\vee \Delta^{-1})}{b}%
\right]^{1/2}$, ${\boldsymbol{\omega}}_i(t_k)=\left[\omega_{i1}(t_k),\cdots,%
\omega_{ip}(t_k) \right]^{^\intercal}$, and ${\boldsymbol{\omega}}%
_{i,\ast}(t_k)={\boldsymbol{\omega}}_i(t_k)/\left\Vert {\boldsymbol{\omega}}%
_i(t_k)\right\Vert$. We first consider $\Pi_{i,l}(1)$. Define 
\begin{equation}  \label{eqB.17}
\xi_{i,k}^{\star}={\boldsymbol{\omega}}_i^{^\intercal}(t_k){\boldsymbol{\xi}}%
_{k}^\ast I\left( |{\boldsymbol{\omega}}_{i,\ast}^{^\intercal}(t_k){%
\boldsymbol{\xi}}_{k}^\ast|\leq \Delta^{-\iota}\right),\quad
\xi_{i,k}^{\diamond}={\boldsymbol{\omega}}_i^{^\intercal}(t_k){\boldsymbol{%
\xi}}_{k}^\ast I\left( |{\boldsymbol{\omega}}_{i,\ast}^{^\intercal}(t_k){%
\boldsymbol{\xi}}_{k}^\ast|> \Delta^{-\iota}\right),
\end{equation}
where $\iota$ is defined in Assumption 4(ii). Note that 
\[
\sum_{k=1}^n L_b(t_k-\tau_l){\boldsymbol{\omega}}%
_i^{^\intercal}(t_k){\boldsymbol{\xi}}_{k}^\ast  
=\sum_{k=1}^n L_b(t_k-\tau_l)\left[\xi_{i,k}^\star-\mathsf{E}%
(\xi_{i,k}^\star)\right]+\sum_{k=1}^n L_b(t_k-\tau_l)\left[%
\xi_{i,k}^\diamond-\mathsf{E}(\xi_{i,k}^\diamond)\right] 
\]
as $\mathsf{E}(\xi_{i,k}^{\star})+\mathsf{E}(\xi_{i,k}^{\diamond})=0$. By
the noise moment condition in Assumption 3(i) and the uniform boundedness
condition on $\left\Vert {\boldsymbol{\omega}}_i(t_k)\right\Vert$ in
Assumption 3(ii), we have 
\begin{equation*}
\mathsf{E}\left( \left| \xi_{i,k}^\diamond\right|\right) \leq C_\omega\cdot%
\mathsf{E}\left[\left\vert {\boldsymbol{\omega}}_{i,\ast}^{^\intercal}(t_k){%
\boldsymbol{\xi}}_{k}^\ast\right\vert I\left( |{\boldsymbol{\omega}}%
_{i,\ast}^{^\intercal}(t_k){\boldsymbol{\xi}}_{k}^\ast|>
\Delta^{-\iota}\right) \right] =O\left(\Delta^{\iota
M_{\xi}^\diamond}\right)=o\left(\nu_{\Delta,p}^\ast\right),
\end{equation*}
where $M_{\xi}^\diamond>0$ is arbitrarily large. Then, by Assumptions 3(i),
4(ii) and the Bonferroni and Markov inequalities, we have, for any $%
\epsilon>0$, 
\begin{eqnarray}
&&\mathsf{P}\left( \max_{1\leq i\leq p}\max_{0\leq l\leq N}\left| \frac{T}{n}%
\sum_{k=1}^n L_b(t_k-\tau_l)\left[\xi_{i,k}^\diamond-\mathsf{E}%
(\xi_{i,k}^\diamond)\right]\right| >\epsilon\nu_{\Delta,p}^\ast\right) 
\notag \\
&\leq &\mathsf{P}\left( \max_{1\leq i\leq p}\max_{0\leq l\leq N}\left| \frac{%
T}{n}\sum_{k=1}^n L_b(t_k-\tau_l)\xi_{i,k}^\diamond\right| >\frac{1}{2}%
\epsilon\nu_{\Delta,p}^\ast \right)  \notag \\
&\leq&\mathsf{P}\left( \max_{1\leq i\leq p}\max_{1\leq k\leq n}\left|
\xi_{i,k}^\diamond\right| >0\right)  \notag \\
&\leq&\mathsf{P}\left( \max_{1\leq i\leq p}\max_{1\leq k\leq n} \left| {%
\boldsymbol{\omega}}_{i,\ast}^{^\intercal}(t_k){\boldsymbol{\xi}}%
_{k}^\ast\right| >\Delta^{-\iota}\right)  \notag \\
&\leq&\sum_{i=1}^p\sum_{k=1}^n \mathsf{P}\left( \left|{\boldsymbol{\omega}}%
_{i,\ast}^{^\intercal}(t_k){\boldsymbol{\xi}}_{k}^\ast\right|
>\Delta^{-\iota}\right)  \notag \\
&\leq&pn\exp\{-s\Delta^{-\iota}\}C_\xi=o(1)  \notag
\end{eqnarray}
for $0<s<s_{0}$, where $C_\xi$ is defined in Assumption 3(i). Hence, we have 
\begin{equation}  \label{eqB.18}
\max_{1\leq i\leq p}\max_{0\leq l\leq N}\left| \frac{T}{n}\sum_{k=1}^n
L_b(t_k-\tau_l)\omega_{i}(t_k)\left[\xi_{i,k}^\diamond-\mathsf{E}%
(\xi_{i,k}^\diamond)\right]\right|=o_P\left(\nu_{\Delta,p}^\ast\right).
\end{equation}
On the other hand, by Assumptions 3 and 4(i)(ii) as well as the Bernstein
inequality for the independent sequence \citep[e.g., Proposition 2.14
in][]{W19}, we may show that 
\begin{eqnarray}
&&\mathsf{P}\left( \max_{1\leq i\leq p}\max_{0\leq l\leq N}\left| \frac{T}{n}%
\sum_{k=1}^n L_b(t_k-\tau_l)\left[\xi_{i,k}^\star-\mathsf{E}(\xi_{i,k}^\star)%
\right]\right| >M\nu_{\Delta,p}^\ast\right)  \notag \\
&\leq & \ \sum_{i=1}^{p}\sum_{l=1}^{N}\mathsf{P}\left( \left| \frac{T}{n}%
\sum_{k=1}^n L_b(t_k-\tau_l)\left[\xi_{i,k}^\star-\mathsf{E}(\xi_{i,k}^\star)%
\right]\right| >M\nu_{\Delta,p}^\ast\right)  \notag \\
&= &O\left( pN \exp\left\{ -C_\star(M)\log(p\vee \Delta^{-1})\right\}
\right)=o(1),  \notag
\end{eqnarray}
where $N$ diverges to infinity at a polynomial rate of $n$, $C_\star(M)$ is
positive and could be sufficiently large by letting $M$ be large enough.
Therefore, we have 
\begin{equation}  \label{eqB.19}
\max_{1\leq i\leq p}\max_{0\leq l\leq N}\left| \frac{T}{n}\sum_{k=1}^n
L_b(t_k-\tau_l)\left[\xi_{i,k}^\star-\mathsf{E}(\xi_{i,k}^\star)\right]%
\right|=O_P\left(\nu_{\Delta,p}^\ast\right).
\end{equation}
By (\ref{eqB.18}) and (\ref{eqB.19}), and noting that $\int_0^TL_b(s-\tau)ds$ is strictly larger than a positive constant uniformly over $\tau$, we readily have that 
\begin{equation}  \label{eqB.20}
\max_{1\leq i\leq p}\max_{0\leq l\leq N}\left|
\Pi_{i,l}^\dagger(1)\right|=O_P\left(\nu_{\Delta,p}^\ast\right).
\end{equation}

Write
\[
\Pi_{i,l}(2)=\sum_{k=1}^n
L_b(t_k-\tau_l)\int_{(k-1)\Delta}^{k\Delta}(X_{i,t_k}-X_{i,s})ds
\]
note that
\begin{eqnarray}
\Pi_{i,l}(2)&=&\sum_{k=1}^n
L_b(t_k-\tau_l)\int_{(k-1)\Delta}^{k\Delta}\left(\int_{s}^{k\Delta}%
\mu_{i,u}du\right)ds+ \sum_{k=1}^n
L_b(t_k-\tau_l)\int_{(k-1)\Delta}^{k\Delta}\left(\int_{s}^{k\Delta}%
\sum_{j=1}^p\sigma_{ij,u}dW_{j,u}\right)ds  \notag \\
&=&\Pi_{i,l}(2,1)+\Pi_{i,l}(2,2).  \notag
\end{eqnarray}
By (\ref{bound}) and Assumption 4(i), we have 
\begin{equation}  \label{eqB.21}
\max_{1\leq i\leq p}\max_{0\leq l\leq N}\left|
\Pi_{i,l}(2,1)\right|=O_P(\Delta)=o_P\left(\nu_{\Delta,p}^\ast\right).
\end{equation}
By the Bonferroni inequality, we may show that, for any $\epsilon>0$ 
\begin{eqnarray}
&&\mathsf{P}\left(\max_{1\leq i\leq p}\sup_{(k-1)\Delta\leq s\leq
k\Delta}\left\vert\int_{s}^{k\Delta}\sum_{j=1}^p\sigma_{ij,u}dW_{j,u}\right%
\vert>\epsilon\nu_{\Delta,p}^\ast\right)  \notag \\
&\leq&\sum_{i=1}^p\mathsf{P}\left(\sup_{(k-1)\Delta\leq s\leq
k\Delta}\left\vert\int_{s}^{k\Delta}\sum_{j=1}^p\sigma_{ij,u}dW_{j,u}\right%
\vert>\epsilon\nu_{\Delta,p}^\ast\right)  \notag \\
&\leq&\sum_{i=1}^p\mathsf{P}\left(\sup_{(k-1)\Delta\leq s\leq
k\Delta}\left\vert\int_{(k-1)\Delta}^{s}\sum_{j=1}^p\sigma_{ij,u}dW_{j,u}%
\right\vert>\frac{1}{2}\epsilon\nu_{\Delta,p}^\ast\right).  \label{eqB.22}
\end{eqnarray}
By the conditional Jensen inequality, we may verify that both $%
\left\{\left\vert\int_{(k-1)\Delta}^{s}\sum_{j=1}^p\sigma_{ij,u}dW_{j,u}%
\right\vert\right\}_{s\geq (k-1)\Delta}$ and $\left\{\exp\left(\psi\left%
\vert\int_{(k-1)\Delta}^{s}\sum_{j=1}^p\sigma_{ij,u}dW_{j,u}\right\vert%
\right)\right\}_{s\geq (k-1)\Delta}$ are sub-martingales, where $\psi>0$.
Using the moment generating function for the folded normal random variable
and (\ref{bound}), we have 
\begin{equation*}
\mathsf{E}\left[\exp\left(\psi\left\vert\int_{(k-1)\Delta}^{k\Delta}%
\sum_{j=1}^p\sigma_{ij,u}dW_{j,u}\right\vert\right)\right]\leq \exp\left(%
\frac{\psi^2\Delta C_\Sigma}{2}\right), 
\end{equation*}
where $C_\Sigma$ is defined in (\ref{bound}). Combining the above
arguments and using Doob's inequality for sub-martingales, we may show that 
\begin{eqnarray}
&&\mathsf{P}\left(\sup_{(k-1)\Delta\leq s\leq
k\Delta}\left\vert\int_{(k-1)\Delta}^{s}\sum_{j=1}^p\sigma_{ij,u}dW_{j,u}%
\right\vert>\frac{1}{2}\epsilon\nu_{\Delta,p}^\ast\right)  \notag \\
&=&\mathsf{P}\left(\sup_{(k-1)\Delta\leq s\leq
k\Delta}\exp\left\{\psi\left\vert\int_{(k-1)\Delta}^{s}\sum_{j=1}^p%
\sigma_{ij,u}dW_{j,u}\right\vert\right\}>\exp\left\{\frac{1}{2}%
\psi\epsilon\nu_{\Delta,p}^\ast\right\}\right)  \notag \\
&\leq&\exp\left(-\frac{\psi\epsilon\nu_{\Delta,p}^\ast}{2}\right)\mathsf{E}%
\left[\exp\left(\psi\left\vert\int_{(k-1)\Delta}^{k\Delta}\sum_{j=1}^p%
\sigma_{ij,u}dW_{j,u}\right\vert\right)\right]  \notag \\
&\leq&\exp\left(\frac{\psi^2\Delta C_\Sigma}{2}-\frac{\psi\epsilon\nu_{%
\Delta,p}^\ast}{2}\right).  \label{eqB.23}
\end{eqnarray}
Then, choosing $\psi=\epsilon\nu_{\Delta,p}^\ast/(2\Delta C_\Sigma)$, by (%
\ref{eqB.22}) and(\ref{eqB.23}), we have 
\begin{eqnarray}
&&\mathsf{P}\left(\max_{1\leq i\leq p}\sup_{(k-1)\Delta\leq s\leq
k\Delta}\left\vert\int_{s}^{k\Delta}\sum_{j=1}^p\sigma_{ij,u}dW_{j,u}\right%
\vert>\epsilon\nu_{\Delta,p}^\ast\right)  \notag \\
&\leq& p\exp\left\{-\frac{(\epsilon\nu_{\Delta,p}^\ast)^2}{8\Delta C_\Sigma}%
\right\}=O\left(p\exp\left\{-\frac{\epsilon^2}{8 C_\Sigma}\cdot\frac{%
\log(p\vee \Delta^{-1})}{b}\right\}\right)=o(1)  \notag
\end{eqnarray}
for any $\epsilon>0$, which indicates that 
\begin{equation}  \label{eqB.24}
\max_{1\leq i\leq p}\max_{0\leq l\leq N}\left|
\Pi_{i,l}(2,2)\right|=o_P\left(\nu_{\Delta,p}^\ast\right).
\end{equation}
By (\ref{eqB.21}) and (\ref{eqB.24}), we readily have that 
\begin{equation}  \label{eqB.25}
\max_{1\leq i\leq p}\max_{0\leq l\leq N}\left|
\Pi_{i,l}(2)\right|=o_P\left(\nu_{\Delta,p}^\ast\right),\ \ \max_{1\leq i\leq p}\max_{0\leq l\leq N}\left|
\Pi_{i,l}^\dagger(2)\right|=o_P\left(\nu_{\Delta,p}^\ast\right).
\end{equation}

For $\Pi_{i,l}^\dagger(3)$, we note that 
\begin{equation*}
\vert\Pi_{i,l}^\dagger(3)\vert\leq\sup_{0\leq u\leq T}|X_{i,u}|\cdot \sum_{k=1}^n
\int_{(k-1)\Delta}^{k\Delta}\left\vert
L_b^\dagger(t_k-\tau_l)-L_b^\dagger(s-\tau_l)\right\vert ds. 
\end{equation*}
By Assumption 4(i), we have 
\begin{equation}  \label{eqB.26}
\max_{0\leq l\leq N}\sum_{k=1}^n \int_{(k-1)\Delta}^{k\Delta}\left\vert
L_b^\dagger(t_k-\tau_l)-L_b^\dagger(s-\tau_l)\right\vert ds=O\left(\Delta b^{-1}\right).
\end{equation}
On the other hand, by (\ref{bound}), 
\begin{equation*}
\sup_{0\leq u\leq T}|X_{i,u}|=\sup_{0\leq u\leq T} \int_{0}^u
|\mu_{i,u}|du+\sup_{0\leq u\leq T} \left\vert\int_{0}^u
\sum_{j=1}^p\sigma_{ij,u}dW_{j,u}\right\vert=\sup_{0\leq u\leq T}
\left\vert\int_{0}^u \sum_{j=1}^p\sigma_{ij,u}dW_{j,u}\right\vert+O_P(1). 
\end{equation*}
Following the proof of (\ref{eqB.24}), we may show that 
\begin{equation*}
\sup_{0\leq u\leq T} \left\vert\int_{0}^u
\sum_{j=1}^p\sigma_{ij,u}dW_{j,u}\right\vert=O_P\left(\sqrt{\log (p\vee
\Delta^{-1})}\right), 
\end{equation*}
indicating that 
\begin{equation}  \label{eqB.27}
\sup_{0\leq u\leq T}|X_{i,u}|=O_P\left(\sqrt{\log (p\vee \Delta^{-1})}%
\right).
\end{equation}
By virtue of (\ref{eqB.26}) and (\ref{eqB.27}), we prove that 
\begin{equation}  \label{eqB.28}
\max_{1\leq i\leq p}\max_{0\leq l\leq N}\left|
\Pi_{i,l}^\dagger(3)\right|=O_P\left(\Delta b^{-1}\sqrt{\log (p\vee \Delta^{-1})}%
\right)=o_P\left(\nu_{\Delta,p}^\ast\right).
\end{equation}

Finally, for $\Pi_{i,l}^\dagger(4)$, we write it as 
\begin{eqnarray}
\Pi_{i,l}^\dagger(4)&=&\left\{\int_{0}^{T}L_b^\dagger(s-\tau_l)\int_0^s\mu_{i,u}duds-%
\int_0^{\tau_l}\mu_{i,u}du\right\}+  \notag \\
&&\left\{\int_{0}^{T}L_b^\dagger(s-\tau_l)\int_0^s\sum_{j=1}^p%
\sigma_{ij,u}dW_{j,u}ds-\int_0^{\tau_l}\sum_{j=1}^p\sigma_{ij,u}dW_{j,u}%
\right\}  \notag \\
&=:&\Pi_{i,l}^\dagger(4,1)+\Pi_{i,l}^\dagger(4,2).  \notag
\end{eqnarray}
By Assumptions 1(i) and 4(i), we readily have that 
\begin{equation}  \label{eqB.29}
\max_{1\leq i\leq p}\max_{0\leq l\leq N}\left|
\Pi_{i,l}^\dagger(4,1)\right|=O_P\left(b\right).
\end{equation}
Following the proof of (\ref{eqB.24}), we may show that 
\begin{equation*}
\mathsf{P}\left(\max_{1\leq i\leq p}\max_{1\leq l\leq N}\sup_{\tau_l\leq
s\leq
\tau_l+b}\left\vert\int_{\tau_l}^{s}\sum_{j=1}^p\sigma_{ij,u}dW_{j,u}\right%
\vert>M\sqrt{b\log(p\vee \Delta^{-1})}\right)\rightarrow0 
\end{equation*}
and 
\begin{equation*}
\mathsf{P}\left(\max_{1\leq i\leq p}\max_{1\leq l\leq N}\sup_{\tau_l-b\leq
s\leq
\tau_l}\left\vert\int_{s}^{\tau_l}\sum_{j=1}^p\sigma_{ij,u}dW_{j,u}\right%
\vert>M\sqrt{b\log(p\vee \Delta^{-1})}\right)\rightarrow0 
\end{equation*}
when $M>0$ is sufficiently large. Consequently, we have 
\begin{equation}  \label{eqB.30}
\max_{1\leq i\leq p}\max_{0\leq l\leq N}\left|
\Pi_{i,l}^\dagger(4,2)\right|=O_P\left(\sqrt{b\log(p\vee \Delta^{-1})}\right).
\end{equation}
Combining (\ref{eqB.29}) and (\ref{eqB.30}), 
\begin{equation}  \label{eqB.31}
\max_{1\leq i\leq p}\max_{0\leq l\leq N}\left| \Pi_{i,l}^\dagger(4)\right|=O_P\left(%
\sqrt{b\log(p\vee \Delta^{-1})}\right).
\end{equation}
The proof of (\ref{eqB.15}) in Lemma \ref{le:B.1} is completed with (\ref{eqB.20}), (%
\ref{eqB.25}), (\ref{eqB.28}) and (\ref{eqB.31}).\hfill$\blacksquare$

\smallskip

\noindent\textbf{Proof of Proposition A.2}.\ \ By (3.3), we have 
\begin{eqnarray}
\widetilde{\Sigma}_{ij,t}-\Sigma_{ij,t}&=&\sum_{l=1}^N K_h^\dagger(\tau_l-t)\Delta%
\widetilde{X}_{i,l}\Delta\widetilde{X}_{j,l}-\Sigma_{ij,t}  \notag \\
&=&\sum_{l=1}^N K_h^\dagger(\tau_l-t)\Delta X_{i,l}\Delta
X_{j,l}-\Sigma_{ij,t}+\sum_{k=1}^3\Xi_{ij,t}(k),  \notag
\end{eqnarray}
where 
\begin{eqnarray}
\Xi_{ij,t}(1)&=&\sum_{l=1}^N K_h^\dagger(\tau_l-t)\Delta X_{i,l}\left(\Delta%
\widetilde{X}_{j,l}-\Delta X_{j,l}\right),  \notag \\
\Xi_{ij,t}(2)&=&\sum_{l=1}^N K_h^\dagger(\tau_l-t)\left(\Delta\widetilde{X}%
_{i,l}-\Delta X_{i,l}\right)\Delta X_{j,l},  \notag \\
\Xi_{ij,t}(3)&=&\sum_{l=1}^N K_h^\dagger(\tau_l-t)\left(\Delta\widetilde{X}%
_{i,l}-\Delta X_{i,l}\right)\left(\Delta\widetilde{X}_{j,l}-\Delta
X_{j,l}\right).  \notag
\end{eqnarray}
By Proposition A.1, we have 
\begin{equation}  \label{eqB.32}
\max_{1\leq i,j\leq p}\sup_{0\leq t\leq T}\left\vert \sum_{l=1}^N
K_h^\dagger(\tau_l-t)\Delta X_{i,l}\Delta
X_{j,l}-\Sigma_{ij,t}\right\vert=O_P\left(h^{\gamma}+\left[\frac{\log
(p\vee N)}{Nh}\right]^{1/2}\right).
\end{equation}
By Lemma \ref{le:B.1} and Assumption 2(i), we have 
\begin{equation}  \label{eqB.33}
\max_{1\leq i,j\leq p}\sup_{0\leq t\leq T}\left\vert
\Xi_{ij,t}(3)\right\vert=O_P\left(N\log(p\vee \Delta^{-1})\left[%
b^{1/2}+\left(\Delta^{-1}b\right)^{1/2}\right]^2\right).
\end{equation}
By Proposition A.1, (\ref{eqB.33}) and the Cauchy-Schwarz inequality, we have 
\begin{equation}  \label{eqB.34}
\max_{1\leq i,j\leq p}\sup_{0\leq t\leq T}\left(\left\vert
\Xi_{ij,t}(1)\right\vert+\left\vert \Xi_{ij,t}(2)\right\vert\right)=O_P\left(%
\sqrt{N\log(p\vee \Delta^{-1})}\left[b^{1/2}+\left(\Delta^{-1}b\right)^{1/2}%
\right]\right).
\end{equation}
The proof of (A.2) in Proposition A.2 is completed by virtue of (\ref%
{eqB.32})--(\ref{eqB.34}).\hfill$\blacksquare$

\smallskip

\noindent\textbf{Proof of Proposition A.3}.\ \ By (3.1) and (3.7), we write 
\begin{eqnarray}
\widehat{\Omega}_{ij}(t)&=&\frac{\Delta}{2}\sum_{k=1}^{n}K_{h_1}^\ast(t_{k}-t)%
\Delta X_{i,k} \Delta X_{j,k}+ \frac{\Delta}{n}%
\sum_{k=1}^{n}K_{h_1}^\ast(t_{k}-t)\Delta X_{i,k} (\xi_{j,k}-\xi_{j,k-1})+  \notag
\\
&&\frac{\Delta}{2}\sum_{k=1}^{n}K_{h_1}^\ast(t_{k}-t)(\xi_{i,k}-\xi_{i,k-1})%
\Delta X_{j,k} +\frac{\Delta}{2}\sum_{k=1}^{n}K_{h_1}^\ast(t_{k}-t)(\xi_{i,k}-%
\xi_{i,k-1})(\xi_{j,k}-\xi_{j,k-1})  \notag \\
&=:&\widehat{\Omega}_{ij,1}(t)+\widehat{\Omega}_{ij,2}(t)+\widehat{\Omega}%
_{ij,3}(t)+\widehat{\Omega}_{ij,4}(t).  \notag
\end{eqnarray}
By Proposition A.1, we have 
\begin{equation}  \label{eqB.35}
\max_{1\leq i,j\leq p}\sup_{0\leq t\leq T}\left\vert\widehat{\Omega}%
_{ij,1}(t)\right\vert=O_P\left(\Delta\right).
\end{equation}
To complete the proof of (A.3), it is sufficient to show 
\begin{equation}  \label{eqB.36}
\max_{1\leq i,j\leq p}\sup_{0\leq t\leq T}\left\vert\widehat{\Omega}%
_{ij,4}(t)-\Omega_{ij}(t)\right\vert=O_P\left(\delta_{\Delta,p}\right).
\end{equation}
In fact, combining (\ref{eqB.35}) and (\ref{eqB.36}), and using the Cauchy-Schwarz
inequality, we 
\begin{equation}  \label{eqB.37}
\max_{1\leq i,j\leq p}\sup_{0\leq t\leq T}\left[\left\vert\widehat{%
\Omega}_{ij,2}(t)\right\vert+\left\vert\widehat{\Omega}_{ij,3}(t)\right\vert%
\right]=O_P\left(\Delta^{1/2}\right).
\end{equation}
By virtue of (\ref{eqB.35})--(\ref{eqB.37}), we readily have (A.3).

It remains to prove (\ref{eqB.36}). We aim to show that 
\begin{eqnarray}
&&\max_{1\leq i,j\leq p}\sup_{0\leq t\leq
T}\left\vert\Delta\sum_{k=1}^{n}K_{h_1}^\ast(t_{k}-t)\xi_{i,k}\xi_{j,k}-%
\Omega_{ij}(t)\right\vert=O_P\left(\delta_{\Delta,p}\right),  \label{eqB.38}
\\
&&\max_{1\leq i,j\leq p}\sup_{0\leq t\leq
T}\left\vert\Delta\sum_{k=1}^{n}K_{h_1}^\ast(t_{k}-t)\xi_{i,k-1}\xi_{j,k-1}-%
\Omega_{ij}(t)\right\vert=O_P\left(\delta_{\Delta,p}\right),  \label{eqB.39}
\\
&&\max_{1\leq i,j\leq p}\sup_{0\leq t\leq T}\left\vert
\Delta\sum_{k=1}^{n}K_{h_1}^\ast(t_{k}-t)\left(\xi_{i,k}\xi_{j,k-1}+\xi_{i,k-1}%
\xi_{j,k}\right)\right\vert=O_P\left(\delta_{\Delta,p}^\ast\right),
\label{eqB.40}
\end{eqnarray}
where $\delta_{\Delta,p}^\ast=\left[\frac{\Delta\log (p\vee \Delta^{-1})}{h_1%
}\right]^{1/2}$. To save the space, we only provide the detailed proof of (%
\ref{eqB.38}) as the proofs of (\ref{eqB.39}) and (\ref{eqB.40}) are similar
(with minor modifications).

Note that 
\begin{eqnarray}
&&\Delta\sum_{k=1}^{n}K_{h_1}^\ast(t_{k}-t)\xi_{i,k}\xi_{j,k}-\Omega_{ij}(t) 
\notag \\
&=&\left\{\Delta\sum_{k=1}^{n}K_{h_1}^\ast(t_{k}-t)\left[\xi_{i,k}\xi_{j,k}-%
\Omega_{ij}(t_k)\right]\right\}
+\left\{\Delta\sum_{k=1}^{n}K_{h_1}^\ast(t_{k}-t)\Omega_{ij}(t_k)-\Omega_{ij}(t)%
\right\}  \notag \\
& =:&\Upsilon_{ij,1}(t)+\Upsilon_{ij,2}(t).  \label{eqB.41}
\end{eqnarray}
Let $\chi_{ij,k}=\xi_{i,k}\xi_{j,k}-\Omega_{ij}(t_k)$, 
\begin{equation*}
\chi_{ij,k}^\star=\chi_{ij,k}I\left( |\chi_{ij,k}|\leq
\Delta^{-\iota_\star}\right) \quad\mathrm{and}\quad\chi_{ij,k}^{\diamond}=%
\chi_{ij,k}-\chi_{ij,k}^{\star}, 
\end{equation*}
where $\iota_\star$ is defined in Assumption 5(iii). Observe that 
\begin{equation}  \label{eqB.42}
\sum_{k=1}^{n}K_{h_1}(t_{k}-t)
\chi_{ij,k}=\sum_{k=1}^{n}K_{h_1}(t_{k}-t)\left[
\chi_{ij,k}^{\star}-\mathsf{E}(\chi_{ij,k}^{\star})\right]
+\sum_{k=1}^{n}K_{h_1}(t_{k}-t)\left[ \chi_{ij,k}^{\diamond}-\mathsf{E}%
(\chi_{ij,k}^{\diamond})\right].
\end{equation}
By Assumptions 3(ii) and 5(i), we have $\mathsf{E}\left[\left|\chi_{ij,k}^{%
\diamond}\right|\right] =O\left( \Delta^{\iota_\star M_{\chi}}\right)$ with $%
M_{\chi}>0$ being arbitrarily large. Then, by Assumption 5(i)(ii) and the
Markov inequality, we have that, for any $\epsilon>0$, 
\begin{eqnarray}
&&\mathsf{P}\left( \max_{1\leq i,j\leq p}\sup_{0\leq t\leq T0}\left|
\Delta\sum_{k=1}^{n}K_{h_1}(t_{k}-t)\left[ \chi_{ij,k}^{\diamond}-\mathsf{E}%
(\chi_{ij,k}^{\diamond})\right]\right| >\epsilon\delta_{\Delta,p}^\ast\right)
\notag \\
&\leq&\mathsf{P}\left( \max_{1\leq i,j\leq p}\sup_{0\leq t\leq
T}\left|
\Delta\sum_{k=1}^{n}K_{h_1}(t_{k}-t)\chi_{ij,k}^{\diamond}\right| >\frac{1}{2%
}\epsilon\delta_{\Delta,p}^\ast\right)  \notag \\
&\leq&\mathsf{P}\left( \max_{1\leq i,j\leq p}\max_{1\leq k\leq n}\left|
\chi_{ij,k}^{\diamond}\right| >0\right)\leq\mathsf{P}\left( \max_{1\leq
i,j\leq p}\max_{1\leq k\leq n}\left| \chi_{ij,k}\right|
>\Delta^{-\iota_\star}\right)  \notag \\
&\leq&\mathsf{P}\left( \max_{1\leq i,j\leq p}\max_{1\leq k\leq
n}\left|\xi_{i,k}\xi_{j,k}\right| >\Delta^{-\iota_\star}-M_\Omega\right) \leq%
\mathsf{P}\left( \max_{1\leq i,j\leq p}\max_{1\leq k\leq n}\left(
\xi_{i,k}^{2}+\xi_{j,k}^{2}\right) >2(\Delta^{-\iota_\star}-M_\Omega)\right)
\notag \\
&\leq&2\mathsf{P}\left( \max_{1\leq i\leq p}\max_{1\leq k\leq
n}\xi_{i,k}^{2}>\Delta^{-\iota_\star}-M_\Omega\right)
\leq2\sum_{i=1}^{p}\sum _{k=1}^{n}\mathsf{P}\left(
\xi_{i,k}^{2}>\Delta^{-\iota_\star}-M_\Omega\right)  \notag \\
&\leq&
2pn\exp\{-sC_\omega^{-1}\left(\Delta^{-\iota_\star}-M_\Omega\right)\}C_\xi^%
\star=o(1)  \label{eqB.43}
\end{eqnarray}
for $0<s<s_{0}$, where $M_\Omega=\max_{1\leq i,j\leq p}\sup_{0\leq t\leq
T}|\Omega_{ij}(t)|\leq C_\omega$, $C_\omega$ is defined in Assumption 3(ii)
and $C_\xi^\star$ is defined in Assumption 5(i).

Cover the closed interval $[0,T]$ by some disjoint intervals $\mathcal{%
T}_{l}^\star$, $l=1,\cdots,V_\star$, with the center $t_{l}^\star$ and
length $d_\star=h_{1}^{2}\delta_{\Delta,p}^\ast \Delta^{\iota_\star}$. By
the Lipschitz continuity of $K(\cdot)$ in Assumption 2(i), we have 
\begin{eqnarray}
& &\max_{1\leq i,j\leq p}\sup_{0\leq t\leq T}\left\vert
\Delta\sum_{k=1}^{n}K_{h_1}(t_{k}-t)\left[ \chi_{ij,k}^{\star}-\mathsf{E}%
(\chi_{ij,k}^{\star})\right] \right\vert  \notag \\
&\leq & \max_{1\leq i,j\leq p}\max_{1\leq l\leq
V_\star}\left\vert\Delta\sum_{k=1}^{n}K_{h_1}(t_{k}-t_l^\star)\left[
\chi_{ij,k}^{\star}-\mathsf{E}(\chi_{ij,k}^{\star})\right] \right\vert + 
\notag \\
&& \max_{1\leq i,j\leq p}\max_{1\leq l\leq V_\star}\sup_{t\in\mathcal{T}%
_l^\star}\left\vert \Delta\sum_{k=1}^{n}\left[
K_{h_1}(t_{k}-t)-K_{h_1}(t_{k}-t_l^\star)\right] \left[ \chi_{ij,k}^{\star}-%
\mathsf{E}(\chi_{ij,k}^{\star})\right] \right\vert  \notag \\
&\leq & \max_{1\leq i,j\leq p}\max_{1\leq l\leq
V_\star}\left\vert\Delta\sum_{k=1}^{n}K_{h_1}(t_{k}-t_l^\star)\left[
\chi_{ij,k}^{\star}-\mathsf{E}(\chi_{ij,k}^{\star})\right] \right\vert + 
\notag \\
& &O\left(\Delta^{-\iota_\star}\right) \max_{1\leq l\leq V_\star}\sup_{t\in%
\mathcal{T}_l^\star}\Delta\sum_{k=1}^{n}\left\vert
K_{h_1}(t_{k}-t)-K_{h_1}(t_{k}-t_l^\star)\right\vert  \notag \\
&\leq & \max_{1\leq i,j\leq p}\max_{1\leq l\leq
V_\star}\left\vert\Delta\sum_{k=1}^{n}K_{h_1}(t_{k}-t_l^\star)\left[
\chi_{ij,k}^{\star}-\mathsf{E}(\chi_{ij,k}^{\star})\right] \right\vert
+O_{P}\left( \delta_{\Delta,p}^\ast\right).  \label{eqB.44}
\end{eqnarray}
On the other hand, by the Bernstein inequality, we may show that 
\begin{eqnarray}
&& \mathsf{P}\left( \max_{1\leq i,j\leq p}\max_{1\leq l\leq
V_\star}\left\vert\Delta\sum_{k=1}^{n}K_{h_1}(t_{k}-t_l^\star)\left[
\chi_{ij,k}^{\star}-\mathsf{E}(\chi_{ij,k}^{\star})\right] \right\vert
>M\delta_{\Delta,p}^\ast\right)  \notag \\
&\leq&\sum_{i=1}^{p}\sum_{j=1}^{p}\sum_{l=1}^{V_\star}\mathsf{P}\left(
\left\vert\Delta\sum_{k=1}^{n}K_{h_1}(t_{k}-t_l^\star)\left[
\chi_{ij,k}^{\star}-\mathsf{E}(\chi_{ij,k}^{\star})\right] \right\vert
>M\delta_{\Delta,p}^\ast\right)  \notag \\
&= &O\left( p^2V_\star \exp\left\{ -C_\diamond(M)\log(p\vee
\Delta^{-1})\right\} \right)=o(1),  \notag
\end{eqnarray}
where $C_\diamond(M)$ is positive and becomes sufficiently large by choosing 
$M$ to be large enough, and $V_\star$ diverges at a polynomial rate of $n$.
Therefore, we have 
\begin{equation}  \label{eqB.45}
\max_{1\leq i,j\leq p}\max_{1\leq l\leq
V_\star}\left\vert\Delta\sum_{k=1}^{n}K_{h_1}(t_{k}-t_l^\star)\left[
\chi_{ij,k}^{\star}-\mathsf{E}(\chi_{ij,k}^{\star})\right] \right\vert
=O_P(\delta_{\Delta,p}^\ast).
\end{equation}
With (\ref{eqB.42})--(\ref{eqB.45}), we can prove that 
\begin{equation}  \label{eqB.46}
\max_{1\leq i,j\leq p}\sup_{0\leq t\leq T}\left\vert
\Upsilon_{ij,1}(t)\right\vert=O_P(\delta_{\Delta,p}^\ast).
\end{equation}

Finally, by the smoothness condition in Assumption 5(ii), we have 
\begin{equation}  \label{eqB.47}
\max_{1\leq i,j\leq p}\sup_{0\leq t\leq T}\left\vert
\Upsilon_{ij,2}(t)\right\vert=O\left(h_1^{\gamma_1}\right).
\end{equation}
By virtue of (\ref{eqB.41}), (\ref{eqB.46}) and (\ref{eqB.47}), we complete
the proof of (\ref{eqB.38}).\hfill$\blacksquare$}

\smallskip

\noindent{\bf Proof of Proposition A.4}.\ \ With Assumption 6 replacing Assumption 1, proofs of the uniform convergence results in Proposition A.4 are the same as the proof of Proposition A.1. Details are omitted here to save the space.\hfill$\blacksquare$


\section*{Appendix C: Further discussion and extension}
\renewcommand{\theequation}{C.\arabic{equation}} \setcounter{equation}{0}

{\small In this appendix, we discuss estimation of the spot precision matrix and address the asynchronicity issue which is common when multiple asset returns are collected.

\subsection*{Appendix C.1: Estimation of the spot precision matrix}

The spot precision matrix of high-frequency data defined as inverse of the
spot volatility matrix, plays an important role in dynamic optimal portfolio
choice. In the low-frequency data setting, estimation of large precision
matrices has been extensively studied in the literature and various
estimation techniques such as penalised likelihood \citep{LF09}, graphical
Danzig selector \citep{Y10} and CLIME \citep{CLL11} have been introduced. In
the high-frequency data setting, \cite{CHLZ20} estimate the precision matrix
defined as inverse of the integrated volatility matrix, derive the relevant
asymptotic properties under various scenarios and apply the estimated
precision matrix to minimum variance portfolio estimation. We next consider estimating the large spot precision matrix under a uniform
sparsity assumption which is different from (2.3). Specifically, assume that model (3.1) holds and
that the large spot precision matrix ${\boldsymbol{\Lambda}}_t:={\boldsymbol{%
\Sigma}}_t^{-1}$ satisfies $\left\{{\boldsymbol{\Lambda}}_t: 0\leq t\leq
T\right\}\in\mathcal{S}_\ast(q, \varpi_\ast(p),T)$, where
\[
\mathcal{S}_\ast(q, \varpi_\ast(p),T)=\left\{{\boldsymbol{\Lambda}}_t=\left[\Lambda_{ij,t}\right]_{p\times p}, t\in%
[0,T]\ \big|\ {\boldsymbol{\Lambda}}_t\succ0,\ \sup_{0\leq t\leq T}\Vert{%
\boldsymbol{\Lambda}}_t\Vert_1\leq C_\Lambda,\ \sup_{0\leq t\leq T}\Vert{%
\boldsymbol{\Lambda}}_t\Vert_{\infty,q}\le \varpi_\ast(p)\right\},
\]
where ``${\boldsymbol{\Lambda}}\succ0$" denotes that ${\boldsymbol{\Lambda}}$
is positive definite and $C_\Lambda$ is a positive constant.

\smallskip

We next apply \cite{CLL11}'s constrained $\ell_1$ minimisation or CLIME
method to estimate the spot precision matrix ${\boldsymbol{\Lambda}}_t$. The
estimate is defined as 
\begin{equation*}
\widetilde{\boldsymbol{\Lambda}}_{t}=\argmin_{\boldsymbol{\Lambda}}\vert{%
\boldsymbol{\Lambda}}\vert_1\ \ \ \ \ \mathrm{subject\ to}\ \ \left\Vert 
\widetilde{\boldsymbol{\Sigma}}_t{\boldsymbol{\Lambda}}-{\mathbf{I}}%
_p\right\Vert_{\mathrm{max}}\leq \rho_4(t), 
\end{equation*}
where $\widetilde{\boldsymbol{\Sigma}}_t=\left(\widetilde{\Sigma}%
_{ij,t}\right)_{p\times p}$ with $\widetilde{\Sigma}_{ij,t}$ defined in (3.3), ${\mathbf{I}}_p$ is a $p\times p$ identity matrix, and $\rho_4(t)$
is a time-varying tuning parameter. The final CLIME estimate of ${%
\boldsymbol{\Lambda}}_t$ is obtained by further symmetrising $\widetilde{%
\boldsymbol{\Lambda}}_{t}$. Suppose that Assumptions 1, 2(i), 3 and 4(i)(ii)
are satisfied and Assumption 4(iii) holds with $\rho_2(t)$ replaced by $%
\rho_4(t)$. Using Proposition A.2 in Appendix A and following the proof of
Theorem 6 in \cite{CLL11}, we may show that 
\[
\sup_{0\leq t\leq T}\left\Vert\widetilde{\boldsymbol{\Lambda}}_{t}-{%
\boldsymbol{\Lambda}}_t\right\Vert=O_P\left(\varpi_\ast(p) \left[%
\zeta_{N,p}^\ast+\nu_{\Delta,p,N}\right]^{1-q}\right).
\]

\subsection*{Appendix C.2: The asynchronicity issue}

In the main text of the paper, we consider a special sampling
scheme: the high-frequency data are synchronised with equally spaced time
points between $0$ and $T$. Such a setting simplifies exposition and
facilitates proofs of the uniform consistency properties. However, in
practice, it is often the case that a large number of assets are traded at
times that are not synchronised. This may induce volatility matrix
estimation bias and possibly result in the so-called Epps effect %
\citep[e.g.,][]{Ep79}. We next deal with the asynchronicity problem and
discuss modifications of the estimation techniques and theory developed in
the previous sections.

\smallskip

Assume that the $i$-th asset price is collected at $t_1^i,\cdots,t_{n_i}^i$,
which are non-equidistant time points over $[0,T]$. To address this
asynchronicity issue, we may adopt a synchronisation scheme before
implementing the large spot volatility matrix estimation method proposed in
the main text. Commonly-used synchronisation schemes
include the generalised sampling time \citep{AFX10}, refresh time %
\citep{BHLS11} and previous tick \citep{Zh11}. We next propose an
alternative technique by slightly amending the localised pre-averaging
estimation in (3.2) to jointly tackle the asynchronicity and noise
contamination issues. Replace the kernel filter in (3.2) by 
\[
\widetilde{\mathbf{X}}_\tau^\ast=\left(\widetilde{X}_{1,\tau}^\ast,\cdots, 
\widetilde{X}_{p,\tau}^\ast\right)^{^\intercal}\ \ \mathrm{with}\ \ 
\widetilde{X}_{i,\tau}^\ast=\sum_{k=1}^{n_i}
L_b(t_k^i-\tau)Z_{i,t_k^i}\left(t_k^i-t_{k-1}^i\right),
\]
and then use $\widetilde{\mathbf{X}}_\tau^\ast$ in the kernel smoothing (3.3). Some mild restrictions need to be imposed on the 
data collection times. For example, let $t_j^i-t_{j-1}^i=c_{j}^in_i^{-1}$, where 
\[0<\underline c\leq\min_{1\leq i\leq p}\min_{1\leq j\leq n_i}c_{j}^i\leq \max_{1\leq i\leq p}\max_{1\leq j\leq n_i}c_{j}^i\leq \overline{c}<\infty,\]
and there exists a $\kappa_0>0$ such that $N=O\left(\underline{n}^{\kappa_0}\right)$ with $\underline{n}=\min_{1\leq i\leq p}n_i$. Following the proof of Lemma B.1, we may show that
\[
\max_{1\leq i\leq p}\max_{0\leq l\leq N}\left\vert \widetilde{X}%
_{i,\tau_l}-X_{i,\tau_l}\right\vert=O_P\left(\sqrt{\log(p\vee \underline n)}%
\left[b^{1/2}+\left(\underline {n}b\right)^{-1/2}\right]\right).
\]
Then, following the proofs of Proposition A.2 and Theorem 2, we may prove a similar uniform convergence rate to (3.5) but with $\nu_{\Delta,p,N}$ replaced by $\sqrt{N\log(p\vee \underline{n})}\left[b^{1/2}+(\underline{n}b)^{-1/2}\right]$.

\smallskip

The time-varying noise covariance matrix estimation also needs to be
modified when large high-frequency data are non-synchronised. As in \cite%
{CHLT21}, we let $\mathcal{T}_i=\left\{t_1^i,t_2^i,\cdots,t_{n_i}^i\right\}$
be the set of time points at which we observe the contaminated asset prices,
and denote 
\begin{equation*}
\mathcal{T}_{ij}=\mathcal{T}_i\cap\mathcal{T}_j=\left\{t_1^{ij},t_2^{ij},%
\cdots,t_{n_{ij}}^{ij}\right\},
\end{equation*}
where $n_{ij}$ is the cardinality of $\mathcal{T}_{ij}$. Then, we modify the
kernel estimate in (3.7) as follows, 
\begin{equation*}
\widetilde\Omega_{ij}(t)=\frac{1}{2}\sum_{k=1}^{n_{ij}}K_{h_1}%
\left(t_k^{ij}-t\right)\Delta Z_{i,t_k^{ij}}\Delta Z_{j,
t_k^{ij}}\left(t_k^{ij}-t_{k-1}^{ij}\right), 
\end{equation*}
where $t_0^{ij}=0$. In contrast to $\widehat\Omega_{ij}(t)$, $t_k$, $%
Z_{i,t_k}$ and $\Delta$ in (3.7) are now replaced by $t_k^{ij}$, $%
Z_{i,t_k^{ij}}$ and $t_k^{ij}-t_{k-1}^{ij}$, respectively. We subsequently
apply the shrinkage to $\widetilde\Omega_{ij}(t)$ when $i\neq j$ and obtain
the final estimate of ${\boldsymbol{\Omega}}(t)$. Assuming $\max_{1\leq
i,j\leq p}\max_{1\leq k\leq
n_{ij}}\left(t_{k}^{ij}-t_{k-1}^{ij}\right)\rightarrow0$ and letting $%
n_\circ=\min_{1\leq i,j\leq p}n_{ij}$, we may similarly derive the uniform
consistency property as in (3.9) but with $\Delta$ replaced by $%
n_\circ^{-1}$. 

}


\section*{Appendix D: Additional simulation results}
\renewcommand{\theequation}{D.\arabic{equation}} \setcounter{equation}{0}

{\small In this appendix, we first consider the asynchronous high-frequency data using the technique discussed in Appendix C.2. We use the same simulation setup as in Section 5.1.1. To generate the asynchronous data, we follow \cite{WZ10} by randomly deleting 2 observations from every consecutive block of 3 synchronous 15-second observations. Consequently, the average number of asynchronous observations for each asset is equal to one third of the number of synchronous observations. The number of assets is set as $p=200$ and $500$ and the replication number is $R=200$. We consider the following two volatility matrix estimates.

\begin{itemize}

\item Noise-contaminated spot volatility matrix estimate $\widetilde{\boldsymbol\Sigma}_{t}^\ast$, extending $\widetilde{\boldsymbol\Sigma}_{t}$ defined in Section 3.1 to the asynchronous high-frequency data with the modification technique introduced in Appendix C.2.

\item Time-varying noise volatility matrix estimate $\widehat{\boldsymbol\Omega}^\ast(t)$, extending $\widehat{\boldsymbol\Omega}(t)$ defined in Section 3.2 to the asynchronous high-frequency data with the modification technique introduced in Appendix C.2.

\end{itemize}

As in Section 5.1, we compute the Mean Frobenius Loss (MFL) and Mean Spectral Loss (MSL) over $200$ repetitions for the estimated volatility matrices (under the sparsity restriction). Tables D.1 and D.2 report the simulation results when $p=200$ and $p=500$, respectively. As shown in Section 5.1.3, the shrinkage volatility matrix estimation significantly outperforms the naive estimation. Comparing with Tables 1 and 2 in the main document, we note that the finite-sample convergence is slowed down when the high-frequency data are not synchronised. 

\medskip

\begin{center}
{\footnotesize Table D.1: Simulation results of the volatility matrix estimation for asynchronous data when $p=200$ }

{\footnotesize 
\begin{tabular}{lllllllclllll}
\hline\hline
&  & \multicolumn{11}{c}{``Banding"} \\ \cline{3-7}\cline{9-13}
&  & {\ Naive} & {\ Hard} & {\ Soft} & {\ AL} & {\ SCAD} & \multicolumn{1}{l}{
} & {\ Naive} & {\ Hard} & {\ Soft} & {\ AL} & {\ SCAD} \\ 
\cline{3-7}\cline{9-13}
$\widetilde{\boldsymbol\Sigma}_{t}^{\ast }$ & {\ MFL} & \multicolumn{1}{r}{21.180} & 
\multicolumn{1}{r}{13.234} & \multicolumn{1}{r}{13.723} & \multicolumn{1}{r}{
13.392} & \multicolumn{1}{r}{13.768} & \multicolumn{1}{l}{MSL} & 
\multicolumn{1}{r}{6.174} & \multicolumn{1}{r}{2.375} & \multicolumn{1}{r}{
2.458} & \multicolumn{1}{r}{2.385} & \multicolumn{1}{r}{2.474} \\ 
$\widehat{\boldsymbol\Omega}^{\ast}(t)$ & {\ MFL} & \multicolumn{1}{r}{38.072} & 
\multicolumn{1}{r}{4.640} & \multicolumn{1}{r}{4.647} & \multicolumn{1}{r}{
4.640} & \multicolumn{1}{r}{4.646} & \multicolumn{1}{l}{MSL} & 
\multicolumn{1}{r}{6.624} & \multicolumn{1}{r}{0.663} & \multicolumn{1}{r}{
0.666} & \multicolumn{1}{r}{0.663} & \multicolumn{1}{r}{0.665} \\  \hline
&  & \multicolumn{11}{c}{``Block-diagonal"} \\ \cline{3-7}\cline{9-13}
&  & {\ Naive} & {\ Hard} & {\ Soft} & {\ AL} & {\ SCAD} & \multicolumn{1}{l}{
} & {\ Naive} & {\ Hard} & {\ Soft} & {\ AL} & {\ SCAD} \\ 
\cline{3-7}\cline{9-13}
$\widetilde{\boldsymbol\Sigma}_{t}^{\ast }$ & {\ MFL} & \multicolumn{1}{r}{21.143} & 
\multicolumn{1}{r}{13.141} & \multicolumn{1}{r}{13.648} & \multicolumn{1}{r}{
13.310} & \multicolumn{1}{r}{13.693} & \multicolumn{1}{l}{MSL} & 
\multicolumn{1}{r}{6.275} & \multicolumn{1}{r}{2.805} & \multicolumn{1}{r}{
2.821} & \multicolumn{1}{r}{2.804} & \multicolumn{1}{r}{2.827} \\ 
$\widehat{\boldsymbol\Omega}^{\ast}(t)$ & {\ MFL} & \multicolumn{1}{r}{38.066} & 
\multicolumn{1}{r}{4.520} & \multicolumn{1}{r}{4.528} & \multicolumn{1}{r}{
4.520} & \multicolumn{1}{r}{4.526} & \multicolumn{1}{l}{MSL} & 
\multicolumn{1}{r}{6.634} & \multicolumn{1}{r}{0.736} & \multicolumn{1}{r}{
0.738} & \multicolumn{1}{r}{0.736} & \multicolumn{1}{r}{0.737} \\  \hline
&  & \multicolumn{11}{c}{``Exponentially decaying"} \\ \cline{3-7}\cline{9-13}
&  & {\ Naive} & {\ Hard} & {\ Soft} & {\ AL} & {\ SCAD} & \multicolumn{1}{l}{
} & {\ Naive} & {\ Hard} & {\ Soft} & {\ AL} & {\ SCAD} \\ 
\cline{3-7}\cline{9-13}
$\widetilde{\boldsymbol\Sigma}_{t}^{\ast }$ & {\ MFL} & \multicolumn{1}{r}{21.454} & 
\multicolumn{1}{r}{13.772} & \multicolumn{1}{r}{14.217} & \multicolumn{1}{r}{
13.914} & \multicolumn{1}{r}{14.258} & \multicolumn{1}{l}{MSL} & 
\multicolumn{1}{r}{6.313} & \multicolumn{1}{r}{2.961} & \multicolumn{1}{r}{
2.968} & \multicolumn{1}{r}{2.958} & \multicolumn{1}{r}{2.972} \\ 
$\widehat{\boldsymbol\Omega}^{\ast}(t)$ & {\ MFL} & \multicolumn{1}{r}{38.098} & 
\multicolumn{1}{r}{4.716} & \multicolumn{1}{r}{4.723} & \multicolumn{1}{r}{
4.717} & \multicolumn{1}{r}{4.722} & \multicolumn{1}{l}{MSL} & 
\multicolumn{1}{r}{6.672} & \multicolumn{1}{r}{0.762} & \multicolumn{1}{r}{
0.764} & \multicolumn{1}{r}{0.762} & \multicolumn{1}{r}{0.764} \\ \hline\hline
\end{tabular}
}
\end{center}

\noindent{\scriptsize The selected bandwidths are $h^{\ast}=90$ and $b^{\ast}=4$ for $\widetilde{\boldsymbol\Sigma}_{t}^{\ast}$ and $h_1^{\ast}=250$ for $\widehat{\boldsymbol\Omega}^{\ast}(t)$, where $h^{\ast}=h/\Delta$, $b^{\ast}=b/\Delta$, and $h_{1}^{\ast}=h_1/\Delta$.}

\begin{center}

{\footnotesize Table D.2: Simulation results of the volatility matrix estimation for asynchronous data when $p=500$ }

{\footnotesize 
\begin{tabular}{lllllllllllll}
\hline\hline
&  & \multicolumn{11}{c}{``Banding"} \\ \cline{3-7}\cline{9-13}
&  & {\ Naive} & {\ Hard} & {\ Soft} & {\ AL} & {\ SCAD} &  & {\ Naive} & {\
Hard} & {\ Soft} & {\ AL} & {\ SCAD} \\ \cline{3-7}\cline{9-13}
$\widetilde{\boldsymbol\Sigma}_{t}^{\ast }$ & {\ MFL} & \multicolumn{1}{r}{32.710} & 
\multicolumn{1}{r}{20.656} & \multicolumn{1}{r}{20.445} & \multicolumn{1}{r}{
20.600} & \multicolumn{1}{r}{20.445} & {\ MSL} & \multicolumn{1}{r}{6.212} & 
\multicolumn{1}{r}{2.440} & \multicolumn{1}{r}{2.427} & \multicolumn{1}{r}{
2.430} & \multicolumn{1}{r}{2.427} \\ 
$\widehat{\boldsymbol\Omega}^{\ast}(t)$ & {\ MFL} & \multicolumn{1}{r}{93.263} & 
\multicolumn{1}{r}{7.348} & \multicolumn{1}{r}{7.348} & \multicolumn{1}{r}{
7.348} & \multicolumn{1}{r}{7.348} & {\ MSL} & \multicolumn{1}{r}{10.724} & 
\multicolumn{1}{r}{0.681} & \multicolumn{1}{r}{0.681} & \multicolumn{1}{r}{
0.681} & \multicolumn{1}{r}{0.681} \\  \hline
&  & \multicolumn{11}{c}{``Block-diagonal"} \\  \cline{3-7}\cline{9-13}
&  & {\ Naive} & {\ Hard} & {\ Soft} & {\ AL} & {\ SCAD} &  & {\ Naive} & {\
Hard} & {\ Soft} & {\ AL} & {\ SCAD} \\ \cline{3-7}\cline{9-13}
$\widetilde{\boldsymbol\Sigma}^{\ast}_t$ & {\ MFL} & \multicolumn{1}{r}{32.928} & 
\multicolumn{1}{r}{21.080} & \multicolumn{1}{r}{20.873} & \multicolumn{1}{r}{
21.026} & \multicolumn{1}{r}{20.873} & {\ MSL} & \multicolumn{1}{r}{6.330} & 
\multicolumn{1}{r}{2.962} & \multicolumn{1}{r}{2.951} & \multicolumn{1}{r}{
2.948} & \multicolumn{1}{r}{2.950} \\ 
$\widehat{\boldsymbol\Omega}^{\ast}(t)$ & {\ MFL} & \multicolumn{1}{r}{93.281} & 
\multicolumn{1}{r}{7.331} & \multicolumn{1}{r}{7.331} & \multicolumn{1}{r}{
7.331} & \multicolumn{1}{r}{7.331} & {\ MSL} & \multicolumn{1}{r}{10.759} & 
\multicolumn{1}{r}{0.773} & \multicolumn{1}{r}{0.773} & \multicolumn{1}{r}{
0.773} & \multicolumn{1}{r}{0.773} \\  \hline
&  & \multicolumn{11}{c}{``Exponentially decaying"} \\  \cline{3-7}\cline{9-13}
&  & {\ Naive} & {\ Hard} & {\ Soft} & {\ AL} & {\ SCAD} &  & {\ Naive} & {\
Hard} & {\ Soft} & {\ AL} & {\ SCAD} \\ \cline{3-7}\cline{9-13}
$\widetilde{\boldsymbol\Sigma}_{t}^{\ast }$ & {\ MFL} & \multicolumn{1}{r}{33.153} & 
\multicolumn{1}{r}{21.524} & \multicolumn{1}{r}{21.371} & \multicolumn{1}{r}{
21.459} & \multicolumn{1}{r}{21.317} & {\ MSL} & \multicolumn{1}{r}{6.341} & 
\multicolumn{1}{r}{3.015} & \multicolumn{1}{r}{3.003} & \multicolumn{1}{r}{
3.001} & \multicolumn{1}{r}{3.003} \\ 
$\widehat{\boldsymbol\Omega}^{\ast}(t)$ & {\ MFL} & \multicolumn{1}{r}{93.287} & 
\multicolumn{1}{r}{7.469} & \multicolumn{1}{r}{7.469} & \multicolumn{1}{r}{
7.469} & \multicolumn{1}{r}{7.469} & {\ MSL} & \multicolumn{1}{r}{10.783} & 
\multicolumn{1}{r}{0.781} & \multicolumn{1}{r}{0.781} & \multicolumn{1}{r}{
0.781} & \multicolumn{1}{r}{0.781} \\  \hline\hline
\end{tabular}
}
\end{center}

\noindent{\scriptsize The selected bandwidths are $h^{\ast }=240$, $b^{\ast }=6$
for $\widetilde{\boldsymbol\Sigma}_{t}^{\ast }$ and $h_1^{\ast}=260$ for $\widehat{\boldsymbol\Omega}^{\ast}(t)$, where $h^{\ast}=h/\Delta$, $b^{\ast}=b/\Delta$, and $h_{1}^{\ast}=h_1/\Delta$.}

\medskip

We next consider estimating the integrated volatility matrix (with normalisation) of the $p$-variate Brownian semi-martingale process $\mathbf{X}_{t}=\left(X_{1,t},X_{2,t},\ldots ,X_{p,t}\right)^{^\intercal}$ over the time interval ${\cal T}$ using high-frequency observations under the sparsity assumption. Define
\begin{equation}
{\boldsymbol\Sigma} _{\cal T}=\left(\Sigma _{{\cal T}, ij}\right)
_{p\times p}=\frac{1}{|{\cal T}|}\int_{\cal T}{\boldsymbol\Sigma}_{t}dt=\frac{1}{|{\cal T}|}\int_{\cal T}\left(\Sigma _{t,ij}\right) _{p\times p}dt,  \label{integrated}
\end{equation}%
where $|{\cal T}|$ denotes the length of ${\cal T}$. Let ${\cal T}=[0,T]$. We use the following two methods to estimate ${\boldsymbol\Sigma} _{\cal T}$ and compare their performance. The first one is the sample analog of the quadratic variation (realised volatility matrix) with shrinkage \citep[e.g.,][]{DLX19}:
\begin{equation}\label{inte-1}
\widehat{\boldsymbol\Sigma}_{\cal T}=\left( \widehat{\Sigma}_{{\cal T}, ij}^{s}\right) _{p\times p}\ \ \text{with }\widehat{\Sigma}_{{\cal T}, ij}^{s}=s_{\rho_\ast}\left( \widehat{\Sigma}_{{\cal T}, ij}\right) I\left( i\neq j\right) +\widehat{\Sigma}_{{\cal T}, ii}I\left( i=j\right), 
\end{equation}%
where $\rho_\ast$ is a user-specified tuning parameter and 
\begin{equation*}
\widehat{\Sigma}_{{\cal T},ij}=\frac{1}{T}\sum_{k=1}^{n}\Delta X_{i,k}\Delta X_{j,k},\quad 1\leq i,j\leq p
\end{equation*}%
with $n=T/\Delta $. Note that the shrinkage is applied to the off-diagonal entries of the estimated integrated matrix which is obtained by summing over the outer product of the $p$-dimensional vector of discrete observations of $\Delta \mathbf{X}$ observed over ${\cal T}=[0,T]$. The second method is to utilise the proposed kernel-weighted spot volatility matrix estimate with shrinkage, i.e.,
\begin{equation}\label{inte-2}
\widehat{\boldsymbol\Sigma}_{{\cal T}}^{\dagger}=\left( \widehat{\Sigma}_{{\cal T}, ij}^\dagger\right) _{p\times p}\quad{\rm with}\quad \widehat{\Sigma}_{{\cal T}, ij}^\dagger=\frac{1}{n}\sum_{k=1}^{n}\widehat{\Sigma}%
_{ij,k\Delta}^{\dagger}
\end{equation}%
where $\widehat{\Sigma}_{ij, k\Delta}^{\dagger}=s_{\rho\left( k\Delta \right) }\left( \widehat{\Sigma}_{ij,k\Delta }\right) I\left(i\neq j\right) +\widehat{\Sigma}_{ii,k\Delta } I\left( i=j\right)$, which is the spot volatility estimate defined in (2.5) of the main text.

\smallskip

We use the same simulation setting as in Section 5.1 of the main text. For simplicity, we only consider the noise-free scenario and $p=500$. We compute the estimation of the integrated covariance matrices ${\boldsymbol\Sigma}_{{\cal T}_j}$ over $20$ equal-length time intervals ${\cal T}_j=\left[ ( j-1)T_\dagger,\  jT_\dagger\right]$, $j=1,2,\cdots ,20$, where $T_\dagger=T/20$ and $T=1/252$. In fact, these intervals are separated by the equidistant time points $t_j$ defined in Section 5.1.2 for assessing the spot volatility matrix estimation. To measure the performance, we define 
\begin{eqnarray*}
\text{MFL}(\widehat{\boldsymbol\Sigma}_{\cal T}) &=&\frac{1}{200}\sum_{m=1}^{200}\left( \frac{1}{20}%
\sum_{j=1}^{20}\left\Vert \widehat{\boldsymbol\Sigma}_{{\cal T}_j}^{(m)}-{\boldsymbol\Sigma}_{{\cal T}_j}^{(m)}\right\Vert _{F}\right) , \\
\text{MSL}(\widehat{\boldsymbol\Sigma}_{\cal T}) &=&\frac{1}{200}\sum_{m=1}^{200}\left( \frac{1}{20}%
\sum_{j=1}^{20}\left\Vert \widehat{\boldsymbol\Sigma}_{{\cal T}_j}^{(m)}-{\boldsymbol\Sigma}_{{\cal T}_j}^{(m)}\right\Vert \right) ,
\end{eqnarray*}%
where $\widehat{\boldsymbol\Sigma}_{{\cal T}_j}^{(m)}$ and ${\boldsymbol\Sigma}_{{\cal T}_j}^{(m)}$ denote the estimated and true integrated volatility matrices in the $m$-th replication. We can similarly define $\text{MFL}(\widehat{\boldsymbol\Sigma}_{\cal T}^\dagger)$ and $\text{MSL}(\widehat{\boldsymbol\Sigma}_{\cal T}^\dagger)$.


\smallskip

As in Section 5.1.3, we consider the four shrinkage methods together with the naive method which does not impose shrinkage. The simulation results are reported in Table D.3. As shown in the previous simulation results, the application of shrinkage substantially improves the estimation accuracy by reducing the MFL and MSL significantly. We note that the integrated volatility matrix estimation defined in (\ref{inte-2}) based on kernel-weighted spot volatility outperforms the standard estimation defined in (\ref{inte-1}) uniformly across the four shrinkage methods and the naive method. This may be partly due to the fact that the standard
integrated volatility matrix estimation (\ref{inte-1}) uses the outer product of only one observation of the $p$-variate vector $\Delta \mathbf{X}$ as an estimate of the integrand in (\ref{integrated}), whereas the estimation (\ref{inte-2}) based on the kernel-weighted spot volatility approximates the integrand by utilising a local sample of size $nh$. Meanwhile, the application of shrinkage to the estimated spot volatility effectively removes small off-diagonal elements in the integrand before calculating the integral.

\begin{center}
\bigskip

Table D.3: Estimation results for the noise-free integrated volatility matrices when $p=500$

{\footnotesize 
\begin{tabular}{llllllclllll}
\hline\hline
 & \multicolumn{11}{c}{\textquotedblleft Banding"} \\ \cline{2-12}
 & \multicolumn{5}{c}{Frobenius Norm} &  & \multicolumn{5}{c}{Spectral Norm
} \\ \cline{2-6}\cline{8-12}
 & \ Naive & \ Hard & \ Soft & \ AL & \ SCAD & \multicolumn{1}{l}{} & \
Naive & \ Hard & \ Soft & \ AL & \ SCAD \\ \cline{2-6}\cline{8-12}
$\text{MFL}(\widehat{\boldsymbol\Sigma}_{\cal T})$ & 
\multicolumn{1}{r}{49.8422} & \multicolumn{1}{r}{18.2364} & 
\multicolumn{1}{r}{11.6644} & \multicolumn{1}{r}{10.1367} & 
\multicolumn{1}{r}{12.0719} & \multicolumn{1}{l}{$\text{MSL}(\widehat{\boldsymbol\Sigma}_{\cal T})$} & \multicolumn{1}{r}{
11.5148} & \multicolumn{1}{r}{1.8991} & \multicolumn{1}{r}{1.4442} & 
\multicolumn{1}{r}{1.3495} & \multicolumn{1}{r}{1.5055} \\ 
$\text{MFL}(\widehat{\boldsymbol\Sigma}_{\cal T}^\dagger)$ & 
\multicolumn{1}{r}{21.7907} & \multicolumn{1}{r}{3.7201} & 
\multicolumn{1}{r}{5.1311} & \multicolumn{1}{r}{4.8468} & \multicolumn{1}{r}{
3.8476} & \multicolumn{1}{l}{$\text{MSL}(\widehat{\boldsymbol\Sigma}_{\cal T}^\dagger)$} & \multicolumn{1}{r}{3.8747} & 
\multicolumn{1}{r}{0.5858} & \multicolumn{1}{r}{0.7116} & \multicolumn{1}{r}{
0.6946} & \multicolumn{1}{r}{0.5593} \\ \hline
 & \multicolumn{11}{c}{\textquotedblleft Block-diagonal"} \\ \cline{2-12}
 & \multicolumn{5}{c}{Frobenius Norm} &  & \multicolumn{5}{c}{Spectral Norm
} \\ \cline{2-6}\cline{8-12}
 & \ Naive & \ Hard & \ Soft & \ AL & \ SCAD & \multicolumn{1}{l}{} & \
Naive & \ Hard & \ Soft & \ AL & \ SCAD \\ \cline{2-6}\cline{8-12}
$\text{MFL}(\widehat{\boldsymbol\Sigma}_{\cal T}) $ & 
\multicolumn{1}{r}{49.8412} & \multicolumn{1}{r}{18.6536} & 
\multicolumn{1}{r}{12.5082} & \multicolumn{1}{r}{11.2693} & 
\multicolumn{1}{r}{12.9696} & \multicolumn{1}{l}{$\text{MSL}(\widehat{\boldsymbol\Sigma}_{\cal T})$} & \multicolumn{1}{r}{
11.6577} & \multicolumn{1}{r}{2.4827} & \multicolumn{1}{r}{1.8710} & 
\multicolumn{1}{r}{1.8116} & \multicolumn{1}{r}{2.0165} \\ 
$\text{MFL}(\widehat{\boldsymbol\Sigma}_{\cal T}^\dagger)$ & 
\multicolumn{1}{r}{21.7919} & \multicolumn{1}{r}{5.5847} & 
\multicolumn{1}{r}{6.3739} & \multicolumn{1}{r}{5.8120} & \multicolumn{1}{r}{
5.4097} & \multicolumn{1}{l}{$\text{MSL}(\widehat{\boldsymbol\Sigma}_{\cal T}^\dagger)$} & \multicolumn{1}{r}{3.9641} & 
\multicolumn{1}{r}{0.8504} & \multicolumn{1}{r}{1.1295} & \multicolumn{1}{r}{
0.8869} & \multicolumn{1}{r}{0.8801} \\ \hline
& \multicolumn{11}{c}{\textquotedblleft Exponentially decaying"} \\ 
\cline{2-12}
 & \multicolumn{5}{c}{Frobenius Norm} & \multicolumn{1}{l}{} & 
\multicolumn{5}{c}{Spectral Norm} \\ \cline{2-6}\cline{8-12}
& \ Naive & \ Hard & \ Soft & \ AL & \ SCAD & \multicolumn{1}{l}{} & \
Naive & \ Hard & \ Soft & \ AL & \ SCAD \\ \cline{2-6}\cline{8-12}
$\text{MFL}(\widehat{\boldsymbol\Sigma}_{\cal T}) $ & 
\multicolumn{1}{r}{49.8468} & \multicolumn{1}{r}{19.2167} & 
\multicolumn{1}{r}{12.8439} & \multicolumn{1}{r}{11.7122} & 
\multicolumn{1}{r}{13.4447} & \multicolumn{1}{l}{$\text{MSL}(\widehat{\boldsymbol\Sigma}_{\cal T})$} & \multicolumn{1}{r}{
11.7491} & \multicolumn{1}{r}{2.5405} & \multicolumn{1}{r}{1.9241} & 
\multicolumn{1}{r}{1.8653} & \multicolumn{1}{r}{2.0720} \\ 
$\text{MFL}(\widehat{\boldsymbol\Sigma}_{\cal T}^\dagger)$ & 
\multicolumn{1}{r}{21.7936} & \multicolumn{1}{r}{5.9783} & 
\multicolumn{1}{r}{6.6726} & \multicolumn{1}{r}{6.0284} & \multicolumn{1}{r}{
5.7007} & \multicolumn{1}{l}{$\text{MSL}(\widehat{\boldsymbol\Sigma}_{\cal T}^\dagger)$} & \multicolumn{1}{r}{4.0020} & 
\multicolumn{1}{r}{0.8927} & \multicolumn{1}{r}{1.1733} & \multicolumn{1}{r}{
0.9228} & \multicolumn{1}{r}{0.9193} \\ \hline\hline
\end{tabular}%
}
\end{center}

}



\end{document}